\documentclass[a4paper,fleqn,usenatbib]{mnras}

\usepackage{txfonts}

\usepackage[T1]{fontenc}
\usepackage{ae,aecompl}

\usepackage{graphicx}	
\usepackage{amstext}	
\usepackage{amssymb}	
 
\title[Probing IGM properties using GRB]{Probing the physical properties of the intergalactic medium using gamma-ray bursts}

\author[T. Dalton et al]{
Tony Dalton,$^{1}$\thanks{E-mail:tonydalton@live.ie}
Simon L. Morris,$^{1}$
and Michele Fumagalli$^{2}$\\
$^{1}$Centre for Extragalactic Astronomy, Durham University, South Road, Durham DH1 3LE, UK\\
$^{2}$Dipartimento di Fisica `G. Occhialini', Universit\`a degli Studi di Milano-Bicocca, Piazza della Scienza 3, I-20126 Milano, Italy\\
}

\date{Accepted 2021 February 02. Received 2021 January 29; in original form 2020 December 14}

\pubyear{2015}

\begin{document}
\label{firstpage}
\pagerange{\pageref{firstpage}--\pageref{lastpage}}
\maketitle

\begin{abstract}
We use Gamma-ray burst (GRB) spectra total continuum absorption to estimate  the  key intergalactic medium (IGM) properties of hydrogen column density ($\mathit{N}\textsc{hxigm}$), metallicity, temperature and ionisation parameter over a redshift range of $1.6 \leq z \leq 6.3$,  using photo-ionisation (PIE) and collisional ionisation equilibrium (CIE)  models for the ionised plasma.  We use more realistic host metallicity, dust corrected where available, in generating the host absorption model, assuming that the host intrinsic hydrogen column density is equal to the measured ionisation corrected intrinsic neutral column from UV spectra ($\textit{N}\textsc{h}\/\ \textsc{i,ic}$). We find that the IGM property results are similar, regardless of whether the model assumes all PIE or CIE. The $\mathit{N}\textsc{hxigm}$ scales as $(1 + z)^{1.0\/\ -\/\ 1.9}$, with equivalent hydrogen mean density at $z = 0$ of $n_0 = 1.8^{+1.5}_{-1.2} \times 10^{-7}$ cm$^{-3}$. The metallicity ranges from $\sim0.1Z\sun$ at $z \sim 2$ to $\sim0.001Z\sun$ at redshift $z > 4$. The PIE model implies a less rapid decline in average metallicity with redshift compared to CIE. Under CIE, the temperature ranges between $5.0 <$ log$(T/$K$)<\/\ 7.1$. For PIE the ionisation parameter ranges between $0.1 <$ log$(\xi) < 2.9$. Using our  model, we conclude that the IGM contributes substantially to the total absorption seen in GRB spectra and that this contribution rises with redshift, explaining why the hydrogen column density inferred from X-rays is substantially in excess of the intrinsic host contribution measured in UV.
\end{abstract}




\section{Introduction}\label{sec:1}
The main objective of this paper is to estimate the key IGM parameters of column density, metallicity, temperature and ionisation, using the latest models for ionised absorbers on the line of sight (LOS) to GRBs. We examine past observations and simulations to set the parameters ranges and priors for our models. Our hypothesis is that the bulk of the excess in observed hydrogen column density in GRB spectra, inferred from X-rays over the intrinsic host contribution measured in the UV, is due to absorption in the IGM, and that this IGM column density increases with redshift.

Most baryonic matter resides in the IGM and in particular, the regions between galaxies. In the early universe, the fraction of baryons in the IGM was even higher, as less material had coalesced gravitationally from it \citep[hereafter M16]{McQuinn2016a}. IGM temperature varies widely over redshift and phase. Recent simulations predict that up to 50\% of the baryons by mass have been shock-heated into a warm-hot phase (WHIM) at low redshift $z < 2$ with $T = 10^5 - 10^7$ K and $n_b = 10^{-6} - 10^{-4}$ cm$^{-3}$ where $n_b$ is the baryon density \citep[e.g][]{Cen1999,Cen2006,Dave2007,   Schaye2015}.  \citet[hereafter M19]{Martizzi2019}, using the IllustrisTNG simulations \footnote{http://www.tng-pro ject.org/}\citep{Piattella2018}, estimated that the cool diffuse IGM constitutes $\sim 39\%$ and the WHIM $\sim 46\%$ of the baryons at redshift $z = 0$.  Observations of the cool diffuse IGM and WHIM are essential for effective tracing of matter across time and to validate the simulations \citep{Danforth2016}. We adopt the common temperature naming convention for IGM plasma: cool is log($T$/K)$ < 5$ and Warm-Hot, log($T$/K$) \sim$ 5 to 7 (M19).  Though we concentrate on the very low density IGM, where relevant we use the common names for systems of different column densities: strong Ly$\alpha$ forest systems (SLFSs): 15 < log$\mathit{N}_{\mathrm{H}\/\ \textsc{i}}$ < 16.2\footnote{Throughout this paper, logarithmic column densities are expressed in units of cm$^{-2}$}; partial Lyman Limit Systems (pLLSs) 16.2 < log$\mathit{N}_{\mathrm{H}\/\ \textsc{i}}$< 17.2; Lyman Limit Systems (LLSs):  17.2 < log$\mathit{N}_{\mathrm{H}\/\ \textsc{i}}$< 19 ; super-LLSs (sLLSs) : 19.0 <log$\mathit{N}_{\mathrm{H}\/\ \textsc{i}}$< 20.3; and Damped Ly$\alpha$ Systems (DLAs) log$\mathit{N}_{\mathrm{H}\/\ \textsc{i}}$ >20.3 \citep[hereafter F14]{Fumagalli2014}.

Over the last several decades, observations of redshifted Ly$\alpha$ absorption in the spectra of quasars has provided a highly sensitive probe of the cool IGM \citep[e.g.][]{Morris1991,York2000,Harris2016,Fumagalli2020a}. In the cool phases of the IGM including voids, $40 - 60\%$ of the universe by mass has $[$O/H$] > - 3$ while, by volume, only $20\%$ of the overdense universe has a metallicity $[$C/H$] > -3$ (F14 and references therein). Both \citet[hereafter S03]{Schaye2003} and \citet[hereafter A08]{Aguirre2008} found virtually no evidence for metallicity evolution in the cool IGM in the range $z = 1.8 - 4.1$, but metallicity did have a strong dependency on density. S03 confirmed that collisional ionisation did not apply to the phases they studied.

A significant fraction of the cool gas probed by SLFSs, pLLSs, and LLSs has been associated with galaxy haloes and the circum-galactic medium (CGM) \citep[hereafter F16]{Pieri2014,Fumagalli2013,Fumagalli2016}. As we move from the diffuse IGM to virialised luminous matter, the metallicity rises from the very low values of [X/H] $\sim -3$ to $-2$, to approximate values of $-1.47$ for SLFS, $-1.3$ for pLLSs, $\leq -2$ for LLSs and $> -1.5$ for DLAs \citep[L19, F16]{Wotta2019}. F16 noted considerable evolution in LLS metallicity. However, these systems contribute only $\sim 4\%$ to the cosmic metal density budget (L19).
The intracluster medium (ICM) mean metallicity in the range $0 \leq z \leq 1.5$ is $Z = 0.23\pm0.01 Z\sun $ \citep[S12;][]{McDonald2016}. At the outer ICM, the metallicity falls to $< 0.01~Z_\odot$ \citep{Mernier2017}, which is the start of the true IGM. Temperature in the ICM are typically log$(T$/K)$  > 7$. However, the ICM only contains $\sim 4\%$ of cosmic metals \citep[M16]{Shull2012}.

At higher temperatures, for some time since the first prediction of substantial baryons at low redshift, the expected baryons were not observed in the WHIM, giving rise to the "missing" baryon problem \citep{Danforth2005,Danforth2008,Shull2012,Shull2014}. Recent literature points to the CGM as reservoir for at least a fraction of this missing matter \citep{Tumlinson2011,Tumlinson2013,Werk2013,Lehner2016}. Other claims to have detected the WHIM include possible detection of $\mathrm{O\/\ \textsc{vii}}$ lines, excess dispersion measure over our Galaxy and the host galaxy in Fast Rasio Bursts (FRB), and stacked X-ray emission from cosmic web filaments using the thermal Sunyaev Zelodovich effect \citep[e.g.][]{Nicastro2018, Macquart2020,Tanimura2020}.

Detection of the WHIM is extremely challenging, as its emission is very weak, it lacks sufficient neutral hydrogen to be seen via Ly$\alpha$ absorption in spectra of distant quasars, and the X-ray absorption signal expected from the WHIM is extremely weak \citep{Nicastro2018,Khabibullin2019}. There appears to be a consensus that, at least for $z < 2$, the predicted mean metallicity of the WHIM from simulations and $\mathrm{O}\/\ \textsc{vi}$ absorption studies is $\sim 0.1~Z\sun$  \citep[e.g.][S12]{Wiersma2011,Danforth2016,Pratt2018}. 

Post reionisation, the vast majority of hydrogen and helium is ionized in the IGM. Therefore, the observation of metals is essential for parameterising the IGM properties including density, temperature and metallicity. Huge work has been completed on individual systems from absorption-line studies that use the ionization states of abundant heavy elements \citep[e.g.][]{Shull2014,Raghunathan2016,Selsing2016,Lusso2015}.  While these surveys have been very successful, most very highly ionised metals are not observed in optical to UV. High resolution X-ray observations are required as they are sensitive to a broad range of  cross-sections over the full integrated LOS. However, tracing individual features of the IGM metals in X-ray with current instruments is very limited. Athena, the proposed European Space Agency X-ray observatory, aims to study the IGM through detailed observations of $\mathrm{O\/\ \textsc{vii}}$ (E = 573 eV) and $\mathrm{O\/\ \textsc{viii}}$ (E = 674 eV) absorption features \citep{Walsh2020}. 
\par While we await this future mission, GRBs are currently one of the most effective observational methods to study the IGM as their X-ray absorption yields information on the total absorbing column density of the matter between the observer and the source \citep[e.g.][]{Galama2001a,Watson2007,Watson2011,Wang2013,Schady2017}.
GRBs are among the most powerful explosions known in the universe. GRBs exist over an extensive range of redshifts and distances, and have high luminosities combined with a broad energy range of observed emissions. Any element that is not fully ionized contributes to the absorption of X-rays. Though Thomson scattering is essentially energy independent, scattering by electrons only becomes important at energies above 10keV \citep[hereafter W00]{Wilms2000}.

Although the X-ray absorption cross-section is mostly dominated by metals, with hydrogen and helium contribution being minimal but not nil (Fig.1 W00), it is typically reported as an equivalent hydrogen column density (hereafter $\textit{N}\textsc{hx}$). $\textit{N}\textsc{hx}$ consists of contributions from the local GRB environment, the IGM, and our own Galactic medium. With current instruments, GRB X-ray absorption cannot generally reveal the redshift of the matter in the column due to a lack of signal to noise and spectral resolution.

 The two main results of earlier studies of the IGM using GRBs are the apparent increase in $\textit{N}\textsc{hx}$ with redshift, and that $\textit{N}\textsc{hx}$ exceeds the host intrinsic neutral hydrogen column density ($\mathit{N}\textsc{h}\/\ \textsc{i}$) in GRB, often by over an order of magnitude  \citep[e.g.][]{Behar2011,Watson2011,Campana2012}. $\mathit{N}\textsc{h}\/\ \textsc{i}$ is generally obtained from observations of strong individual absorbers in the GRB host system. The cause of an $\mathit{N}\textsc{hxigm}$ excess over $\mathit{N}\textsc{h}\/\ \textsc{i}$, and the $\mathit{N}\textsc{hxigm}$ rise with redshift seen in GRBs has been the source of much debate over the last two decades. One school of thought argues that the GRB host accounts for all the excess and evolution e.g. dense Helium (He $\textsc{ii}$) regions close to the GRB \citep{Watson2013}, ultra-ionised gas in the environment of the GRB \citep{Schady2011}, a dense environment near the burst location \citep{Campana2012}, dust extinction bias \citep{Watson2012}, and/or a host galaxy mass $\mathit{N}\textsc{hxigm}$ relation \citep{Buchner2017}. The other school of thought argues that the IGM is the cause of excess absorption and redshift relation e.g. \citep{Starling2013a,Arcodia2016,Rahin2019a}. While we acknowledge that the GRB host may contribute to the excess absorption, it is the IGM that is the focus of this paper.
 
 The convention in prior work using GRB was to use solar metallicity as a device used to place all of the absorbing column density measurements on a comparable scale. These works all noted that the resulting column densities were, therefore, lower limits as GRB typically have much lower metallicities. \citet[hereafter D20]{Dalton2020} used realistic GRB host metallicities to generate improved estimates of $\textit{N}\textsc{hx}$. They confirmed the $\textit{N}\textsc{hx}$ redshift relation and that the revised $\textit{N}\textsc{hx}$ showed an even greater excess over $\mathit{N}\textsc{h}\/\ \textsc{i}$.  

As the bulk of matter in the IGM is ionised and exists outside of gravitationally bound structures, in this paper we use a homogeneity assumption.  We will use tracer objects that have LOS orders of magnitude greater than the large scale structure. 

The sections that follow are: Section \ref{sec:2} describes the data selection and methodology; Section \ref{sec:3} covers the models for the IGM LOS including key assumptions and plausible value ranges for key parameters; Section \ref{sec:4} gives the results of GRB spectra fitting using collisional and photoionisation IGM models with free IGM key parameters; we discuss the results and compare with other studies in Section \ref{sec:5}; and Section \ref{sec:6} gives our conclusions. Appendix A covers model comparisons and investigating the robustness of the IGM model fits. We suggest for readers interested in the key findings on IGM parameters from fittings only, read Sections \ref{sec:4}, \ref{sec:5} and \ref{sec:6}. Readers interested in detailed spectra fitting methodology and model assumptions should also read Sections \ref{sec:2} and \ref{sec:3}. Finally, for readers interested in more detailed examination of key IGM parameters, plus software model comparisons, read Appendix A.

\section{Data selection and methodology}\label{sec:2}
 We used the D20 data for $\textit{N}\textsc{hx}$ which consisted of all observed GRBs with spectroscopic redshift available up to 31 July 2019 from the UK \textit{Swift} Science Data Centre\footnote {http://www.swift.ac.uk/xrtspectra}
repository \citep[hereafter \textit{Swift};][]{Burrows2005}. Spectra from the \textit{Swift} repository were taken from the Photon Counting Late Time mode. D20 investigated the plausibility of assuming that the host intrinsic hydrogen column density is equal to the measured ionisation corrected intrinsic neutral column from UV spectra. D20 used ionisation corrections from F16 who report on the values of the neutral fraction as a function of NHI. We follow their method for GRB host hydrogen column density. The GRB $\textit{N}\textsc{h}\/\ \textsc{i}$ sample is taken from \citet{Tanvir2019}. 
Our sample criteria was that the GRB has detections with quantified uncertainties for $\textit{N}\textsc{hx}$, $\textit{N}\textsc{h}\/\ \textsc{i}$, and spectroscopic redshift. The selection criteria resulted in a total GRB sample of 61. We selected the best S/N sample representative of the range from $1.6 < z < 6.32$. D20 examined if a S/N limited sample would cause a bias by plotting log($\textit{N}\textsc{hx}$) versus both log of total error in $\textit{N}\textsc{hx}$ and total error/$\textit{N}\textsc{hx}$ for all detections. The scatter appeared random so any selection by total error/$\textit{N}\textsc{hx}$ should not result in a bias in $\textit{N}\textsc{hx}$. The total final data sample consists therefore of 32 GRB, details are available in the online supplementary material. 

We refitted the GRB spectra using \textsc{xspec} v12.10.1 \citep[hereafter A96]{Arnaud1996}, assuming an underlying power law in the X-ray band from 0.3 - 10.0 keV, which is suitable for the vast majority of GRB and again is consistent with the \textit{Swift} repository \citep[hereafter S13]{Starling2013a}. The Galactic component is fixed to \textit{Swift} values based on \cite[hereafter W13]{Willingale2013}. \citet{Asplund2009} is generally regarded as providing the most accurate solar abundances. However, we used the solar abundances from W00 which take into account dust and H$_2$ in the interstellar medium in galaxies.

Most works on the WHIM use absorption line observations focusing on oxygen, carbon, nitrogen and neon because of their relatively high abundance, and because the strongest resonance lines in He and H-like ions are in a relatively ‘clean’ wavelength band, compared to typical X-ray spectra resolutions. Due to the small Doppler broadening, ignoring turbulence, the lines rapidly saturate. The challenge for X-ray spectroscopy in the IGM is to detect small equivalent widths \citep[hereafter R08]{Richter2008}, which is only possible currently in the UV. 
Accordingly, we chose to base our work on total absorption by the ionised IGM as opposed to fitting individual line absorption, avoiding misidentifation of absorption features \citep[hereafter G15]{Gatuzz2018}.

When fitting models to spectra, chi-squared ($\chi^2$) is the generally used statistical method. However, deep field X-ray sources such as our sample from \textit{Swift}, only have a small number of photon counts, well into the Poisson regime. The $\chi^2$ regression approach is inappropriate in this circumstance \citep{Buchner2014}. The common practice of rebinning data to use a $\chi^2$ statistic results in loss of energy resolution.
The maximum likelihood C-statistic \citep{Cash1979}, based on the Poisson likelihood, does not suffer from these issues. For a spectrum with many counts per bin the C-statistic $\rightarrow \chi^2$, but where the number of counts per bin is small, the value for C-statistic can be substantially smaller than the $\chi^2$ value \citep{Kaastra2017}. Accordingly, we use the C-statistic (Cstat in \textsc{xspec}) with no rebinning. 

Typically, when using \textsc{xspec} to fit models to spectra, local optimisation algorithms like the Levenberg- Marquardt algorithm are employed to iteratively explore the space from a starting point. However, given we are studying the IGM with X-ray spectra, we can expect some degeneracies between the parameters. Therefore,  there may be multiple, separate, adequate solutions, i.e. local probability maxima. In these circumstances, these algorithms cannot identify them or jump from one local maximum to the other. The \textsc{steppar} function in \textsc{xspec} allows the forcing of parameters to specific ranges. This can overcome the local maximum problem to some degree. 
Markov chain Monte Carlo (MCMC) is a commonly employed integration method for Bayesian parameter estimation. However, MCMC also has difficulty finding and jumping between well-separated maxima \citep{Buchner2014}. Given the issues of goodness of fit and getting out of local probability maxima, we  use a combination of the \textsc{steppar} function and confirmation with MCMC to validate our fitting and to provide confidence intervals on Cstat. We prefer this approach over alternatives such as the Akaike Information Criterion (AIC). The AIC, popular in astrophysics is $AIC = \chi^2 + 2k$ where k is the number of parameters of the model. However, as it is based on $\chi^2$, it suffers from the same problems i.e. based on a Gaussian assumption for errors and requirung a high bin count.

We follow a similar MCMC methodology as \cite{Foreman-Mackey2013} for the number of walkers ($\times$ 10 free parameters), chain length and burn-in period. We use Goodman Weare MCMC \citep{Goodman2010}. 

\section{Models for the GRB LOS}\label{sec:3}
In this section we describe the motivation and expected physical conditions in the IGM that lead to our choice of CIE and PIE models, the priors and parameter ranges. 

In our models, we use different \textsc{xspec} (A96) absorption sub-routines for the absorbers on the LOS. For Galaxy absorption ($\textit{N}\textsc{hxGal})$, we use \textsc{tbabs} (W00) fixed to the values measured by W13. \textsc{tbabs} calculates the cross-section for X-ray absorption by the ISM as the sum of the cross sections for the gas, grain and molecules in the ISM.  For the GRB host galaxy absorption, we use \textsc{tbvarabs} which is the same as \textsc{tbabs} but with metallicity, dust and redshift as free variables. We follow D20 for the metallicity of the GRB host galaxy i.e. using dust corrected actual metallicity where available, and otherwise their average GRB host value of $Z = 0.07Z\sun$. The $\textit{N}\textsc{hx}$ for the GRB host is fixed to the $\textit{N}\textsc{h}\/\ \textsc{i,ic}$ values, following the D20 method. By fixing $\textit{N}\textsc{hx}$ for both our Galaxy and the GRB host, the excess absorption in the models is regarded as being produced by the IGM.
W00 noted that the \textsc{tbabs} model does not include the effects of the warm phase or of the ionized phase of the ISM. We review the photoionisation (PIE) and collisional ionisation equilibrium (CIE) models available in \textsc{xspec} to determine which is best for our purposes of modelling the IGM. These are \textsc{warmabs} (\citealt[hereafter K09]{Kallman2009}) for PIE, \textsc{hotabs} (K09), \textsc{ioneq} (G15) and \textsc{absori} \citep{Done1992} for CIE. In earlier works on using tracers such as GRB and quasars for IGM absorption, \textsc{absori} was generally used \citep[e.g.][]{Behar2011,Starling2013a}. While \textsc{absori} was the best model available when it was developed in 1992, it is not self-consistent as it allows one to have both ionisation parameter and temperature as free parameters which would not occur in either PIE or CIE \citep{Done1992}. Other issues with \textsc{absori} are that it only uses ionisation edges with no line absorption. The metals included are limited to H, He, C, N, O Ne, Mg, Si, S and Fe, and only Fe is allowed as a free parameter. Accordingly, we do not use it for PIE, but do for CIE to compare with earlier studies in Appendix A.

We chose to use \textsc{warmabs} as a more sophisticated PIE model which is designed to determine the physical conditions in partially ionized gases. It calculates the absorption  due to neutral and all ionized species of elements with atomic number Z $\leq$ 30. While \textsc{warmabs} does not account for self-shielding, this effect is negligible at high ionization parameters (K09). For CIE, we chose \textsc{hotabs}, a sophisticated code, similar to \textsc{warmabs} except that it has temperature as the free parameter as opposed to ionisation. An alternative CIE model is \textsc{ioneq}. It is similar to \textsc{hotabs} except that it allows metallicity to vary for O, Ne and Fe only (G15). \textsc{warmabs}, \textsc{hotsabs} and \textsc{ioneq} have turbulent velocity (vturb) as a free parameter. We examined varying vturb to assess impact on fittings. The broad range trialled ($0 - 150$ km s$^{-1}$) showed minimal variation in column densities or other parameters. Thus, we set vturb = 0.

We model the IGM assuming a thin uniform plane parallel slab geometry in thermal and ionization equilibrium. This simplistic approximation is generally used to represent a LOS through a homogenous medium and is appropriate for our model \citep[e.g.][]{Savage2014,Nicastro2017,Khabibullin2019,Lehner2019}. This slab is placed at half the GRB redshift as an approximation of the full LOS medium. We examined placing the slab at half the redshift equivalent distance integral as an approximation of distance, but it did not change our results.

Any LOS to a GRB is likely to encounter many different intervening phases of matter, density, temperature and photoionisation levels. Given the current quality of the GRB spectra, the most pragmatic approach is to define the parameters ranges and priors from the past measurements and observations. The slab fit results will then characterise 'typical' conditions integrated along the LOS. We now review the physical processes and key conditions to pin down the range of parameters and models that are best suited for our analysis.

\subsection{Ionisation processes and equilibrium conditions}
Generally, there are two processes that determine the ionisation state of plasma in the IGM i.e. photoionisation and collisional ionisation. Different physical processes are therefore involved and any assumptions regarding whether collisional, photoionisation or a combination dominates will impact any attempts to model the IGM. 

The photoionisation rate, $\Gamma_{\mathrm{H\/\
 \textsc{i}}}$  depends on the ionising radiation field in the IGM provided by the cosmic ionising background (CIB). For photoionisation IGM modelling, the ionisation parameter $\xi$ (the ratio of ionising photon density to electron density) is a key variable. We set the parameter range as $0 \leq$ log($\xi) \leq 3$ for an ionised IGM \citep{Starling2013a}. 
For IGM modelling with the collisional assumption for ionisation, temperature is a key variable. Collisions by thermal electrons ionise hydrogen to a high degree for gas temperatures $>1.5 \times 10^4$ K. Metals require higher temperatures or ionisation. Therefore, we set the parameter range as $4 \leq$ log$(T$/K) $\leq 8$. 

 We chose to use equilibrium based models. The relation between ionisation state or fraction and gas temperature and ionisation parameter explicitly assumes that the gas is in an ionisation equilibrium (R08). Opinions on the IGM equilibrium state differ greatly \citep[e.g.][]{Branchini2009,Nicastro2018}. In non-equilibrium, the plasma remains over-ionised compared to CIE at any temperature, as recombination lags behind cooling \citep{Gnat2007}. 
While there is still debate on the equilibrium state of the WHIM, it is likely that a substantial part of the baryons in the universe are in regions where extremely low densities and ionisation equilibrium conditions persist (M16). Importantly for IGM modelling, well outside the influence of galaxies and clusters, the radiation background  becomes more important. \cite{Oppenheimer2013} noted that non-equilibrium effects are smaller in the presence of the CIB, and are overestimated in CIE. However, they also showed that in the presence of AGN, a large fraction of the metal-enriched intergalactic medium may consist of non-equilibrium  regions \citep{Oppenheimer2013a}.

In summary, for our IGM models we assume ionisation equilibrium, with the caveat that the equilibrium assumption could result in an underestimation of column density where the IGM plasma remains over ionised in non-equilibrium conditions.

\subsection{Metallicity in the IGM phases}

As we are using ionised metal absorption, we have to allow for a large range in the metallicity parameter as the LOS to the GRBs trace the various IGM phases. In our analysis, we include the highly ionised metals that dominate X-ray absorption given that H and He are relatively unimportant. Below 1 keV, C, N, O, and Ne are the main absorbers, while above 1keV, Si, S, and Fe dominate (W00). $\mathrm{O}\/\ \textsc{vii}\/\ \& \/\ \mathrm{O}\/\ \textsc{viii}$ are the most abundant and Neon species also very important ($\mathrm{Ne}\/\ \textsc{vii}, \mathrm{Ne}\/\ \textsc{viii}, \mathrm{Ne}\/\ \textsc{ix}, \mathrm{Ne}\/\ \textsc{x}$).
 Metallicity currently constitutes the main uncertainty of the IGM models \citep{Branchini2009}. 

 In Section \ref{sec:1}, we noted that in the cool IGM phases, typical metallicity is observed to be at the very low range $-4 < [X/$H$] < -2$ \citep[e.g.][S03,A04]{Simcoe2004}. In the hotter phases including the WHIM, the metallicity has been observed to be $[X/$H$] \sim -1$ \citep[e.g.][S12]{Danforth2016,Pratt2018}. As we are modelling the LOS through the cool, warm and hot diffuse IGM, but noting that some contribution will come from overdense phases, we will set the \textsc{xspec} metallicity parameter range as $-4 < [X/H] < -0.7$ ($0.0001 < Z/Z\sun < 0.2)$.

\subsection{Cosmic ionising background}
The abundance of ionic species is partially dependent on the CIB. Many studies have been completed on the sources of the background radiation such as star forming galaxies and AGN with power laws in the range $1.4 - 2$ \citep[e.g.][]{Haardt96,Haardt2012,DeLuca2004,Luo2011,Moretti2012}. To explore uncertainties in the UV background, \citet{Crighton2015} and \citet{Fumagalli2016} introduced a free parameter $\alpha$UV to account for the  AGN-dominated (hard) to galaxy-dominated (soft) spectrum. A common practice is to adopt a fixed power law for the background radiation. This is a reasonable approach in calculating the ionisation balance. 

\begin{figure}

	\includegraphics[width=\columnwidth]{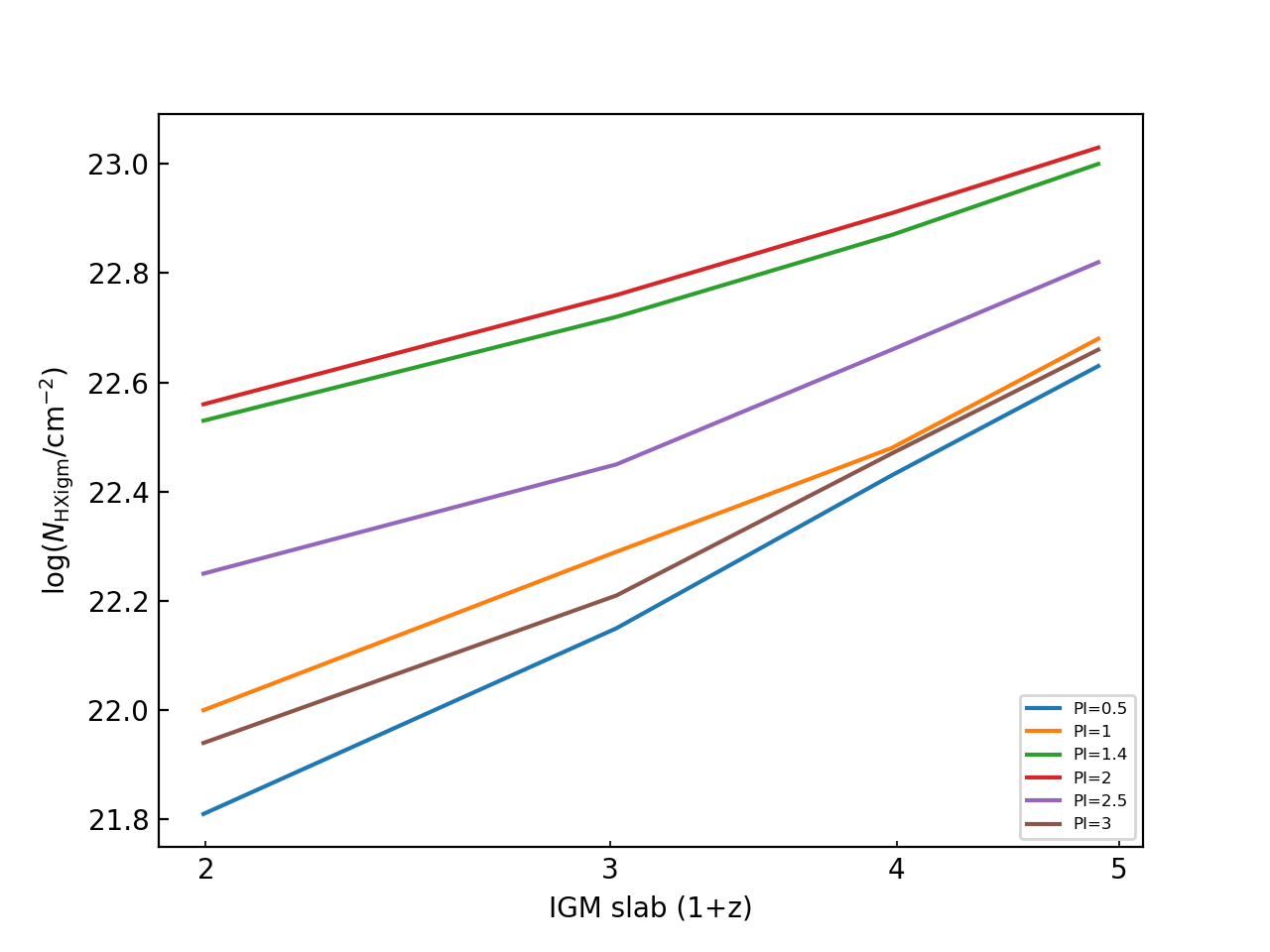}
    \caption{$\textit{N}\textsc{hx}$ and redshift relation for different CIB indices using \textsc{absori} for GRB120909A. A power law of 2 produces the highest estimated column densities at all redshift, but only marginally greater than a photon index of 1.4}
    \label{fig:Figure 2}
\end{figure}

\textsc{absori} is the only \textsc{xspec} model which allows the background CIB as a free parameter. In \textsc{warmabs}, \textsc{hotabs} and \textsc{ioneq}, the CIB photon index is set to 2 which is consistent with the work by \citet{Moretti2012}. In many prior works on the IGM using \textsc{absori}, general practice has been to set CIB photon index to 1.4 following \citep{DeLuca2004}. We examined the impact of different CIB indices on column density using \textsc{absori} on GRB120909A in Fig.~\ref{fig:Figure 2}.

\graphicspath{ {./figurespaper2/}  }

A photon index of 2 results in the highest estimated column density at all redshift. The lowest column densities resulted from a CIB photon index of 0.5 at logarithmic difference $\sim 0.5$ lower than when the index is 2. However, the difference in the commonly observed CIB range of $1.4 - 2$ is minimal.  Accordingly, we set the photon index at 2 for \textsc{absori} to allow comparison with the other models where it is fixed at 2.

\subsection{Full models for the GRB LOS}
  Based on the extensive observations and simulations of the IGM to date, a combined CIE and PIE model for the IGM would be required for optimum fitting of the GRB spectra. However, given the quality of the spectra, we chose instead to examine the two extreme scenarios where all the IGM absorption is either in the CIE or PIE phase. As \textsc{warmabs} and \textsc{hotabs} are the most sophisticated PIE and CIE models, these are used for generating the final results of the IGM parameters of density, metallicity, temperature and ionisation parameter in Section \ref{sec:4}. We then investigated robustness and validity of models and assumptions, together with a comparison of the various ionisation model results (Appendix A).

\begin{table}
	\centering
	\caption{Upper and lower limits for the free parameters in the IGM models. Power law slope and normalisation for the GRB spectrum were also free parameters. The fixed parameters are Galactic and host log($\textit{N}\textsc{hx}$), GRB redshift, the IGM slab at half the GRB redshift, and host metallicity at the observed dust corrected value, or $Z = 0.07Z\sun$. }
	\label{tab:Table_1}
	\begin{tabular}{lccr} 
		\hline
		IGM parameter & equilibrium model &  range in \textsc{xspec} models\\
		\hline
		column density & PIE \& CIE & $20 \leq$ log($\textit{N}\textsc{hx}$) $\leq 23$ \\
		temperature\ & CIE & $4 \leq$ log($T$/K) $\leq 8$ \\
		ionisation & PIE & 0 $\leq$ log($\xi$) $\leq 3$ \\
		metallicity & PIE \& CIE & $-4 \leq [X/$H$] \leq -0.7$ \\
		\hline
	\end{tabular}
\end{table}

The parameters ranges that were applied to the PIE and CIE models are summarised in Table \ref{tab:Table_1}. The full multiplicative models (*) which we trialled for the absorbers on the LOS and including the GRB spectra power law (po) in \textsc{xspec} terminology are:\\PIE: \textsc{tbabs*warmabs*tbvarabs*po}\\
CIE: \textsc{tbabs*hotabs*tbvarabs*po}\\
CIE: \textsc{tbabs*ioneq*tbvarabs*po}\\
CIE: \textsc{tbabs*absori*tbvarabs*po}\\

\begin{figure}
\centering
 \includegraphics[width=\columnwidth]{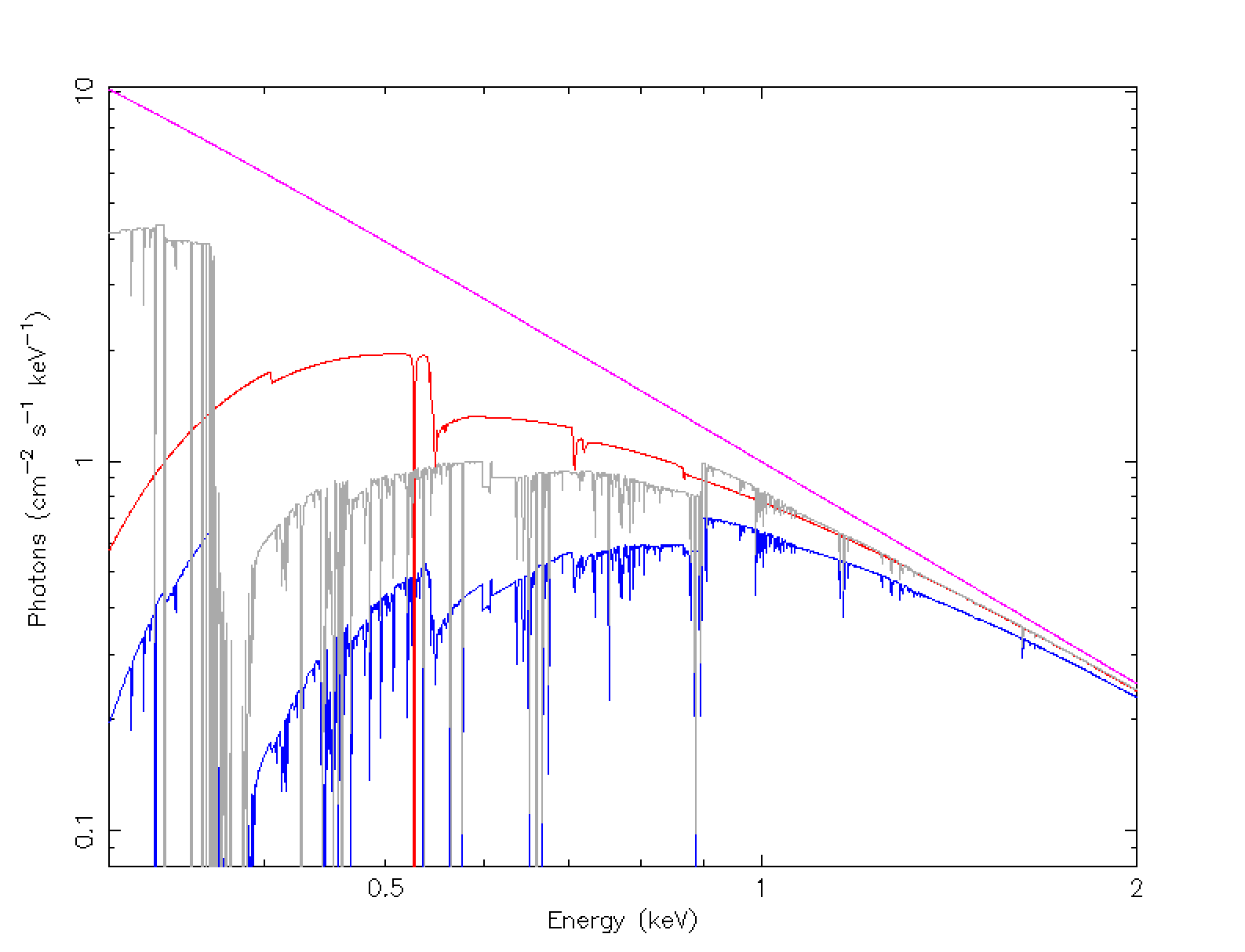}
 \caption{Model components for the LOS absorption using \textsc{hotabs} for IGM CIE absorption in the energy range 0.3 - 2.0keV ($\textit{Swift}$ GRB spectra extend to 10keV). The  model example is for a GRB at redshift $z = 2$, with $[X/$H$] = 0.1$ for the IGM, log($\textit{N}\textsc{hx}) = 21$ for our Galaxy and the GRB host. The IGM log($\textit{N}\textsc{hxigm}) = 22$ approximates the total column density of the IGM LOS to $z=2$. Most absorbtion is due to the IGM (grey) and our Galaxy (red). The GRB host has little contribution due to its redshift and low metallicity (magenta). The total absorption from all three components is the blue line.}
 \label{fig:modelcomponents}
\end{figure}

 Fig.~\ref{fig:modelcomponents} shows an example of the model components for the full LOS absorption using \textsc{hotabs} for IGM CIE absorption. The  model example is for a GRB at redshift $z = 2$, with $[X/$H$] = 0.1$ for the IGM. For our Galaxy and the GRB host, log($\textit{N}\textsc{hx}) = 21$, and log($\textit{N}\textsc{hxigm}) = 22$ (to represent the column density of the IGM LOS cumulatively to $z=2$). The absorption by the GRB host is minimal compared to the ISM of our Galaxy and the IGM. This is because of the redshift $z = 2$ and low metallicity $Z = 0.07Z\sun$ of the GRB host. The sample transmission plots using \textsc{hotabs} and \textsc{warmabs} show the impact of different key parameters and redshift for both PIE and CIE models are available in the online supplementary material.

\graphicspath{ {./figurespaper2/}  }

\begin{figure*}
     \centering
     \begin{tabular}{c|c}
    \includegraphics[scale=0.28]{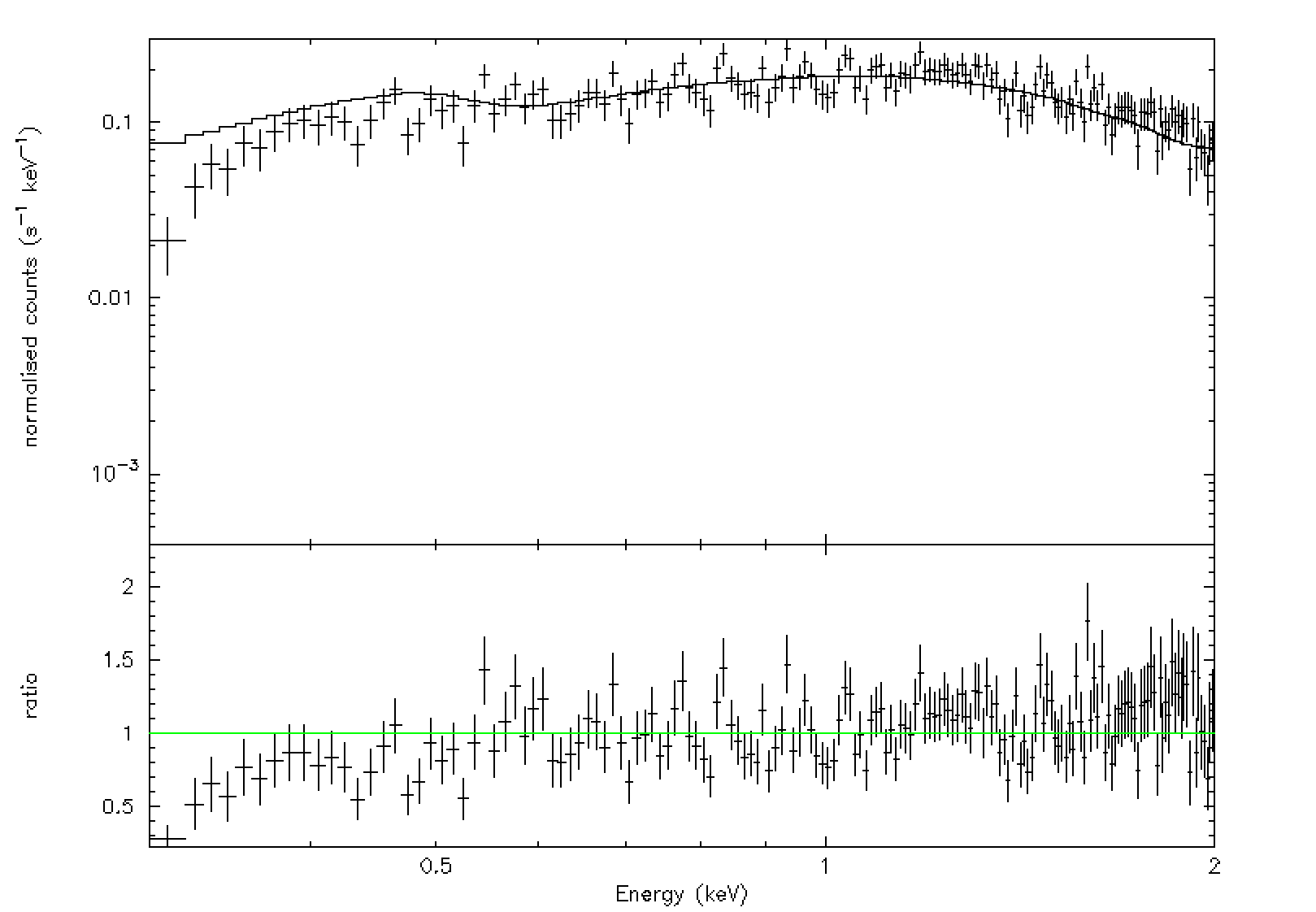} &
    \includegraphics[scale=0.28
    ]{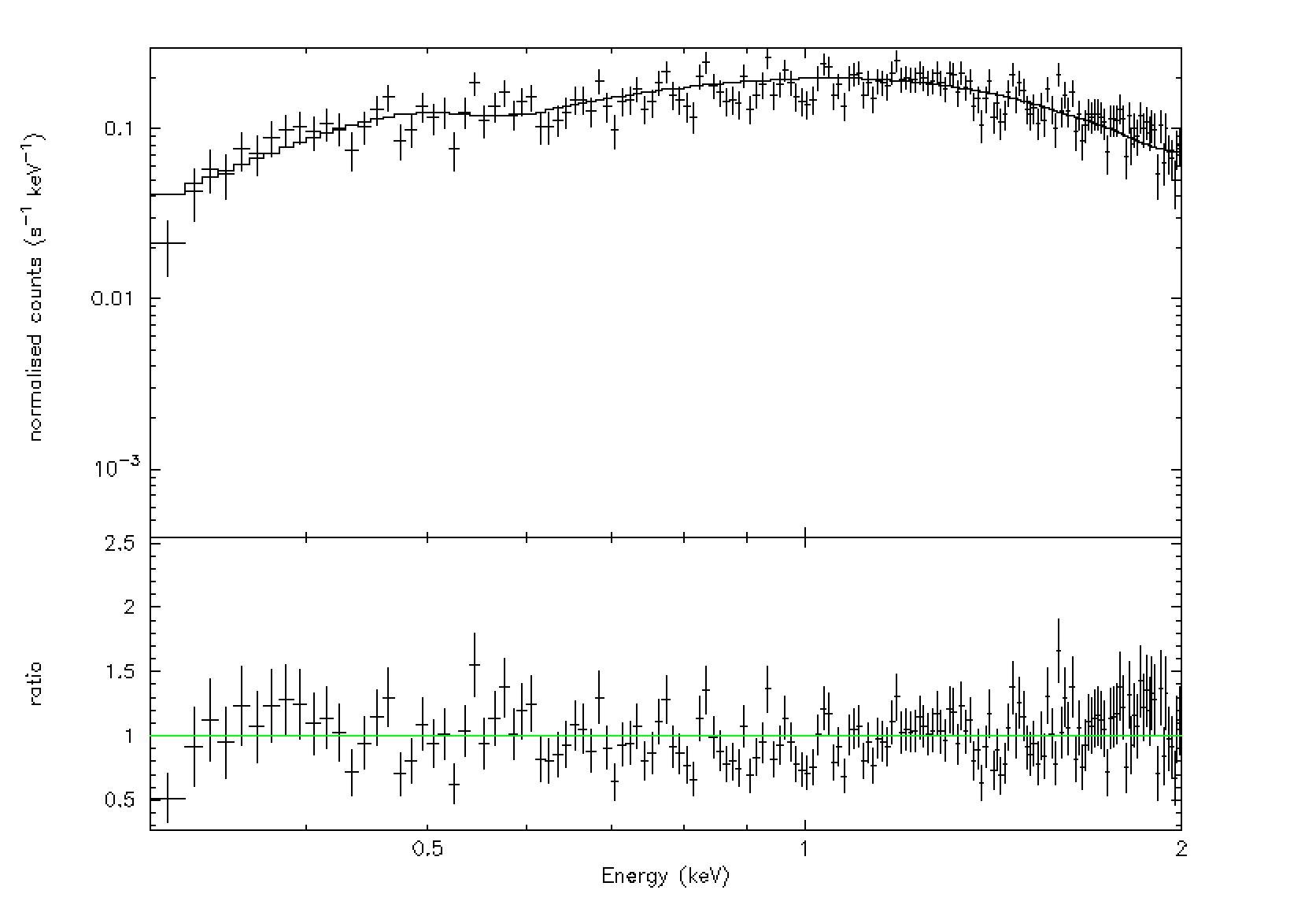}\\
    \includegraphics[scale=0.27]{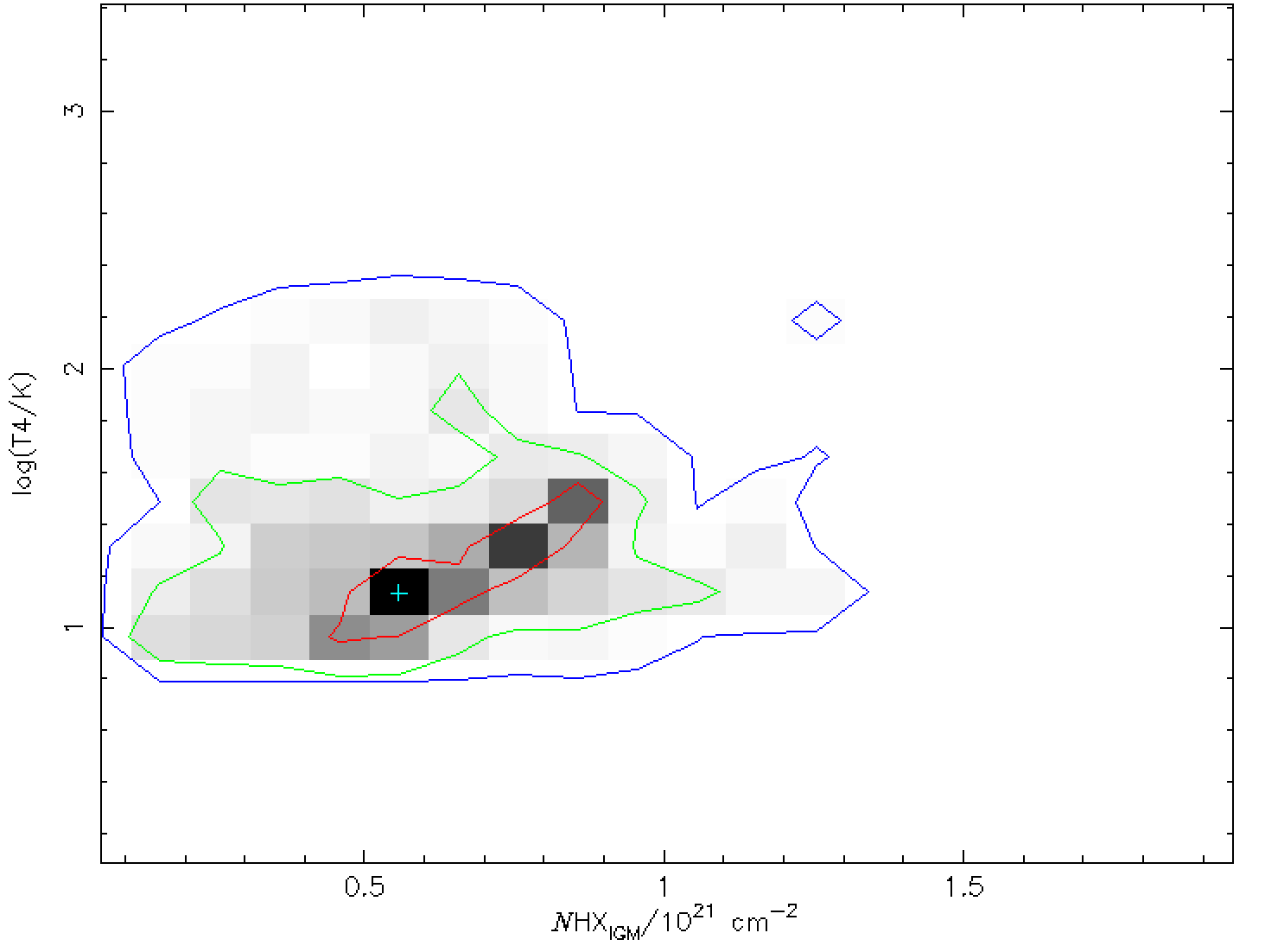} &
    \includegraphics[scale=0.27]{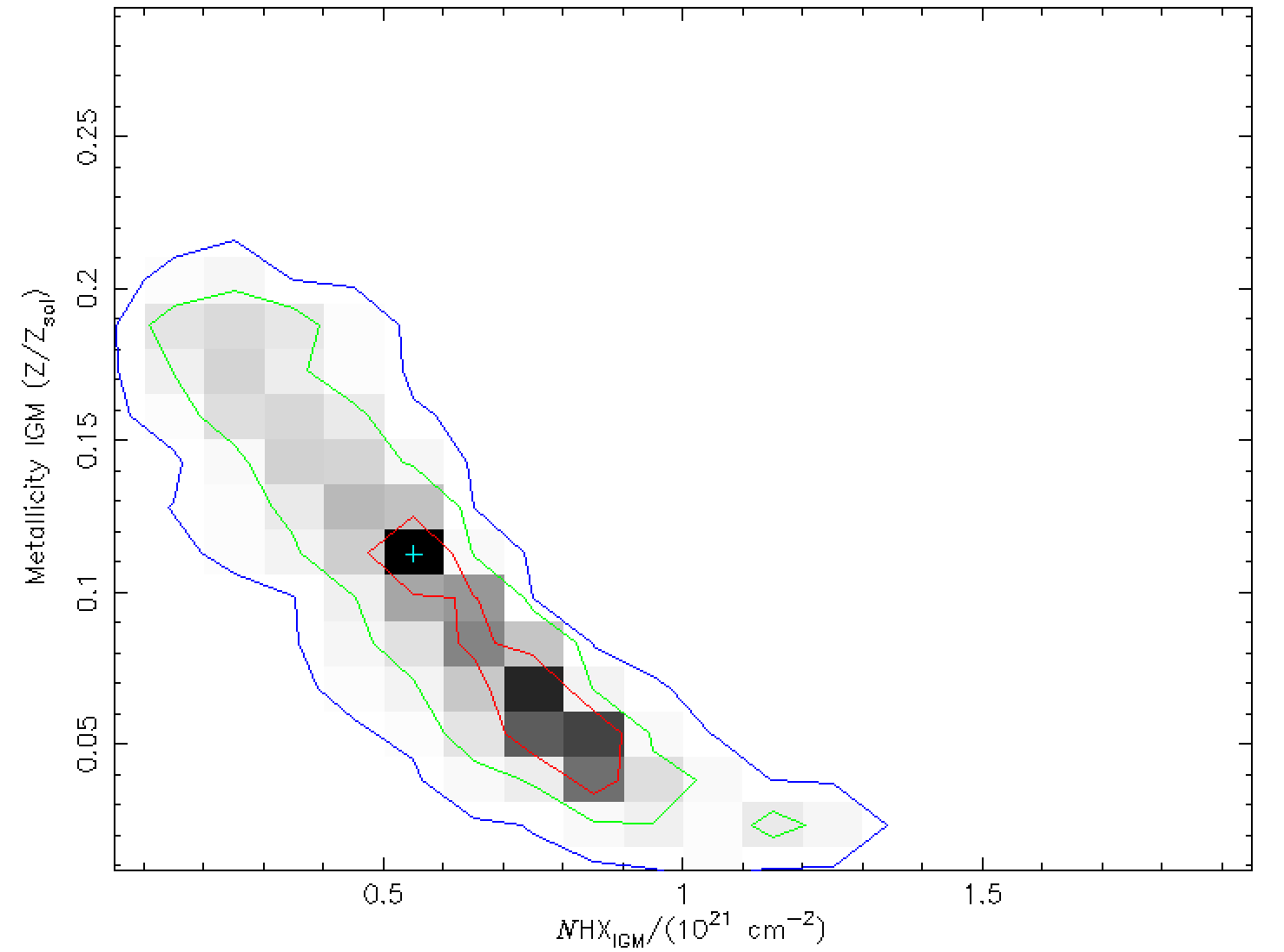}
    
    \\
    \end{tabular}
    \caption{Impact of adding additional model components to a simple power law in fitting GRB150403A. Top-left panel is with $\textit{N}\textsc{hxGal}$ only. Top-right panel is with the addition of a fixed host component and CIE IGM absorption component.  The spectrum fit shows  improvement in low energy absorption over the simple power law fit with $\textit{N}\textsc{hxGal}$. The bottom-left and right panels show the MCMC integrated probability results for $\mathit{N}_{\textsc{hxigm}}$ with temperature and metallicity respectively. The red, green and blue contours represent $68\%, 95\%$ and $99\%$ ranges for the two parameters respectively, with grey-scale showing increasing integrated probability from dark to light. On the y-axis in the bottom-left panel T4 means the log of the temperature is in units of 10$^4$ K.}
\label{fig:GRB150403 fits}
\end{figure*}

\graphicspath{ {./figurespaper2/}  }

\begin{figure*}
     \centering
     \begin{tabular}{c|c}
    \includegraphics[scale=0.5]{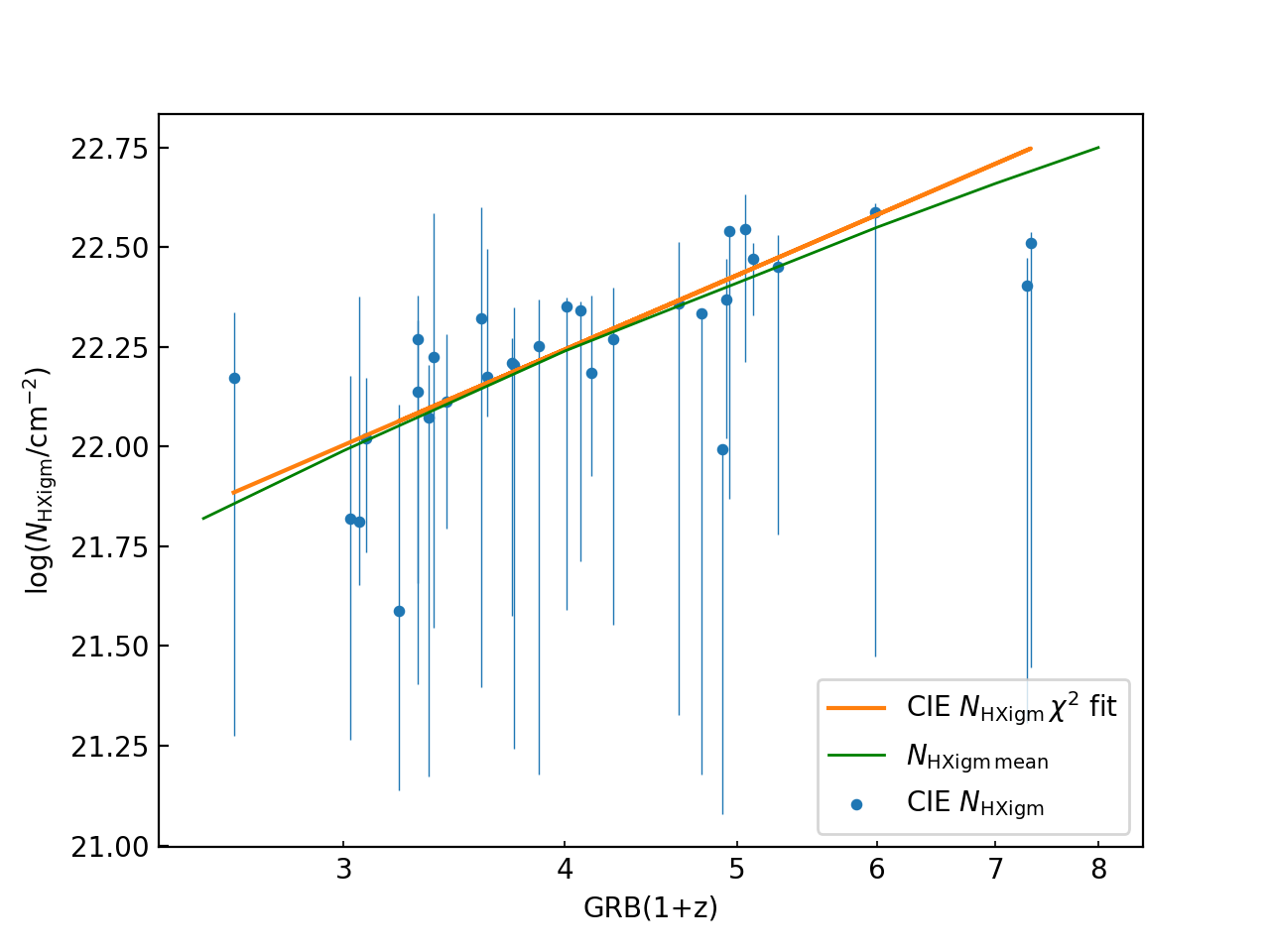} &
    \includegraphics[scale=0.5]{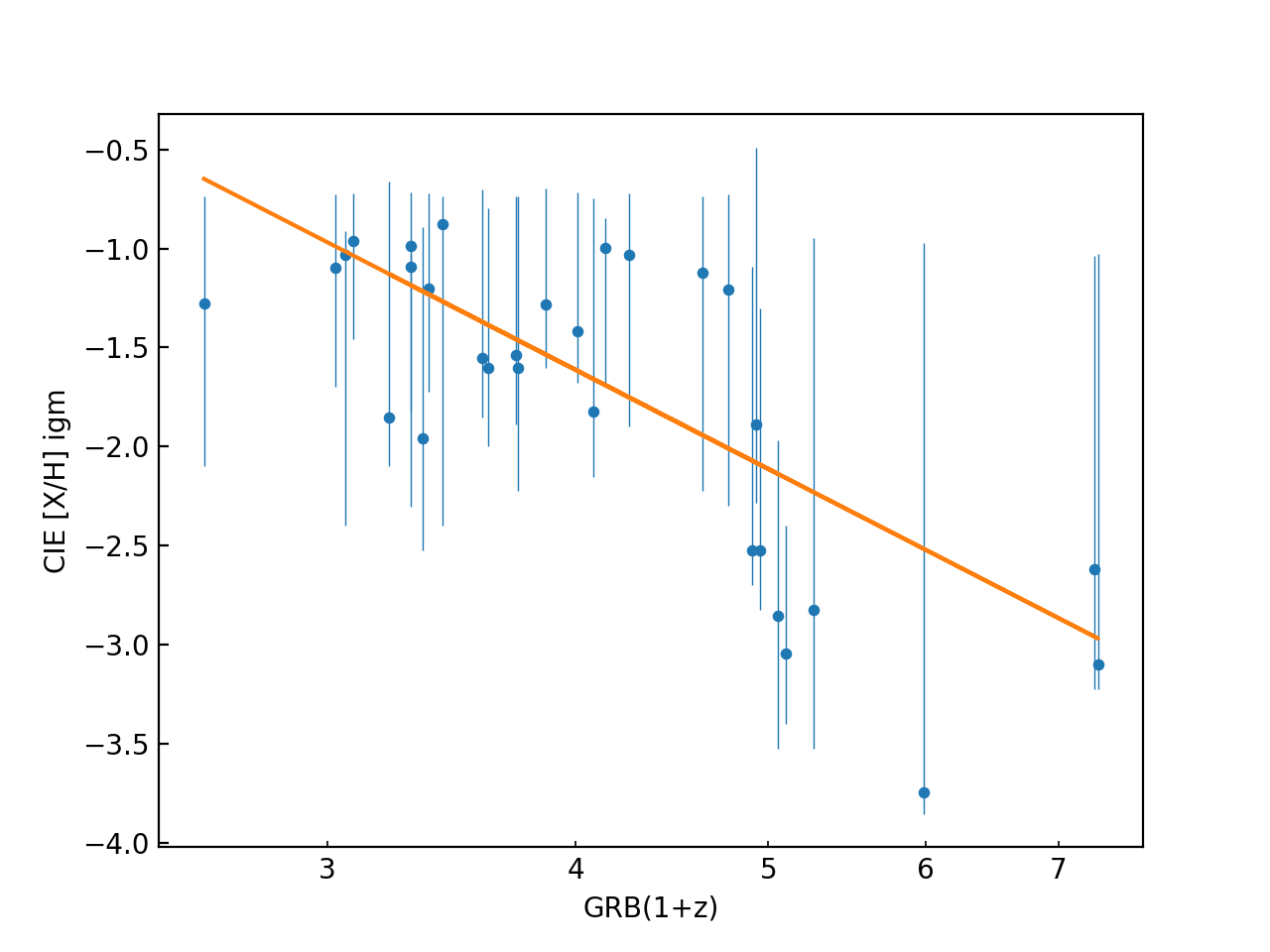}  \\
    \includegraphics[scale=0.5]{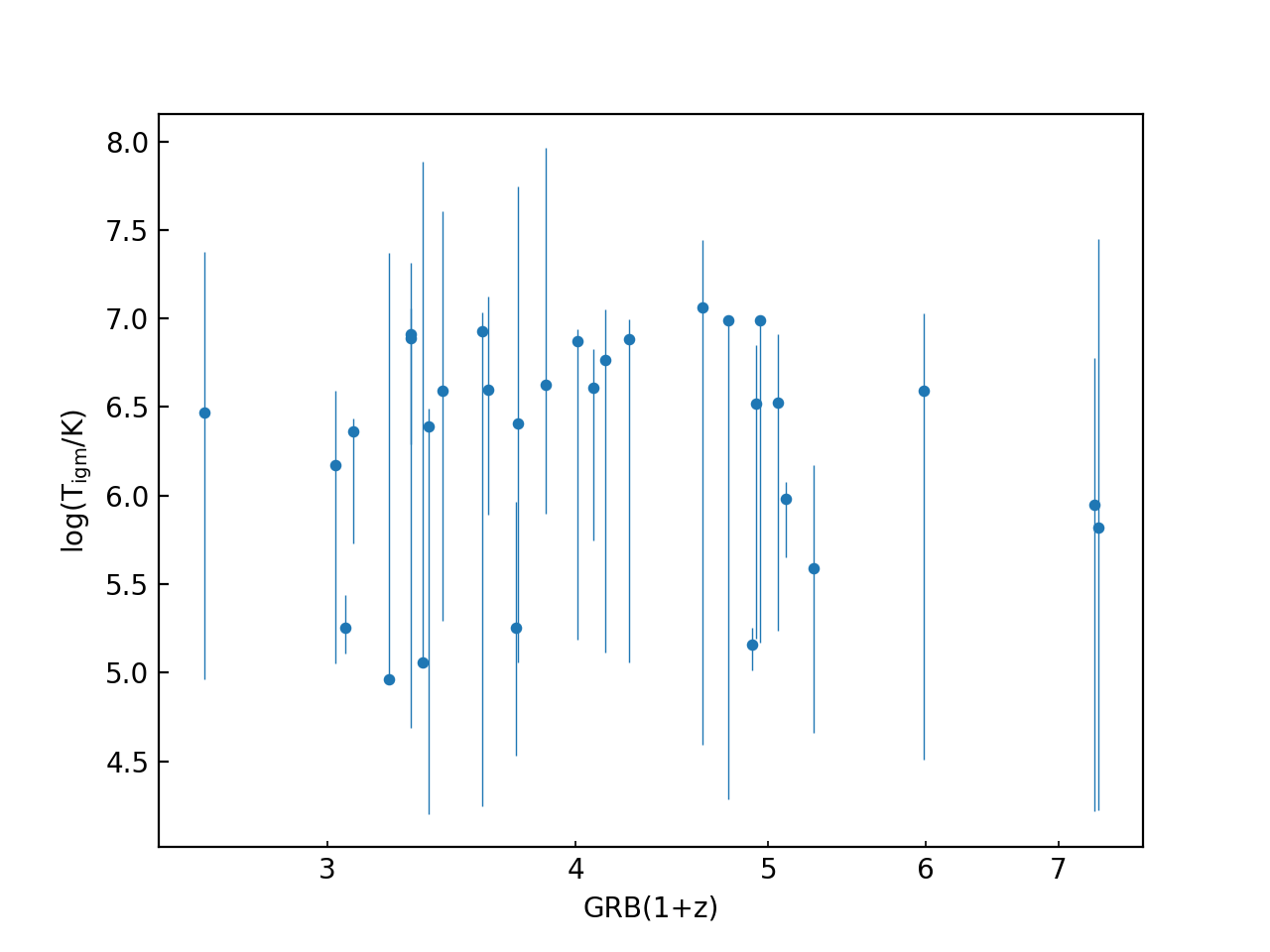} &
    \includegraphics[scale=0.5]{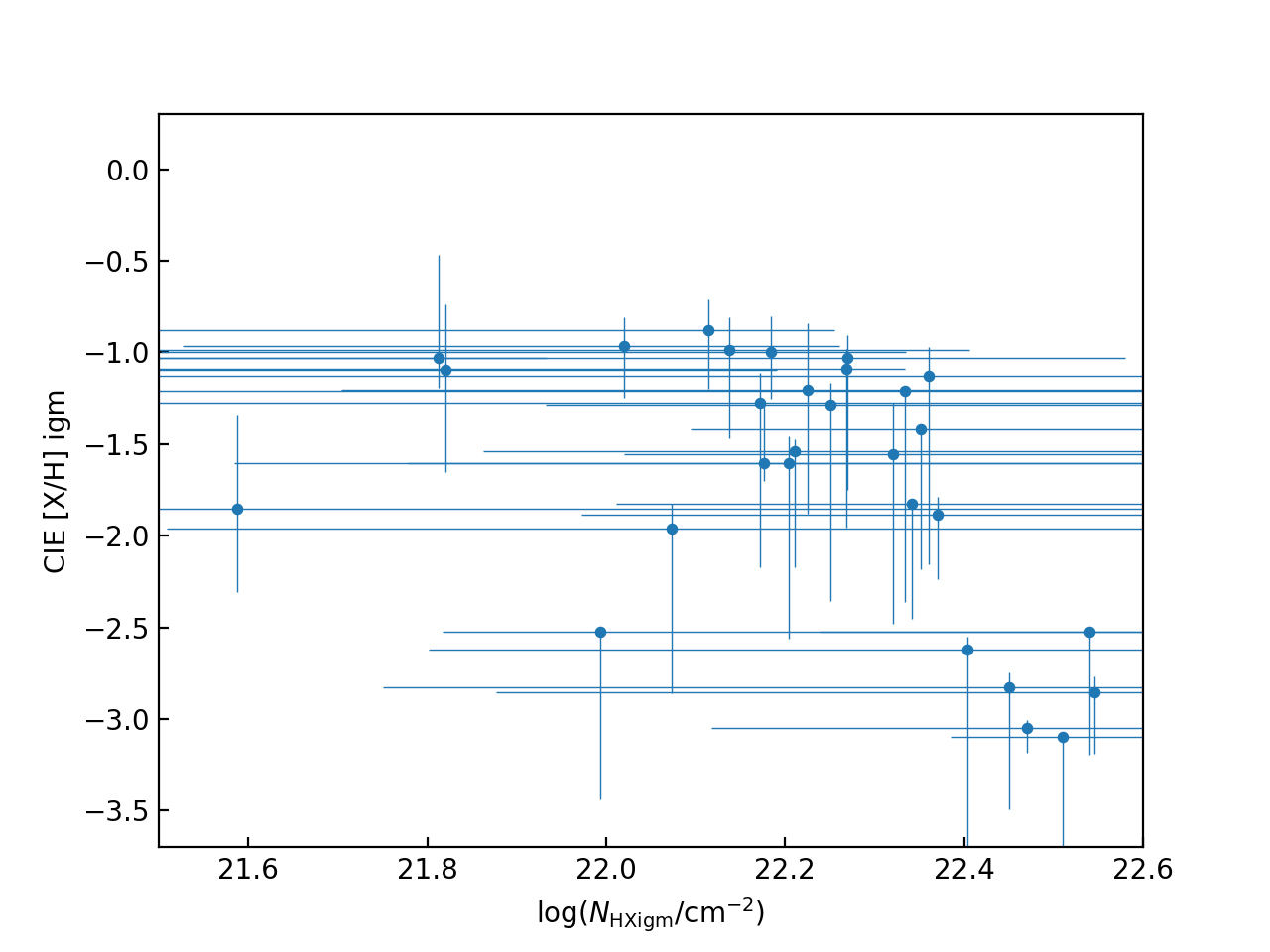}  \\
    \end{tabular}
    \caption{Results of the IGM parameters using the CIE (\textsc{hotabs}) model . The error bars are reported with a $90\%$ confidence interval. The green line is the simple IGM model using a mean IGM density. Top-left panel is $\textit{N}\textsc{hx}$ and redshift. Top-right panel is [X/H] and redshift. Bottom-left panel is temperature and redshift. Bottom-right panel is $[X/$H$]$ and $\textit{N}\textsc{hx}$. The orange line is the 1 sigma $\chi^2$ fit. We do not include a $\chi^2$ curve in the temperature-redshift plot, or the $[X/\mathrm{H}]$ and $\textit{N}\textsc{hx}$ plot as the fit was poor due to a large scatter and error bars.}
        \label{fig:CIEfreeplots}
\end{figure*}

\section{Results using CIE and PIE IGM models with key parameters free}\label{sec:4}

In Section \ref{sec:4}, we firstly discuss the impact of using additional model components on a sub-sample in Section \ref{subsec:4.1}, then give the results for the full sample using the CIE IGM absorption model in Section \ref{subsec:4.2} and using the PIE IGM model in Section \ref{subsec:4.3}.

\subsection{GRB model improvement results}\label{subsec:4.1}

We started by fitting a sub-sample of GRB from our full sample, to examine the impact of adding additional model components leading to the full model including the ionised IGM. We show the fitting results in Fig. \ref{fig:GRB150403 fits} for GRB150403A at redshift $z = 2.06$ as an example of typical results.  We initially fitted with a simple model of power law and absorption only from our Galaxy. The top-left panel in Fig. \ref{fig:GRB150403 fits} shows residuals at low energy with a Cstat of 737.6 for 684 degrees of freedom (dof). We then add a fixed GRB host absorber equal to the measured ionised corrected intrinsic neutral column from UV spectra as detailed in Section \ref{sec:2} plus a variable IGM component. The IGM component is either the CIE or PIE model. The top-right panel of Fig. \ref{fig:GRB150403 fits} shows the result for the model with the CIE IGM component. The spectral fit is visually  improved compared with the Galaxy only model, with much less low energy residual. The Cstat for the full model with the CIE IGM component is 655.0 (680 dof), and for the PIE model 652.8 (680 dof). The  Fig. \ref{fig:GRB150403 fits} bottom-left and right panels show the MCMC integrated probability results for $\mathit{N}_{\textsc{hxigm}}$ with temperature and metallicity respectively. The red, green and blue contours represent $68\%, 95\%$ and $99\%$ ranges for the two parameters respectively. On the y-axis bottom-left panel, log(T4/K) means that 0 is log$(T/$K$) \/\ = \/\ 4$.

All the fittings in the sub-sample for both CIE and PIE model components showed Ctat results as good as or better than the simple models where all the absorption in excess of our Galaxy was assumed to be at the GRB host redshift. Accordingly, we proceeded to fit the full GRB sample with our CIE and PIE models and give the results for the IGM parameters in Sections \ref{subsec:4.2} and \ref{subsec:4.3}. In these scenarios, we use \textsc{hotabs} for CIE and \textsc{warmabs} for PIE  with $\textit{N}\textsc{hx}$, metallicity and temperature or ionisation parameters all free. The error bars for all fits are reported with a $90\%$ confidence interval. In the plots of $\textit{N}\textsc{hx}$ and redshift, the green line is the mean hydrogen density of the IGM based on the simple model used in D20 and references therein.
\begin{equation} 
\label{eq:simpleIGM}
\mathit{N}_{\textsc{hxigm}} = \frac{n_0 c}{H_0} \int_0^z \frac{(1 + z)^2 dz}{[\Omega _M(1 + z)^3 + \Omega _\Lambda ]^\frac{1}{2}}
\end{equation}
where n$_{0}$ is the hydrogen density at redshift zero, taken as $1.7 \times 10^{-7}$ cm$^{-3}$ \citep{Behar2011}.
This value is based on $90\%$ of the baryons being in the IGM. Values for this IGM fraction in the literature vary e.g. $0.5 - 0.7$ (S12) and $0.84$ \citep{Zhang2020a}.

\subsection{Results for IGM parameters using a CIE IGM component}\label{subsec:4.2}

We give the key detailed results for the IGM parameters from fitting our model CIE model to the GRB spectra in our sample.

Modelling the IGM using \textsc{hotabs} for CIE with parameters $\mathit{N}\textsc{hx}$, Z and T free, results in  $\mathit{N}\textsc{hxigm}$ showing similar values and correlation with redshift as the mean IGM density model in Fig. \ref{fig:CIEfreeplots} (top left). A power law fit to the $\mathit{N}\textsc{hxigm}$ versus redshift trend scales as $(1 + z)^{1.9\pm0.2}$. The reduced $\chi^2 = 0.34$ indicates a good fit. The mean hydrogen density at $z = 0$ from the sample is $n_0 = 1.8^{+1.5}_{-1.2} \times 10^{-7}$ cm$^{-3}$, providing a good constraint on this important IGM parameter. Nearly all GRB fits are proximate to both the $\chi^2$ fit and mean IGM density curve. However, there are a few notable outliers, especially the lowest redshift GRB140430A with $z = 1.6$. This fit has much higher $\mathit{N}\textsc{hxigm}$ than both the mean density curve and the $\chi^2$ fit. This could be due to a strong absorber on the LOS. To test this, we removed the cap of $Z =0.2\/\ Z\sun$ as the there is covariance between column density and metallicity parameters i.e. a higher metallicity results in a lower column density. The best fit then for GRB140430A was with $Z \approx 0.4~Z\sun$ with a much lower column density similar to the mean IGM model density at $z = 1.6$. At the highest redshift $z >6$, the two GRB fits are well below the cosmic mean density and $\chi^2$ fit. There is some dispersion in the GRB data points at $z \leq 2$.  This could indicate that at these redshifts, the GRB host contribution is dominant over the IGM absorption, while at higher redshifts, the host contribution becomes diluted by the IGM, therefore showing a smaller dispersion of results. It is notable that most of the GRB fits sit at the high end of the $90\%$ confidence interval error bars. \textsc{xspec} failed to fit two GRB from sample of 32. This was either because the best fit was at the limit of the parameters or the error bars were at one or both parameter limits. This could be due to poor spectra resolution or that the parameter range was too narrow.

 The top-right panel of Fig. \ref{fig:CIEfreeplots} shows the dependence of $[X/$H$]$ with redshift. A power law fit to the $[X/$H$]$ versus redshift trend scales as $(1+z)^{-5.2\pm1.0}$. The reduced $\chi^2 = 1.3$ indicates a plausible linear fit. Metallicity ranges from [X/H] $\sim -1\/\ (0.1~Z\sun)$ at $z = 2$ to [X/H] $\sim -3\/\ (0.001~Z\sun)$ at high redshift ($z > 4$). 

There is a large range in the fitted temperature $5.0 <$ log($T$/K) $< 7.1$, with substantial error bars in the Fig. \ref{fig:CIEfreeplots} bottom-left panel. The mean temperature over the full redshift range is log($T$/K) $= 6.3^{+0.6}_{-1.3}$. These values are consistent with the generally accepted WHIM range. It is interesting to note that even at the highest redshifts $z > 4$, temperatures of log$(T/$K) > 5 appear to give good fits. Further, there is no apparent general decline in temperature with redshift. It should be noted, however, that the fits are for the integrated LOS and not representative of any individual absorber temperature.

In nearly all GRB fits, the Cstat was at least as good as, or better than simple fits with all absorption in addition to that in our Galaxy assumed to be at the GRB host. The MCMC integrated plots were consistent with the $\textsc{steppar}$ results. In conclusion, with the caveats of low GRB X-ray resolution, small data sample and the slab model to represent to full LOS, there are reasonable grounds for arguing that the CIE model using \textsc{hotabs} is plausible for modelling the warm/hot component  of the IGM at all redshifts. The results are consistent with a mean hydrogen density of $ n_{0} = 1.7 \times 10^{-7}$ cm$^{-3}$, providing constraints on this IGM parameter of $n_0 = 1.8^{+1.5}_{-1.2} \times 10^{-7}$ cm$^{-3}$. However, cosmological simulations suggest that the fraction of mass contained in the warm-hot IGM phase is a strong function of redshift being $\sim 49\%$ at $z = 0$, dropping by a factor of 20 by $z = 4$, while the diffuse cooler IGM becomes dominant \citep{Martizzi2019}. Our model indicates a decline in the average metallicity on the LOS, with a significant drop in metallicity at the highest redshifts. The temperature range of log($T$/K) $\sim 5 - 7$ and mean of $6.3^{+0.6}_{-1.3}$ are consistent with the expected values from simulations for a warm/hot phase. We discuss the results further and compare with other studies in Section \ref{sec:6}.

\subsection{Results for IGM parameters using a PIE IGM component}\label{subsec:4.3}

\graphicspath{ {./figurespaper2/}  }

\begin{figure*}
     \centering
     \begin{tabular}{c|c}
    \includegraphics[scale=0.55]{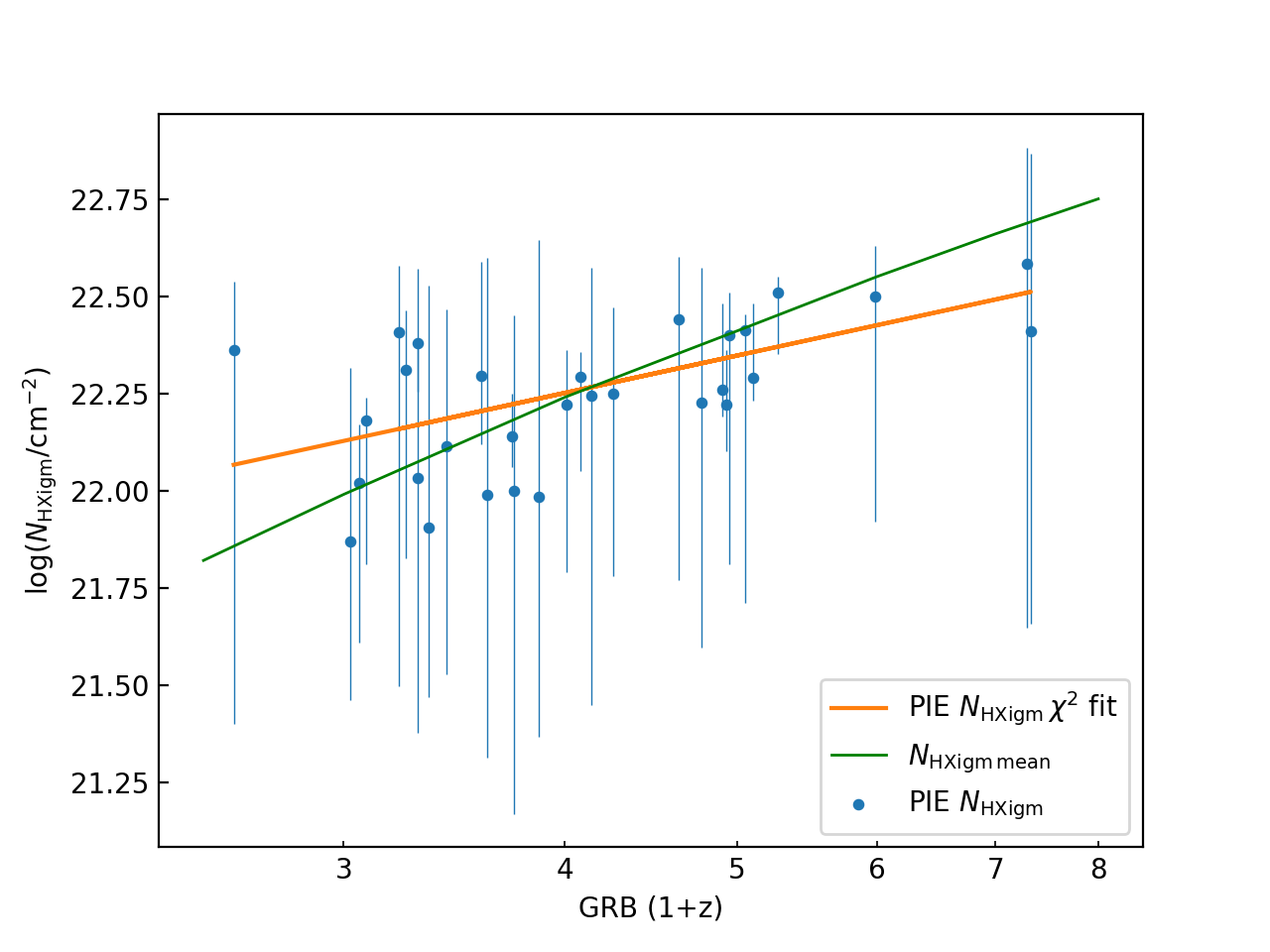} &
    \includegraphics[scale=0.55]{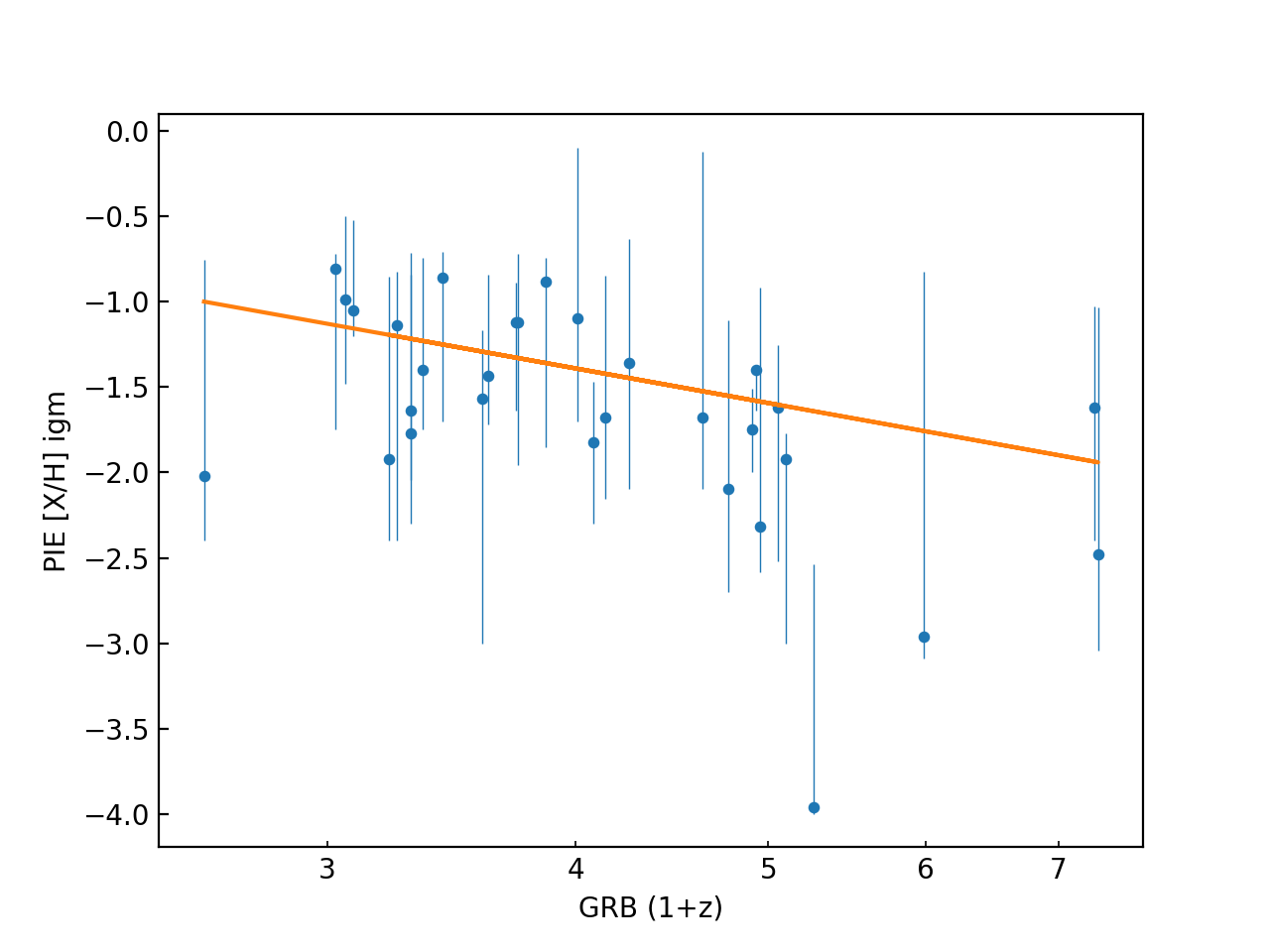}  \\
    \includegraphics[scale=0.55]{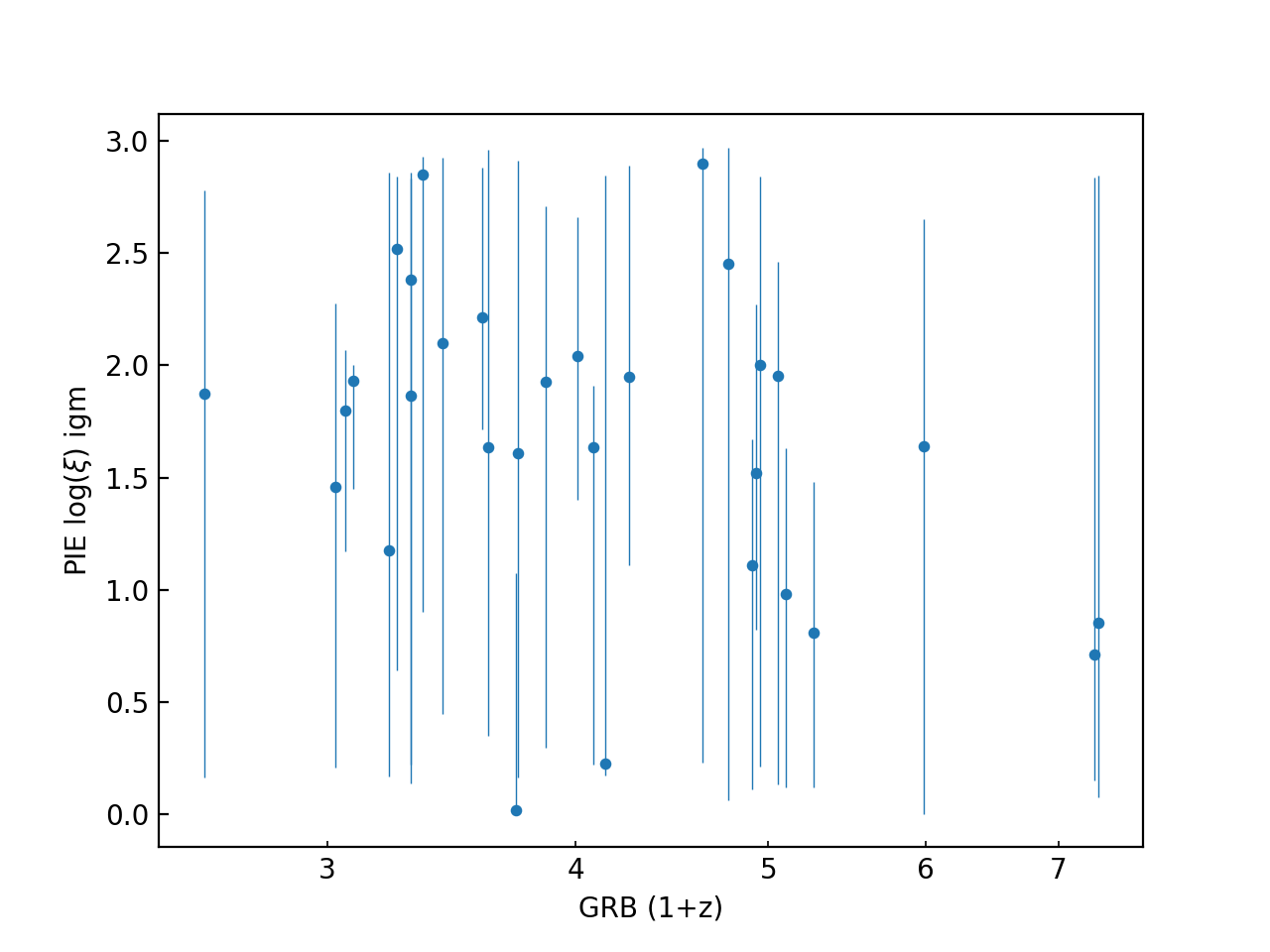} &
    \includegraphics[scale=0.55]{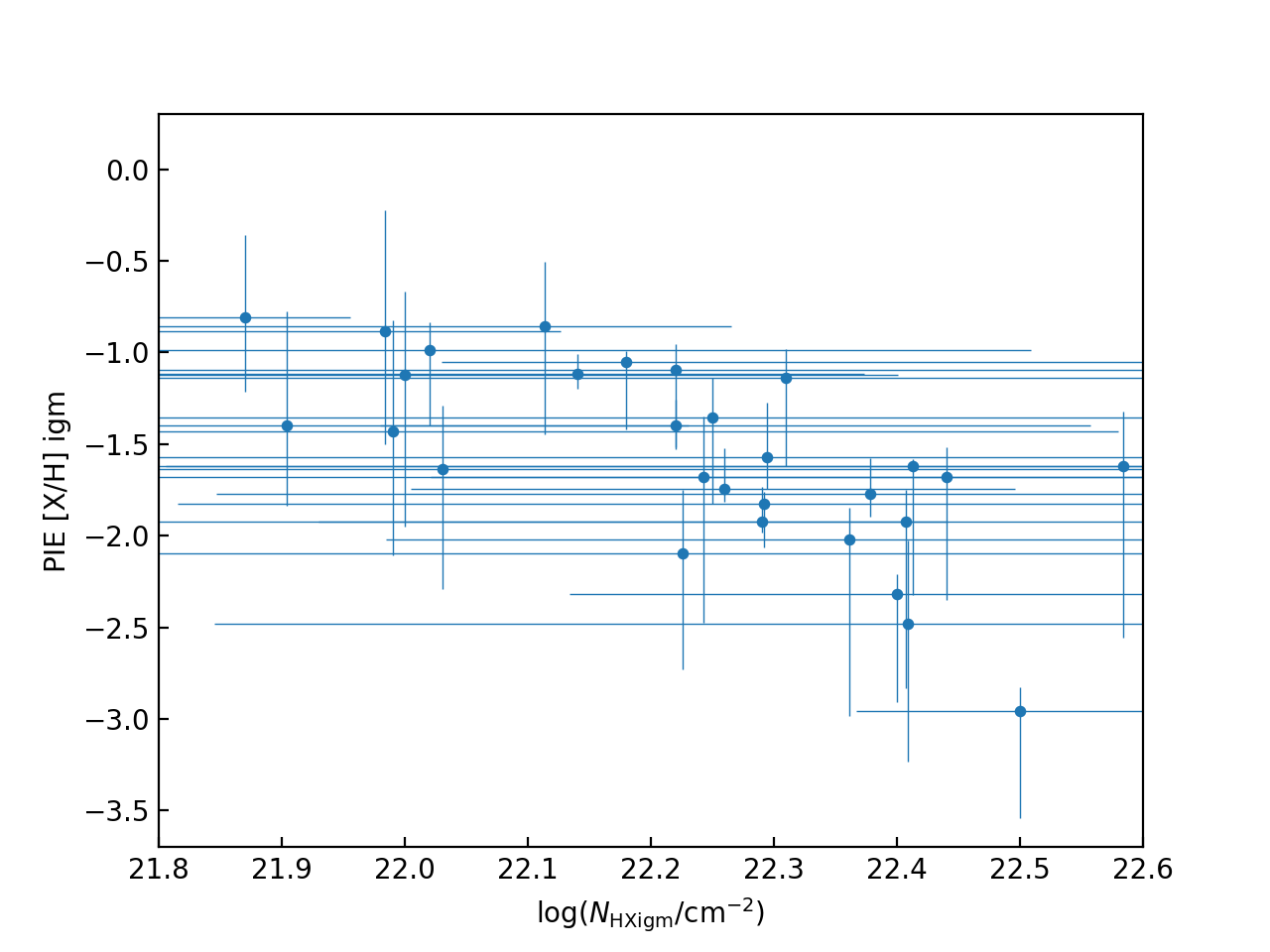}  \\
    \end{tabular}
    \caption{Results for the PIE IGM parameters using PIE (\textsc{warmabs}) models. The error bars  are reported with a $90\%$ confidence interval. The green line in the top-left panel is the simple IGM model using a mean IGM density. Top-left panel is $\textit{N}\textsc{hx}$ and redshift. Top-right panel is $[X/\mathrm{H}]$ and redshift. Bottom-left panel is ionisation parameter and redshift. Bottom-right panel is $[X/\mathrm{H}$ and $\textit{N}\textsc{hx}$. The orange line is the $\chi^2$ fit. We do not include a $\chi^2$ curve in the ionisation-redshift plot , or the $[X/\mathrm{H}]$ and $\textit{N}\textsc{hx}$ plot as the fit was poor due to a large scatter.}
        \label{fig:PIEfreeplots}
\end{figure*}

Modelling the IGM using \textsc{warmabs} for PIE with $\mathit{N}\textsc{hxigm}$, Z and $\xi$ as free parameters results in values for $\mathit{N}\textsc{hxigm}$ and correlation with redshift comparable to the mean IGM density model (the top-left panel of Fig. \ref{fig:PIEfreeplots}). The data points appear to show less dispersion and are marginally below the mean IGM density model at higher redshift $z > 3$, though with large error bars. A power law fit to the $\mathit{N}\textsc{hxigm}$ versus redshift trend scales as $(1 + z)^{1.0\pm0.3}$, flatter than for CIE.  The reduced $\chi^2$ of 0.67 indicates a reasonable fit. Only 2 GRB failed to be fit in \textsc{xspec} out of our sample of 32. The mean hydrogen density at $z = 0$ from the sample is $n_0 = 1.8^{+1.4}_{-1.2} \times 10^{-7}$ cm$^{-3}$, providing a similar constraint on the IGM density parameter to our CIE model. It is notable that the lowest redshift GRB140430A $z = 1.6$ in our sample has a best fit $\mathit{N}\textsc{hxigm}$ again considerably higher than the mean density curve, similar to the CIE model. Again, this could indicate a higher metallicity absorber along the LOS.

Fig. \ref{fig:PIEfreeplots} top-right panel is a plot of $[X/$H$]$ versus redshift with a trend scaling as $(1 + z)^{-2.1\pm0.8}$. The reduced $\chi^2$ of 1.84 indicates a reasonable linear fit. The metallicity is approximately $[X/$H$] \sim -1.0 \/\ (0.1~Z\sun)$ at $z \sim  2$ falling to $[X/$H$] \sim -2.4 \/\ (0.004~Z\sun)$ at $z > 4$. At higher redshift, the average LOS metallicity value appears to decline more rapidly, but with large error bars. This could suggest that there was very little higher metallicity aborption at those redshifts. However, there are very few high redshift GRB in the sample. 

There is a large range in the fitted ionisation parameter $0.0 <$ log($\xi)\/\ < 2.9$, with substantial error bars in Fig. \ref{fig:PIEfreeplots} (bottom left). The mean ionisation parameter over the full redshift range is log($\xi)\/\  = 1.6^{+0.9}_{-1.2}$. We note that these values are the LOS average and not representative of any individual absorber or redshift.

As discussed in Section \ref{sec:2}, we use Cstat minimization due to small number of photon counts, well into the Poisson regime. Thus, $\chi^2$ and AIC are inappropriate. If the model improvement criteria is any improvement in Cstat when compared with the model with all excess absorption over Galactic assumed at the host redshift, then 26/30 CIE, 23/30 PIE, 27/30 combined show improvements. 
Approximating a $\chi^2$ criterion, some authors consider fits to be significantly improved by the addition of a component if Cstat$^2> 2.71$ for each extra free parameter e.g. \citep{Ricci2017}. Our IGM inclusive \textsc{xspec} models have 2 extra free parameters, so the test would be Cstat$^2 > 5.42$. Using this criterion, 12/30 for CIE and 9/30 are significantly improved, combined 15/30.

 In conclusion, with the same caveats as for CIE, there are reasonable grounds for arguing that the PIE model using \textsc{warmabs} is plausible for modelling the cool diffuse IGM. The results are similar to the mean hydrogen density of $ n_{0} = 1.7 \times  10^{-7}$ cm$^{-3}$. The PIE model results show a decline with redshift in the average metallicty values along the integrated LOS. The ionisation range and mean are consistent with the expected values from simulations.  It is not possible to conclude whether PIE or CIE is the better single model for the IGM at all redshift. From the outset, it was noted that a combined model is likely but this requires better data. The results may indicate that we are seeing  different IGM phases along the LOS, though we are examining the extremes of CIE an PIE models separately.

\section{Discussion and comparison with other studies}\label{sec:5}
In both CIE and PIE IGM scenarios, the IGM is highly ionised with either high temperatures or high ionisation parameters. We find that under both scenarios for the IGM, the average metallicity along the LOS declines with redshift, with the caveat that we are using a cumulative LOS absorption. The decline in average metallicity is less in the PIE model. It may be that at lower redshift regions of higher metallicity such as the WHIM may be dominant, while at higher redshift, the diffuse cool IGM becomes dominant, diluting the average metallicity, but we would require a combined CIE/ PIE model to establish this.  Fixing metallicity to any value gave poor and unreliable results (see Appendix A). Most prior studies were based on simplistic assumptions of solar metallicity, and all absorption in excess of our Galaxy being at the host redshift\cite[e.g.][]{Behar2011,Campana2010,Campana2012}. Further, the absorber was assumed to be neutral.  Later studies used \textsc{absori} which has only 10 metals, and only Fe variable \citep{Starling2013a,Campana2015}. The use of solar metalicity leads to underestimate of column density, while the assumption of all excess absorption at the host redshift leads to overestimation of the column density. 

Our analysis, using the more sophisticated models for PIE and CIE, shows that substantially higher metallicity is indicated at lower redshifts compared to higher redshifts. As noted in more detail in Sections  \ref{sec:1} and \ref{sec:3}, there is some consensus for the diffuse cool IGM metallicity with redshifts $z = 2 - 4$ at $Z = 0.001~Z\sun$ ($[X/$H$] = -3)$, but with little or no observed evolution. However, these are either in Ly$\alpha$ forest regions or more dense systems such as LLSs or DLAs (e.g. S03, A08, F14, F16). At lower redshift $z = 0 - 2$, in the WHIM and the ICM, there is some consensus that the predicted mean metallicity is $Z \sim 0.1Z\sun$  \cite[e.g.][S12]{Wiersma2011,Danforth2016}, though it is unlikely that many GRB LOS pierce the ICM. Our CIE and PIE models in Section \ref{sec:4} show that regardless of the IGM model, both CIE and PIE are picking up substantial absorption by highly ionised absorbers. This use of sophisticated ionised absorber models for GRB has not been completed previously.

\cite{Campana2015} completed simulations of IGM absorption using GRBs and quasars. For GRBs, their simulations indicated that the LOS does not contain any absorbers with over-density $\Delta  > 100$, log$(T/$K) $\sim  5 - 7$  and mean metallicity $Z = 0.03Z\sun$. As we are tracing the full LOS and not any individual absorber, we cannot compare our results directly with \citet{Campana2015} in terms of overdensities. Their result for temperature range is consistent with ours for the CIE model. However, we find that a decline in metallicity is observed in both CIE and PIE scenarios for the IGM. In contrast, using AGNs, their simulations showed prevalence of absorption systems with large over-densities $(\Delta > 300)$ at $z \sim 0.5 - 1.2$,  temperature of $\sim (3 - 15) \times 10^6 $ K and mean metallicity in these regions of $Z = 0.3\pm0.1 Z\sun$. We would agree with their speculation that it is unlikely that GRB trace different LOS to AGN through the diffuse IGM and that therefore, these large overdensities and high metallicity may be proximate to the AGNs, e.g. in their CGM.

 \citet{Behar2011} noted that in their GRB sample the observed opacity at low energies, while high at low redshift, tended toward an asymptotic value at $z \sim 2$. They interpreted this as possible evidence for the detection of absorption by a diffuse, highly ionized intergalactic medium. This interpretation would solve the problems of the lack of correlation observed between the $\mathit{N}\textsc{hx}$ and $\mathit{N}\textsc{h i}$  in GRB afterglows, and the very low apparent dust-to-metals ratios. \citet{Rahin2019a} extended this earlier work to include all GRB up to 2019 and found very similar results.

There have been some claims in recent years to have found the missing baryons in the WHIM using different tracers. We now compare our work with some of these studies. \citet{Arcodia2018} used blazars as potential IGM absorption tracers. They modelled IGM absorption using \textsc{igmabs} in \textsc{xspec}. This is based on \textsc{absori}, with solar abundance and limited number of metals. Only 4 blazars were fitted. Their resulting average $n_0 = 1.01^{+0.53}_{-0.72}$ cm$^{-3}$ is lower than our result. Their temperature log($T$/K)$ = 6.45^{+0.51}_{-2.12}$ and ionisation log$(\xi) = 1.47\pm0.27$ (note log$(\xi)$ was tied to $n_0$) are consistent with our results. They derived a value of $Z_{IGM} = 0.59^{+0.31}_{-0.42}Z\sun$ based on an average IGM density from their fittings, as compared to $n_0 = 1.7 \times10^{-7}$ cm$^{-3}$ which is substantially higher than our results but they noted this should only be viewed as a consistency check.

 \citet{Macquart2020} used FRB dispersion measure (DM) to measure the total electron column density on the LOS to the FRB host. Their sample is limited to 5 FRB in the redshift range 0.12 to 0.52. To isolate the possible IGM component, they fix the Galactic DM to values measured by \citet{Cordes2002} with an additional fixed component to represent the Galactic halo of 50 pc cm$^{-3}$. They also assumed a fixed FRB host at DM$_{host}$ $= 50/(1 + z)$ pc cm $^{-3}$. They add that further analysis of their sample mildly favors a median host galaxy contribution of $\sim 100$ pc cm$^{-3}$ with a factor of two dispersion around this value. This is the conventional approach in FRB i.e. to fix the FRB host DM with the assumption that all excess DM is then due to the IGM. This is very different to the traditional GRB approach of assuming all absorption in excess of our Galaxy is at the host redshift. Their resulting baryon fraction is a median value of $\Omega_b h_{70} = 0.056$ and a $68\%$ confidence interval spanning $[0.046, 0.066]$. Based on this result they claim that their FRB DM measurements confirm the presence of baryons with the density estimated from the CMB and Big Bang Nucleosynthesis, and are consistent with all the missing baryons being present in the ionized intergalactic medium. Our median value for the baryon fraction for both CIE and PIE models is $\Omega_b h_{70} = 0.043$ and a $90\%$ confidence interval spanning $[0.014, 0.079]$ (CIE) and $[0.014,0.077]$ (PIE) which is consistent with their results. Their approach to fixing the host DM component is also analogous to our approach with GRB.

\citet{Nicastro2018} claim to have observed the WHIM in absorption. Only 1 to 2 strong $\mathrm{O\/\ \textsc{vii}}$ absorbers are predicted to exist per unit redshift. \citet{Nicastro2018} reported observations of two $\mathrm{O\/\ \textsc{vii}}$ systems. System 1 ($z = 0.43$) had $T = 6.8^{+9.9}_{-3.6} \times 10^{5}$ K, $\mathit{N}\textsc{hxigm} = 1.6^{+0.8}_{-0.5}/ (Z/Z\sun) \times 10^{19}$ cm$^{-2}$. System 2 ($z = 0.36$) had $T = 5.4^{+9.0}_{-1.7} \times 10^{5}$ K, $\mathit{N}\textsc{hxigm} = 0.9^{+1.5}_{-0.9}/(Z/Z\sun) \times 10^{19}$ cm$^{-2}$. With $Z=0.1Z\sun$, and with an average of 1.5 systems per unit redshift, this gives $\mathit{N}\textsc{hxigm} \sim 1.9 \times 10^{20}$ cm$^{-2}$ for $z \sim 1$. While the temperatures are consistent with our results, the column densities are an order of magnitude lower than our results and that of \citep{Arcodia2018} and \citep{Macquart2020} for blazars and FRB respectively. This could be interpreted as supporting the contribution of the IGM to absorption over and above individual strong WHIM absorbers.

It is not possible at present to detect the individual filaments using the thermal Sunyaev–Zeldovich (tSZ) effect as the signal is much smaller than both the noise in the latest CMB experiments, and compared to the sensitivity of Planck. \citet[hereafter T20]{Tanimura2020} used gas filaments between the Luminous Red Galaxy (LRG) pairs relying on stacking the individual frequency maps for 88000 pairs in the low redshift range $z < 0.4$.  The stacking removes the CMB component while the dust foreground becomes homogeneous. Their stacked tSQZ signal, with an assumed temperature of $5 \times 10^6 $ K for the filaments, gives an electron overdensity of $\sim 13$ , based on electron number density today to be $n_{e0} = 2.2 \times 10^{-7}$ cm$^{-3}$. This is consistent with \citep{Cen2006} and M19 simulations for the WHIM. Our results for the mean density and temperature ranges in the CIE scenario are consistent with these results, with the caveat that the IGM slab in our model is placed at half the GRB redshift i.e. at much higher redshifts than the T20 sample.

While there are GRB at low redshift with high $\mathit{N}\textsc{hx}$, the bulk of low redshift GRB are consistent with our work. The mean NHX (revised for $Z = 0.07Z\sun$ or actual metallicity, dust-corrected if available) for $z < 1$ taken from D20 is  $5.3 \times 10^{21}$ cm$^{-2}$. GRB $\mathit{N}\textsc{h i}$ do not show any relation with redshift. The mean for a sample again from D20 with $z < 2$ is $4.9 \times 10^{21}$ cm$^{-2}$. Following our method of approximating the GRB host column density as equal to $\mathit{N}\textsc{h i}$, this leaves only a small absorption difference $0.4 \times 10^{21}$ cm$^{-2}$. At $z \sim 0.5$, the mean IGM from our work is $\sim 1.5 \times 10^{21}$ cm$^{-2}$.

Some GRB at low redshift have very high X-ray absorption e.g. GRB190114C. The host-galaxy system of GRB190114C is composed of two galaxies, a close pair merger system \citep{Postigo2019}. Drawing from their observations, there are several possible factors which may explain the very high intrinsic $\mathit{N}\textsc{hx}$ in this low redshift GRB. The GRB exploded within the central cluster of the host galaxy, where the density is higher, at a projected distance of $\sim 170$ pc from the core. The GRB location is indicative of a denser environment than typically observed for GRBs.The host system stellar mass is an order of magnitude higher than the median value of GRB hosts at $0 < z < 1$ as measured for the BAT6 host sample. 
Finally,the GRB host has a much higher metallicity at 0.43 than the average GRB host at 0.07 from D20.

Our results show that substantial absorption probably occurs in the IGM in both the PIE and CIE scenarios. Most fits have consistently, if marginally better Cstat results compared to the simple model with all excess absorption occurring at the GRB host redshift. We would argue that while some excess absorption is attributable to the GRB host, and that better host models may identify this host excess absorption, the IGM  contributes  substantially  to  the  total  absorption  seen in GRB spectra, and that it indeed rises with redshift. This IGM absorption at least partly explains why $\mathit{N}\textsc{hxigm}$ seen in GRB full LOS afterglow spectra is substantially in excess of the intrinsic $\mathit{N}\textsc{h}\/\ \textsc{i}$ in GRB hosts. However, the CGM in the GRB host may also contribute to the $\mathit{N}\textsc{hxigm}$, and future models incorporating more advanced modelling for a warm/hot CGM component in GRB hosts are needed to explore the relative contribution of the IGM and the host CGM to the observed absorption.

\section{Conclusions}\label{sec:6}
  The main aim of this paper is to probe the key parameters of density, metallicity, temperature and photo-ionisation of the IGM using sophisticated software models for the ionised plasma. We use spectra from $\mathit{Swift}$ for GRBs as our tracers with a redshift range of $1.6 \leq z \leq 6.3$. We isolated the IGM LOS contribution to the total absorption for the GRBs by assuming that the GRB host absorption is equal to ionised corrected intrinsic neutral column $\textit{N}\textsc{h}\/\ \textsc{i,ic}$ estimated from the Ly$\alpha$ host absorption. We use more realistic host metallicity, dust corrected where available in generating the host absorption model. We  model  the  IGM  assuming  a  thin  uniform  plane  parallel  slab  geometry  in  thermal  and  ionization  equilibrium to represent a LOS  through  a  homogeneous  isothermal medium. We use \textsc{xspec} fitting with \textsc{steppar} and MCMC to generate best fits to the GRB spectra.  Our work uses the continuum total absorption to model plasma as opposed to fitting individual line absorption as the required resolution is not available currently in X-ray. We set the \textsc{xspec} metallicity parameter range as $-4 < [X/$H$] < -0.7$ ($0.0001 < Z/Z\sun < 0.2)$, with temperature for CIE at $4 < $ log$(T/$K)$ < 8$ and ionisation parameter between $0 \leq $ log($\xi$) $\leq 3$. The CIB photon index is fixed at 2.

Our main findings and conclusions are:
\begin{enumerate}
  \item Modelling the IGM using \textsc{hotabs} for CIE with parameters $\mathit{N}\textsc{hx}$, Z and T free appears to present plausible results for $\mathit{N}\textsc{hxigm}$ with an equivalent mean hydrogen density at $z = 0$ of $n_0 = 1.8^{+1.5}_{-1.2} \times 10^{-7}$ cm$^{-3}$. It shows similar values and correlation with redshift as the mean IGM density model, Fig. \ref{fig:CIEfreeplots} top-left panel. A power law fit to the $\mathit{N}\textsc{hxigm}$ versus redshift trend scales as $(1 + z)^{1.9\pm0.2}$.    
 \item  A power law fit to the $[X/$H$]$ and redshift trend for CIE scales as $(1+z)^{-5.2\pm1.0}$, Fig. \ref{fig:CIEfreeplots} top-right panel. Metallicity ranges from $[X/$H] $= -1\/\ (Z = 0.1Z\sun)$ at $z \sim 2$ to $[X/$H] $ \sim -3\/\ (Z = 0.001Z\sun)$ at high redshift $z > 4$. This could suggest that at low redshift, the higher metallicity warm-hot phase is dominant with $Z \sim0.1~Z\sun$, while at higher redshift the low metallicity IGM away from knots and filaments is dominant. 
 \item The CIE temperature range is $5.0 <$ log$(T/$K$)<\/\ 7.1$, Fig. \ref{fig:CIEfreeplots} bottom-left panel indicating that very highly ionised metals are prominent absorbers over the LOS. The mean temperature over the full redshift range is log$(T/$K$)$ = $6.3^{+0.6}_{-1.3}$. These values are consistent with the generally accepted WHIM range and with the latest simulations.  
 
\item Modelling the IGM using \textsc{warmabs} for PIE with $\mathit{N}\textsc{hxigm}$, Z and $\xi$ as free parameters appears to present plausible results though with more scatter at lower redshift compared to our CIE model . The PIE $\mathit{N}\textsc{hxigm}$ shows values and rise with redshift comparable to the mean IGM hydrogen density model in Fig. \ref{fig:PIEfreeplots} top-left panel. A power law fit to the $\mathit{N}\textsc{hxigm}$ versus redshift trend scales as $(1 + z)^{1.0\pm0.3}$, a much flatter power law than for CIE. The mean hydrogen density equivalent from this model at $z = 0$ is $n_0 = 1.8^{+1.4}_{-1.2} \times 10^{-7}$ cm$^{-3}$, very similar to the CIE result.
\item In the PIE scenario, there is a power law fit to the $[X/$H$]$ and redshift trend scaling as $(1 + z)^{-2.1\pm0.8}$, a slower decline than under the CIE IGM model. The metallicity is approximately $[X/$H$]\/\ = -1.0 \/\ (Z = 0.1Z\sun)$ at $z \sim 2$ falling to $[X/$H]$ = -2.4 \/\ (Z = 0.004Z\sun)$ at $z > 4$. 
\item The PIE ionisation parameter range from fits is $0.1 <$ log$(\xi) \/\ < 2.9$, Fig. \ref{fig:PIEfreeplots} bottom-left panel. The mean ionisation parameter over the full redshift range is log$(\xi) \/\ = \/\ 1.6^{+0.9}_{-1.2}$. 
\item  Regardless of the assumed ionisation state of the IGM, both models pick up considerable highly ionised absorption. 
\item We compared our CIE model with \textsc{absori} in Appendix A which was generally used in prior studies using GRBs as IGM tracers. \textsc{absori} is limited with only 10 metals, all fixed to solar metallicity except Fe. Our CIE and PIE IGM models use software which include all metals and ionisation species up to $Z \leq 30$, with variable metallicity. In conclusion, \textsc{absori} is no longer a preferred model for IGM absorption and the results of earlier studies using it for IGM modelling may not be reliable.
\item All our GRB spectra have fits as good as or better than the model with all excess absorption assumed to occur at the GRB host redshift. While some excess absorption may be attributable to the GRB host and its CGM, in our models the IGM contributes substantially to the total absorption seen in GRB spectra, and it rises with redshift. We provide clear evidence that a complete model should also account for a (possibly dominant) fraction of intervening IGM material.

This study is based on observations of GRB X-ray spectra, and provides results on the IGM parameters. The constraints will only be validated when observations are available from instruments with large effective area, high energy resolution, and a low energy threshold in the soft X-ray energy band e.g. Athena which will study the IGM through detailed observations of $\mathrm{O\/\ \textsc{vii}}$ and $\mathrm{O\/\ \textsc{viii}}$ absorption features with equivalent width $> 0.13$ eV and $> 0.09$ eV respectively.

\end{enumerate}

\section*{Acknowledgements}
We thank the reviewer, Darach Watson for his valuable and constructive feedback, which contributed to improving the quality of this paper. We thank T. Kallman and K. Arnaud for assistance with \textsc{warmabs}, \textsc{hotabs} and \textsc{xspec}, and E. Gatuzz with \textsc{ioneq}. S.L. Morris also acknowledges support from STFC (ST/P000541/1).
This project has received funding from the European Research Council (ERC) under the European Union's Horizon 2020 research and innovation programme (grant agreement No 757535) and by Fondazione Cariplo (grant No 2018-2329).

\section*{Data Availability}
Supplementary data containing the results for the IGM properties  from fitting the GRB spectra with the  PIE and CIE model  components,  and  the  transmission  plots  for  the  CIE and PIE models with different IGM parameter examples are available in the online supplementary material. For access to other data in this work, please contact Tony Dalton.



\bibliographystyle{mnras}
\bibliography{references.bib} 

\begin{thebibliography}{}
\makeatletter
\relax
\def\mn@urlcharsother{\let\do\@makeother \do\$\do\&\do\#\do\^\do\_\do\%\do\~}
\def\mn@doi{\begingroup\mn@urlcharsother \@ifnextchar [ {\mn@doi@}
  {\mn@doi@[]}}
\def\mn@doi@[#1]#2{\def\@tempa{#1}\ifx\@tempa\@empty \href
  {http://dx.doi.org/#2} {doi:#2}\else \href {http://dx.doi.org/#2} {#1}\fi
  \endgroup}
\def\mn@eprint#1#2{\mn@eprint@#1:#2::\@nil}
\def\mn@eprint@arXiv#1{\href {http://arxiv.org/abs/#1} {{\tt arXiv:#1}}}
\def\mn@eprint@dblp#1{\href {http://dblp.uni-trier.de/rec/bibtex/#1.xml}
  {dblp:#1}}
\def\mn@eprint@#1:#2:#3:#4\@nil{\def\@tempa {#1}\def\@tempb {#2}\def\@tempc
  {#3}\ifx \@tempc \@empty \let \@tempc \@tempb \let \@tempb \@tempa \fi \ifx
  \@tempb \@empty \def\@tempb {arXiv}\fi \@ifundefined
  {mn@eprint@\@tempb}{\@tempb:\@tempc}{\expandafter \expandafter \csname
  mn@eprint@\@tempb\endcsname \expandafter{\@tempc}}}

\bibitem[\protect\citeauthoryear{Aguirre, Dow‐Hygelund, Schaye  \&
  Theuns}{Aguirre et~al.}{2008}]{Aguirre2008}
Aguirre A.,  Dow‐Hygelund C.,  Schaye J.,   Theuns T.,  2008, \mn@doi [ApJ]
  {10.1086/592554}, 689, 851

\bibitem[\protect\citeauthoryear{Arcodia, Campana  \& Salvaterra}{Arcodia
  et~al.}{2016}]{Arcodia2016}
Arcodia R.,  Campana S.,   Salvaterra R.,  2016, \mn@doi [A{\&}A]
  {10.1051/0004-6361/201628326}, 590, 1

\bibitem[\protect\citeauthoryear{Arcodia, Campana, Salvaterra  \&
  Ghisellini}{Arcodia et~al.}{2018}]{Arcodia2018}
Arcodia R.,  Campana S.,  Salvaterra R.,   Ghisellini G.,  2018, A{\&}A, 590,
  A82

\bibitem[\protect\citeauthoryear{Arnaud}{Arnaud}{1996}]{Arnaud1996}
Arnaud K.,  1996, Astron. Data Anal. Softw. Syst., 101, 17

\bibitem[\protect\citeauthoryear{Asplund, Grevesse, Sauval  \& Scott}{Asplund
  et~al.}{2009}]{Asplund2009}
Asplund M.,  Grevesse N.,  Sauval A.~J.,   Scott P.,  2009, \mn@doi [Annual
  Review of Astronomy and Astrophysics] {10.1007/s10509-010-0288-z}, 47

\bibitem[\protect\citeauthoryear{Behar, Dado, Dar  \& Laor}{Behar
  et~al.}{2011}]{Behar2011}
Behar E.,  Dado S.,  Dar A.,   Laor A.,  2011, \mn@doi [ApJ]
  {10.1088/0004-637X/734/1/26}, 734, 26

\bibitem[\protect\citeauthoryear{Branchini et~al.,}{Branchini
  et~al.}{2009}]{Branchini2009}
Branchini E.,  et~al., 2009, \mn@doi [ApJ] {10.1088/0004-637X/697/1/328}, 697,
  328

\bibitem[\protect\citeauthoryear{Buchner et~al.,}{Buchner
  et~al.}{2014}]{Buchner2014}
Buchner J.,  et~al., 2014, \mn@doi [A{\&}A] {10.1051/0004-6361/201322971}, 564,
  1

\bibitem[\protect\citeauthoryear{Buchner, Schulze  \& Bauer}{Buchner
  et~al.}{2017}]{Buchner2017}
Buchner J.,  Schulze S.,   Bauer F.~E.,  2017, \mn@doi [MNRAS]
  {10.1093/mnras/stw2423}, 464, 4545

\bibitem[\protect\citeauthoryear{Burrows et~al.,}{Burrows
  et~al.}{2005}]{Burrows2005}
Burrows D.~N.,  et~al., 2005, \mn@doi [Space Science Reviews]
  {10.1007/s11214-005-5097-2}, 120, 165

\bibitem[\protect\citeauthoryear{Campana, Th{\"{o}}ne, {de Ugarte Postigo},
  Tagliaferri, Moretti  \& Covino}{Campana et~al.}{2010}]{Campana2010}
Campana S.,  Th{\"{o}}ne C.~C.,  {de Ugarte Postigo} A.,  Tagliaferri G.,
  Moretti A.,   Covino S.,  2010, \mn@doi [MNRAS]
  {10.1111/j.1365-2966.2009.16006.x}, 402, 2429

\bibitem[\protect\citeauthoryear{Campana et~al.,}{Campana
  et~al.}{2012}]{Campana2012}
Campana S.,  et~al., 2012, \mn@doi [MNRAS] {10.1111/j.1365-2966.2012.20428.x},
  421, 1697

\bibitem[\protect\citeauthoryear{Campana, Salvaterra, Ferrara  \&
  Pallottini}{Campana et~al.}{2015}]{Campana2015}
Campana S.,  Salvaterra R.,  Ferrara A.,   Pallottini A.,  2015, \mn@doi
  [A{\&}A] {10.1051/0004-6361/201425083}, 575, A43

\bibitem[\protect\citeauthoryear{Cash}{Cash}{1979}]{Cash1979}
Cash W.,  1979, ApJ, 228, 939

\bibitem[\protect\citeauthoryear{Cen \& Ostriker}{Cen \&
  Ostriker}{1999}]{Cen1999}
Cen R.,  Ostriker j.~p.,  1999, AJ, 514, 1

\bibitem[\protect\citeauthoryear{Cen \& Ostriker}{Cen \&
  Ostriker}{2006}]{Cen2006}
Cen R.,  Ostriker J.~P.,  2006, \mn@doi [ApJ] {10.1086/506505}, 650, 560

\bibitem[\protect\citeauthoryear{Cordes \& Lazio}{Cordes \&
  Lazio}{2002}]{Cordes2002}
Cordes J.~M.,  Lazio T. J.~W.,  2002, arXiv preprint astro-ph/0207156.

\bibitem[\protect\citeauthoryear{Crighton, Hennawi, Simcoe, Cooksey, Murphy,
  Fumagalli, {Xavier Prochaska}  \& Shanks}{Crighton
  et~al.}{2015}]{Crighton2015}
Crighton N.~H.,  Hennawi J.~F.,  Simcoe R.~A.,  Cooksey K.~L.,  Murphy M.~T.,
  Fumagalli M.,  {Xavier Prochaska} J.,   Shanks T.,  2015, \mn@doi [MNRAS]
  {10.1093/mnras/stu2088}, 446, 18

\bibitem[\protect\citeauthoryear{Dalton \& Morris}{Dalton \&
  Morris}{2020}]{Dalton2020}
Dalton T.,  Morris S.~L.,  2020, \mn@doi [MNRAS] {10.1093/mnras/staa1321}, 495,
  2342

\bibitem[\protect\citeauthoryear{Danforth \& Shull}{Danforth \&
  Shull}{2005}]{Danforth2005}
Danforth C.~W.,  Shull J.~M.,  2005, \mn@doi [ApJ] {10.1086/429285}, 624, 555

\bibitem[\protect\citeauthoryear{Danforth \& Shull}{Danforth \&
  Shull}{2008}]{Danforth2008}
Danforth C.~W.,  Shull J.~M.,  2008, ApJ, 679, 194

\bibitem[\protect\citeauthoryear{Danforth et~al.,}{Danforth
  et~al.}{2016}]{Danforth2016}
Danforth C.~W.,  et~al., 2016, \mn@doi [ApJ] {10.3847/0004-637X/817/2/111},
  817, 111

\bibitem[\protect\citeauthoryear{Dav{\'{e}} \& Oppenheimer}{Dav{\'{e}} \&
  Oppenheimer}{2007}]{Dave2007}
Dav{\'{e}} R.,  Oppenheimer B.~D.,  2007, \mn@doi [MNRAS]
  {10.1111/j.1365-2966.2006.11177.x}, 374, 427

\bibitem[\protect\citeauthoryear{{De Luca} \& Molendi}{{De Luca} \&
  Molendi}{2004}]{DeLuca2004}
{De Luca} A.,  Molendi S.,  2004, \mn@doi [A{\&}A]
  {10.1051/0004-6361:20034421}, 419, 837

\bibitem[\protect\citeauthoryear{Done, Mulchaey, Mushotzky  \& Arnaud}{Done
  et~al.}{1992}]{Done1992}
Done C.,  Mulchaey J.,  Mushotzky R.,   Arnaud K.,  1992, ApJ, 395, 275

\bibitem[\protect\citeauthoryear{Foreman-Mackey, Hogg, Lang  \&
  Goodman}{Foreman-Mackey et~al.}{2013}]{Foreman-Mackey2013}
Foreman-Mackey D.,  Hogg D.~W.,  Lang D.,   Goodman J.,  2013, \mn@doi [PASP]
  {10.1086/670067}, 125, 306

\bibitem[\protect\citeauthoryear{Fumagalli}{Fumagalli}{2014}]{Fumagalli2014}
Fumagalli M.,  2014, Memorie della Societa Astronomica Italiana, 85, 355

\bibitem[\protect\citeauthoryear{Fumagalli, O'Meara, Prochaska  \&
  Worseck}{Fumagalli et~al.}{2013}]{Fumagalli2013}
Fumagalli M.,  O'Meara J.~M.,  Prochaska J.~X.,   Worseck G.,  2013, \mn@doi
  [ApJ] {10.1088/0004-637X/775/1/78}, 775

\bibitem[\protect\citeauthoryear{Fumagalli, O'Meara  \& {Xavier
  Prochaska}}{Fumagalli et~al.}{2016}]{Fumagalli2016}
Fumagalli M.,  O'Meara J.~M.,   {Xavier Prochaska} J.,  2016, \mn@doi [MNRAS]
  {10.1093/mnras/stv2616}, 455, 4100

\bibitem[\protect\citeauthoryear{Fumagalli, Fotopoulou, Avenue  \&
  Bs}{Fumagalli et~al.}{2020}]{Fumagalli2020a}
Fumagalli M.,  Fotopoulou S.,  Avenue T.,   Bs B.,  2020, \mn@doi [MNRAS]
  {10.1093/mnras/staa2388}, 498, 1951

\bibitem[\protect\citeauthoryear{Galama \& Wijers}{Galama \&
  Wijers}{2001}]{Galama2001a}
Galama T.~J.,  Wijers R. A. M.~J.,  2001, \mn@doi [ApJ] {10.1086/319162}, 549,
  L209

\bibitem[\protect\citeauthoryear{Gatuzz \& Churazov}{Gatuzz \&
  Churazov}{2018}]{Gatuzz2018}
Gatuzz E.,  Churazov E.,  2018, \mn@doi [MNRAS] {10.1093/mnras/stx2776}, 474,
  696

\bibitem[\protect\citeauthoryear{Gnat \& Sternberg}{Gnat \&
  Sternberg}{2007}]{Gnat2007}
Gnat O.,  Sternberg A.,  2007, \mn@doi [ApJS] {10.1086/509786}, 168, 213

\bibitem[\protect\citeauthoryear{Goodman \& Weare}{Goodman \&
  Weare}{2010}]{Goodman2010}
Goodman J.,  Weare J.,  2010, Communications in Applied Mathematics and
  Computational Science, 5, 65

\bibitem[\protect\citeauthoryear{Haardt \& Madau}{Haardt \&
  Madau}{1996}]{Haardt96}
Haardt F.,  Madau P.,  1996, ApJ, 461, 20

\bibitem[\protect\citeauthoryear{Haardt \& Madau}{Haardt \&
  Madau}{2012}]{Haardt2012}
Haardt F.,  Madau P.,  2012, \mn@doi [ApJ] {10.1088/0004-637X/746/2/125}, 746

\bibitem[\protect\citeauthoryear{Harris et~al.,}{Harris
  et~al.}{2016}]{Harris2016}
Harris D.~W.,  et~al., 2016, \mn@doi [AJ] {10.3847/0004-6256/151/6/155}, 151, 1

\bibitem[\protect\citeauthoryear{Kaastra}{Kaastra}{2017}]{Kaastra2017}
Kaastra J.~S.,  2017, \mn@doi [A{\&}A] {10.1051/0004-6361/201629319}, 605, 2

\bibitem[\protect\citeauthoryear{Kallman, Bautista, Goriely, Mendoza, Miller,
  Palmeri, Quinet  \& Raymond}{Kallman et~al.}{2009}]{Kallman2009}
Kallman T.~R.,  Bautista M.~A.,  Goriely S.,  Mendoza C.,  Miller J.~M.,
  Palmeri P.,  Quinet P.,   Raymond J.,  2009, \mn@doi [ApJ]
  {10.1088/0004-637X/701/2/865}, 701, 865

\bibitem[\protect\citeauthoryear{Khabibullin \& Churazov}{Khabibullin \&
  Churazov}{2019}]{Khabibullin2019}
Khabibullin I.,  Churazov E.,  2019, \mn@doi [MNRAS] {10.1093/mnras/sty2992},
  482, 4972

\bibitem[\protect\citeauthoryear{Lehner, O'Meara, Howk, Prochaska  \&
  Fumagalli}{Lehner et~al.}{2016}]{Lehner2016}
Lehner N.,  O'Meara J.~M.,  Howk J.~C.,  Prochaska J.~X.,   Fumagalli M.,
  2016, \mn@doi [ApJ] {10.3847/1538-4357/833/2/283}, 833, 283

\bibitem[\protect\citeauthoryear{Lehner, Wotta, Howk, O'Meara, Oppenheimer  \&
  Cooksey}{Lehner et~al.}{2019}]{Lehner2019}
Lehner N.,  Wotta C.~B.,  Howk J.~C.,  O'Meara J.~M.,  Oppenheimer B.~D.,
  Cooksey K.~L.,  2019, \mn@doi [ApJ] {10.3847/1538-4357/ab41fd}, 887, 5

\bibitem[\protect\citeauthoryear{Luo et~al.,}{Luo et~al.}{2011}]{Luo2011}
Luo B.,  et~al., 2011, \mn@doi [ApJ] {10.1088/0004-637X/740/1/37}, 740

\bibitem[\protect\citeauthoryear{Lusso, Worseck, Hennawi, Prochaska, Vignali,
  Stern  \& O'Meara}{Lusso et~al.}{2015}]{Lusso2015}
Lusso E.,  Worseck G.,  Hennawi J.~F.,  Prochaska J.~X.,  Vignali C.,  Stern
  J.,   O'Meara J.~M.,  2015, \mn@doi [MNRAS] {10.1093/mnras/stv516}, 449, 4204

\bibitem[\protect\citeauthoryear{Macquart et~al.,}{Macquart
  et~al.}{2020}]{Macquart2020}
Macquart J.~P.,  et~al., 2020, \mn@doi [https://arxiv.org/abs/2005.13161]
  {10.1038/s41586-020-2300-2}, pp 1--54

\bibitem[\protect\citeauthoryear{Martizzi et~al.,}{Martizzi
  et~al.}{2019}]{Martizzi2019}
Martizzi D.,  et~al., 2019, \mn@doi [MNRAS] {10.1093/mnras/stz1106}, 486, 3766

\bibitem[\protect\citeauthoryear{McDonald et~al.,}{McDonald
  et~al.}{2016}]{McDonald2016}
McDonald M.,  et~al., 2016, \mn@doi [ApJ] {10.3847/0004-637x/826/2/124}, 826,
  124

\bibitem[\protect\citeauthoryear{McQuinn}{McQuinn}{2016}]{McQuinn2016a}
McQuinn M.,  2016, \mn@doi [Annual Review of Astronomy and Astrophysics]
  {10.1146/annurev-astro-082214-122355}, 54, 313

\bibitem[\protect\citeauthoryear{Mernier et~al.,}{Mernier
  et~al.}{2017}]{Mernier2017}
Mernier F.,  et~al., 2017, \mn@doi [A{\&}A] {10.1051/0004-6361/201630075}, 603,
  1

\bibitem[\protect\citeauthoryear{Moretti, Vattakunnel, Tozzi, Salvaterra,
  Severgnini, Fugazza, Haardt  \& Gilli}{Moretti et~al.}{2012}]{Moretti2012}
Moretti A.,  Vattakunnel S.,  Tozzi P.,  Salvaterra R.,  Severgnini P.,
  Fugazza D.,  Haardt F.,   Gilli R.,  2012, \mn@doi [A{\&}A]
  {10.1051/0004-6361/201219921}, 548, 1

\bibitem[\protect\citeauthoryear{Morris, Weymann, Savage  \& Gilliland}{Morris
  et~al.}{1991}]{Morris1991}
Morris S.,  Weymann R.,  Savage B.,   Gilliland R.,  1991, \mn@doi [ApJ]
  {10.1017/CBO9781107415324.004}, 377, L21

\bibitem[\protect\citeauthoryear{Nicastro, Krongold, Mathur  \& Elvis}{Nicastro
  et~al.}{2017}]{Nicastro2017}
Nicastro F.,  Krongold Y.,  Mathur S.,   Elvis M.,  2017, Astronomische
  Nachrichten, 338, 281

\bibitem[\protect\citeauthoryear{Nicastro et~al.,}{Nicastro
  et~al.}{2018}]{Nicastro2018}
Nicastro F.,  et~al., 2018, Nature, 558, 406

\bibitem[\protect\citeauthoryear{Oppenheimer \& Schaye}{Oppenheimer \&
  Schaye}{2013a}]{Oppenheimer2013}
Oppenheimer B.~D.,  Schaye J.,  2013a, \mn@doi [MNRAS] {10.1093/mnras/stt1043},
  434, 1043

\bibitem[\protect\citeauthoryear{Oppenheimer \& Schaye}{Oppenheimer \&
  Schaye}{2013b}]{Oppenheimer2013a}
Oppenheimer B.~D.,  Schaye J.,  2013b, \mn@doi [MNRAS] {10.1093/mnras/stt1150},
  434, 1063

\bibitem[\protect\citeauthoryear{Piattella}{Piattella}{2018}]{Piattella2018}
Piattella O.~F.,  2018, \mn@doi [https://arxiv.org/abs/1803.00070]
  {10.1007/978-3-319-95570-4}

\bibitem[\protect\citeauthoryear{Pieri et~al.,}{Pieri et~al.}{2014}]{Pieri2014}
Pieri M.,  et~al., 2014, \mn@doi [MNRAS] {10.1093/mnras/stu577}, 441, 1718

\bibitem[\protect\citeauthoryear{Postigo et~al.,}{Postigo
  et~al.}{2020}]{Postigo2019}
Postigo A. D.~U.,  et~al., 2020, A{\&}A, 633, A68

\bibitem[\protect\citeauthoryear{Pratt, Stocke, Keeney  \& Danforth}{Pratt
  et~al.}{2018}]{Pratt2018}
Pratt C.~T.,  Stocke J.~T.,  Keeney B.~A.,   Danforth C.~W.,  2018, \mn@doi
  [ApJ] {10.3847/1538-4357/aaaaac}, 855, 18

\bibitem[\protect\citeauthoryear{Raghunathan, Clowes, Campusano,
  S{\"{o}}chting, Graham  \& Williger}{Raghunathan
  et~al.}{2016}]{Raghunathan2016}
Raghunathan S.,  Clowes R.~G.,  Campusano L.~E.,  S{\"{o}}chting I.~K.,  Graham
  M.~J.,   Williger G.~M.,  2016, \mn@doi [MNRAS] {10.1093/mnras/stw2095}, 463,
  2640

\bibitem[\protect\citeauthoryear{Rahin \& Behar}{Rahin \&
  Behar}{2019}]{Rahin2019a}
Rahin R.,  Behar E.,  2019, \mn@doi [ApJ] {10.3847/1538-4357/ab3e34}, 885, 47

\bibitem[\protect\citeauthoryear{Ricci et~al.,}{Ricci et~al.}{2017}]{Ricci2017}
Ricci C.,  et~al., 2017, \mn@doi [ApJS] {10.3847/1538-4365/aa96ad}, 233, 17

\bibitem[\protect\citeauthoryear{Richter, Paerels  \& Kaastra}{Richter
  et~al.}{2008}]{Richter2008}
Richter P.,  Paerels F.~B.,   Kaastra J.~S.,  2008, \mn@doi [Space Science
  Reviews] {10.1007/s11214-008-9325-4}, 134, 25

\bibitem[\protect\citeauthoryear{Savage, Kim, Wakker, Keeney, Shull, Stocke  \&
  Green}{Savage et~al.}{2014}]{Savage2014}
Savage B.~D.,  Kim T.~S.,  Wakker B.~P.,  Keeney B.,  Shull J.~M.,  Stocke
  J.~T.,   Green J.~C.,  2014, \mn@doi [ApJ] {10.1088/0067-0049/212/1/8}, 212

\bibitem[\protect\citeauthoryear{Schady}{Schady}{2017}]{Schady2017}
Schady P.,  2017, Royal Society open science, 4, 170304

\bibitem[\protect\citeauthoryear{Schady, Savaglio, Kr{\"{u}}hler, Greiner,   \&
  Rau}{Schady et~al.}{2011}]{Schady2011}
Schady P.,  Savaglio S.,  Kr{\"{u}}hler T.,  Greiner J.,    Rau A.,  2011,
  A{\&}A, 525, A113

\bibitem[\protect\citeauthoryear{Schaye, Aguirre, Kim, Theuns, Rauch  \&
  Sargent}{Schaye et~al.}{2003}]{Schaye2003}
Schaye J.,  Aguirre A.,  Kim T.,  Theuns T.,  Rauch M.,   Sargent W. L.~W.,
  2003, \mn@doi [ApJ] {10.1086/378044}, 596, 768

\bibitem[\protect\citeauthoryear{Schaye et~al.,}{Schaye
  et~al.}{2015}]{Schaye2015}
Schaye J.,  et~al., 2015, \mn@doi [MNRAS] {10.1093/mnras/stu2058}, 446, 521

\bibitem[\protect\citeauthoryear{Selsing, Fynbo, Christensen  \&
  Krogager}{Selsing et~al.}{2016}]{Selsing2016}
Selsing J.,  Fynbo J. P.~U.,  Christensen L.,   Krogager J.-K.,  2016, \mn@doi
  [A{\&}A] {10.1051/0004-6361/201527096}, 585, a87

\bibitem[\protect\citeauthoryear{Shull, Smith  \& Danforth}{Shull
  et~al.}{2012}]{Shull2012}
Shull J.~M.,  Smith B.~D.,   Danforth C.~W.,  2012, \mn@doi [ApJ]
  {10.1088/0004-637X/759/1/23}, 759, 23

\bibitem[\protect\citeauthoryear{Shull, Danforth  \& Tilton}{Shull
  et~al.}{2014}]{Shull2014}
Shull J.~M.,  Danforth C.~W.,   Tilton E.~M.,  2014, \mn@doi [ApJ]
  {10.1088/0004-637X/796/1/49}, 796

\bibitem[\protect\citeauthoryear{Simcoe, Sargent  \& Rauch}{Simcoe
  et~al.}{2004}]{Simcoe2004}
Simcoe R.~A.,  Sargent W. L.~W.,   Rauch M.,  2004, \mn@doi [ApJ]
  {10.1086/382777}, 606, 92

\bibitem[\protect\citeauthoryear{Starling, Willingale, Tanvir, Scott, Wiersema,
  O'Brien, Levan  \& Stewart}{Starling et~al.}{2013}]{Starling2013a}
Starling R.~L.,  Willingale R.,  Tanvir N.~R.,  Scott A.~E.,  Wiersema K.,
  O'Brien P.~T.,  Levan A.~J.,   Stewart G.~C.,  2013, \mn@doi [MNRAS]
  {10.1093/mnras/stt400}, 431, 3159

\bibitem[\protect\citeauthoryear{Tanimura, Aghanim, Kolodzig, Douspis  \&
  Malavasi}{Tanimura et~al.}{2020}]{Tanimura2020}
Tanimura H.,  Aghanim N.,  Kolodzig A.,  Douspis M.,   Malavasi N.,  2020,
  \mn@doi [A{\&}A] {10.1051/0004-6361/202038521}, 643, 1

\bibitem[\protect\citeauthoryear{Tanvir et~al.,}{Tanvir
  et~al.}{2019}]{Tanvir2019}
Tanvir N.,  et~al., 2019, arXiv preprint arXiv:1903.07835

\bibitem[\protect\citeauthoryear{Tumlinson et~al.,}{Tumlinson
  et~al.}{2011}]{Tumlinson2011}
Tumlinson J.,  et~al., 2011, \mn@doi [Science] {10.1126/science.1209840}, 334,
  948

\bibitem[\protect\citeauthoryear{Tumlinson et~al.,}{Tumlinson
  et~al.}{2013}]{Tumlinson2013}
Tumlinson J.,  et~al., 2013, \mn@doi [ApJ] {10.1088/0004-637X/777/1/59}, 777

\bibitem[\protect\citeauthoryear{Walsh, McBreen, Martin-Carrillo, Dauser,
  Wijers, Wilms, Schaye  \& Barret}{Walsh et~al.}{2020}]{Walsh2020}
Walsh S.,  McBreen S.,  Martin-Carrillo A.,  Dauser T.,  Wijers N.,  Wilms J.,
  Schaye J.,   Barret D.,  2020, https://arxiv.org/abs/2007.10158, pp 1--14

\bibitem[\protect\citeauthoryear{Wang}{Wang}{2013}]{Wang2013}
Wang J.,  2013, \mn@doi [ApJ] {10.1088/0004-637X/776/2/96}, 776, 96

\bibitem[\protect\citeauthoryear{Watson}{Watson}{2011}]{Watson2011}
Watson D.,  2011, \mn@doi [A{\&}A] {10.1051/0004-6361/201117120}, 533, 16

\bibitem[\protect\citeauthoryear{Watson \& Jakobsson}{Watson \&
  Jakobsson}{2012}]{Watson2012}
Watson D.,  Jakobsson P.,  2012, \mn@doi [ApJ] {10.1088/0004-637X/754/2/89},
  754

\bibitem[\protect\citeauthoryear{Watson, Hjorth, Fynbo, Jakobsson, Foley,
  Sollerman  \& Wijers}{Watson et~al.}{2007}]{Watson2007}
Watson D.,  Hjorth J.,  Fynbo J. P.~U.,  Jakobsson P.,  Foley S.,  Sollerman
  J.,   Wijers R. A. M.~J.,  2007, \mn@doi [ApJ] {10.1086/518310}, 660, L101

\bibitem[\protect\citeauthoryear{Watson et~al.,}{Watson
  et~al.}{2013}]{Watson2013}
Watson D.,  et~al., 2013, \mn@doi [ApJ] {10.1088/0004-637X/768/1/23}, 768, 23

\bibitem[\protect\citeauthoryear{Werk, Prochaska, Thom, Tumlinson, Tripp,
  O'Meara  \& Peeples}{Werk et~al.}{2013}]{Werk2013}
Werk J.~K.,  Prochaska J.~X.,  Thom C.,  Tumlinson J.,  Tripp T.~M.,  O'Meara
  J.~M.,   Peeples M.~S.,  2013, \mn@doi [ApJ] {10.1088/0067-0049/204/2/17},
  204

\bibitem[\protect\citeauthoryear{Wiersma, Schaye  \& Theuns}{Wiersma
  et~al.}{2011}]{Wiersma2011}
Wiersma R.~P.,  Schaye J.,   Theuns T.,  2011, \mn@doi [MNRAS]
  {10.1111/j.1365-2966.2011.18709.x}, 415, 353

\bibitem[\protect\citeauthoryear{Willingale, Starling, Beardmore, Tanvir  \&
  O'brien}{Willingale et~al.}{2013}]{Willingale2013}
Willingale R.,  Starling R.~L.,  Beardmore A.~P.,  Tanvir N.~R.,   O'brien
  P.~T.,  2013, \mn@doi [MNRAS] {10.1093/mnras/stt175}, 431, 394

\bibitem[\protect\citeauthoryear{Wilms, Allen  \& McCray}{Wilms
  et~al.}{2000}]{Wilms2000}
Wilms J.,  Allen A.,   McCray R.,  2000, \mn@doi [ApJ] {10.1086/317016}, 542,
  914

\bibitem[\protect\citeauthoryear{Wotta, Lehner, Howk, O'Meara, Oppenheimer  \&
  Cooksey}{Wotta et~al.}{2019}]{Wotta2019}
Wotta C.~B.,  Lehner N.,  Howk J.~C.,  O'Meara J.~M.,  Oppenheimer B.~D.,
  Cooksey K.~L.,  2019, \mn@doi [ApJ] {10.3847/1538-4357/aafb74}, 872, 81

\bibitem[\protect\citeauthoryear{York et~al.,}{York et~al.}{2000}]{York2000}
York D.~G.,  et~al., 2000, AJ, 120, 1579

\bibitem[\protect\citeauthoryear{Zhang, Zhang, Li  \& Lorimer}{Zhang
  et~al.}{2020}]{Zhang2020a}
Zhang R.~C.,  Zhang B.,  Li Y.,   Lorimer D.~R.,  2020,
  https://arxiv.org/abs/2011.06151

\makeatother
\end{thebibliography}

\section*{SUPPORTING INFORMATION}
Supplementary data containing the results for the IGM properties from fitting the GRB spectra with the PIE and CIE model components, and the transmission plots for the CIE and PIE models with different IGM parameter examples are available at MNRAS online.


\clearpage
\appendix

\section{Model comparisons and investigating robustness of CIE and PIE free parameter fits}\label{app:C}
\subsection{Metallicity fixed to $Z = 0.01Z\sun$}

\graphicspath{ {./figurespaper2/}  }

\begin{figure*}
     \centering
     \begin{tabular}{c|c|c}
    \includegraphics[scale=0.3]{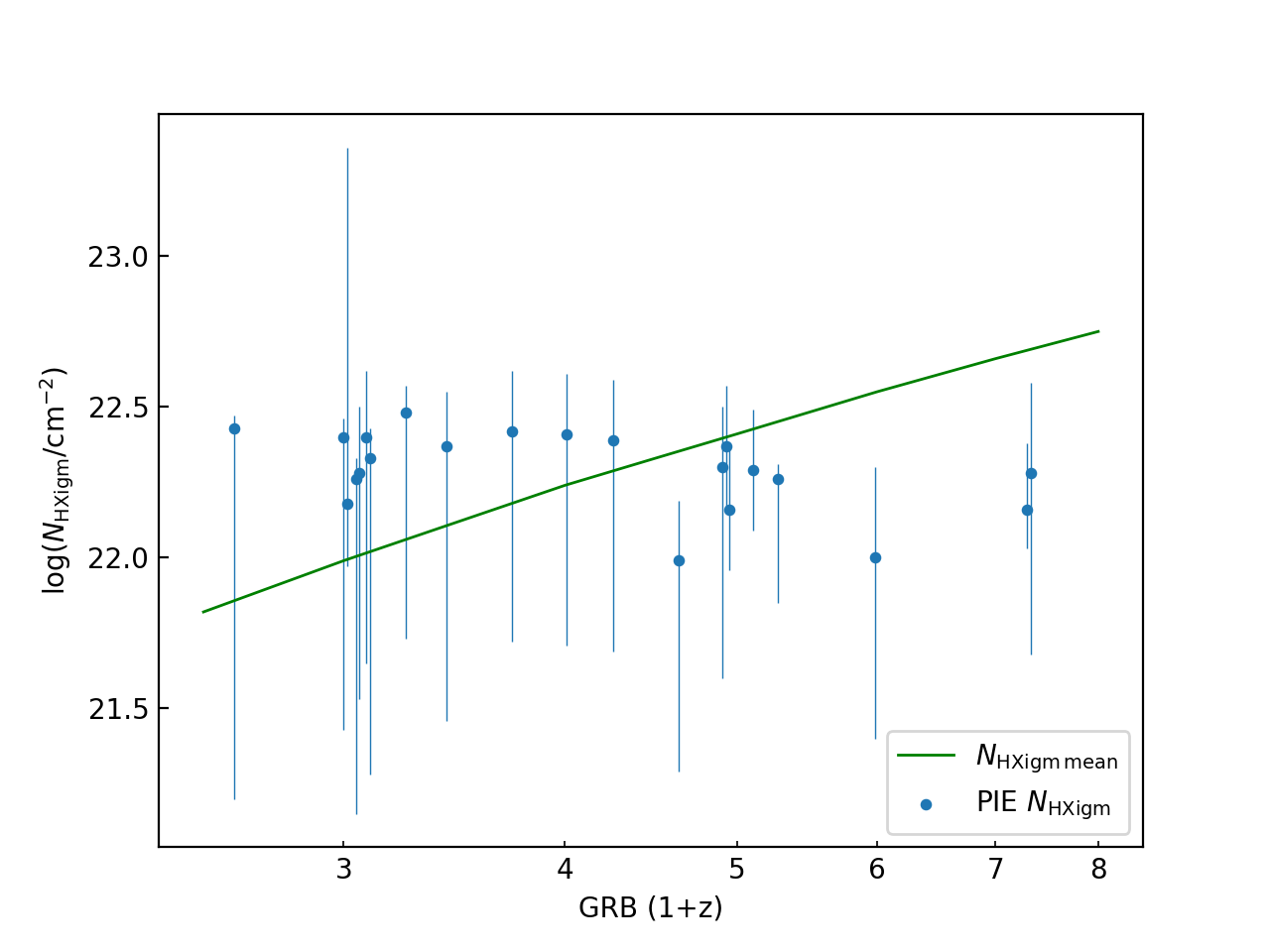} &
    \includegraphics[scale=0.3]{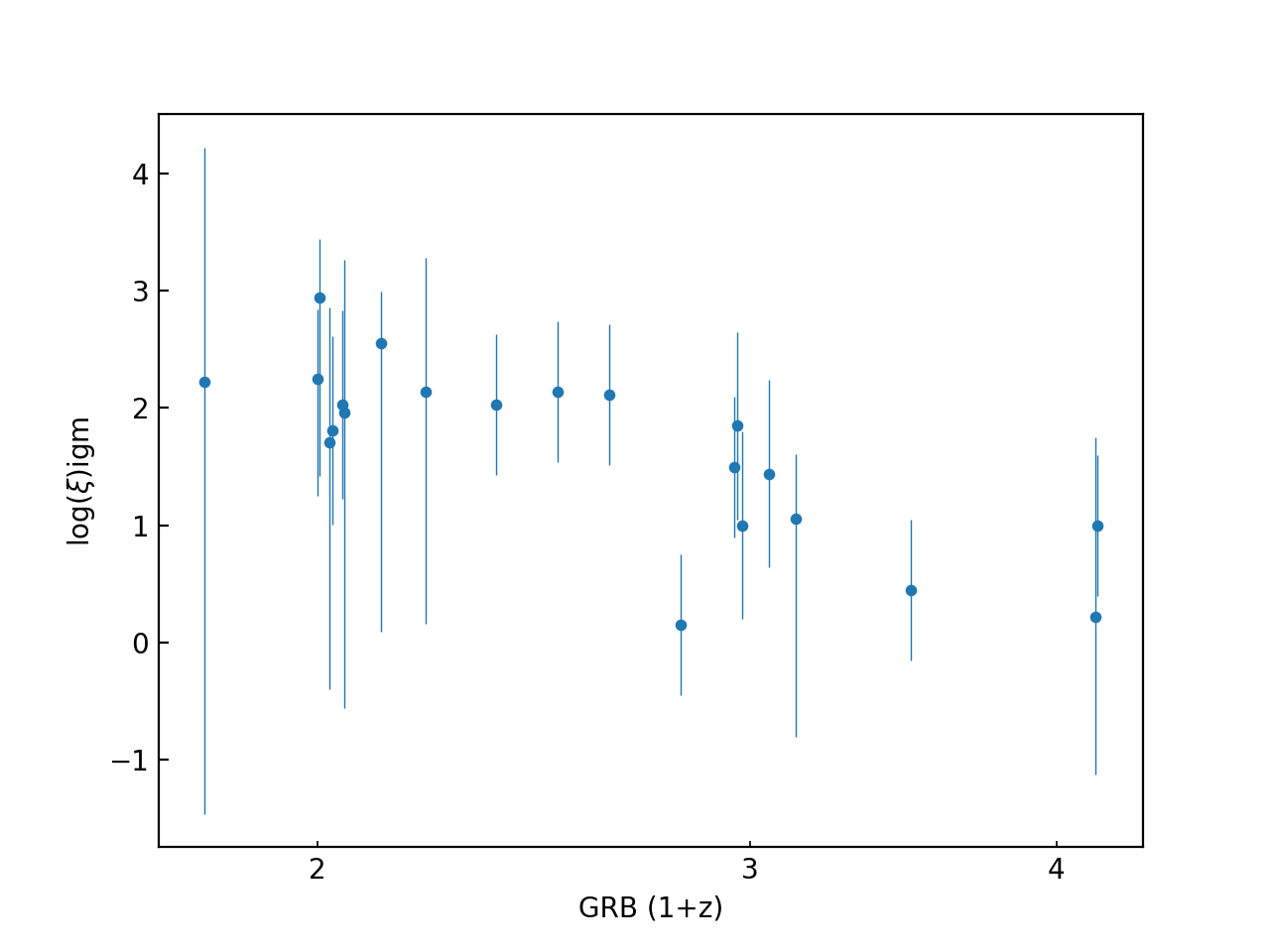} &
    \includegraphics[scale=0.16]{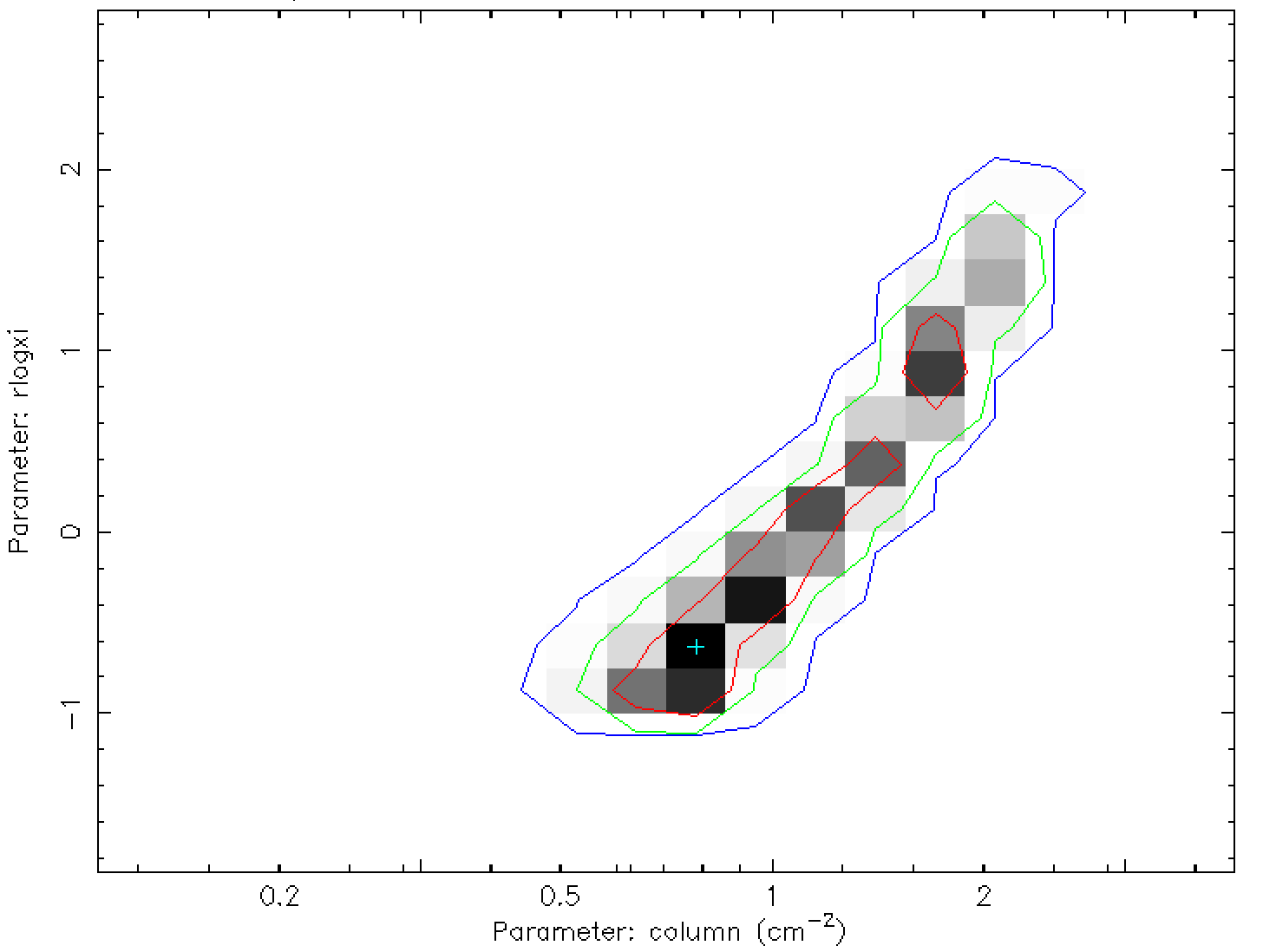}
     \\
    \end{tabular}
    \caption{Results for the IGM parameters using the PIE warmabs model with $Z = 0.01Z\sun$. The error bars are reported with a $90\%$ confidence interval. Left panel is $\textit{N}\textsc{hx}$ and redshift. The green line is the simple IGM model using the mean IGM density. Middle panel is ionisation parameter versus redshift. Right panel is an example of an integrated MCMC plot. The red, green and blue contours represent $68\%, 95\%$ and $99\%$ ranges for the two parameters respectively, with grey-scale showing increasing integrated probability from dark to light. On the y-axis rlogxi = log$(\xi)$.}
        \label{fig:PIEZ.01}
\end{figure*}
\graphicspath{ {./figurespaper2/}  }

\begin{figure*}
     \centering
     \begin{tabular}{c|c|c}
    \includegraphics[scale=0.3]{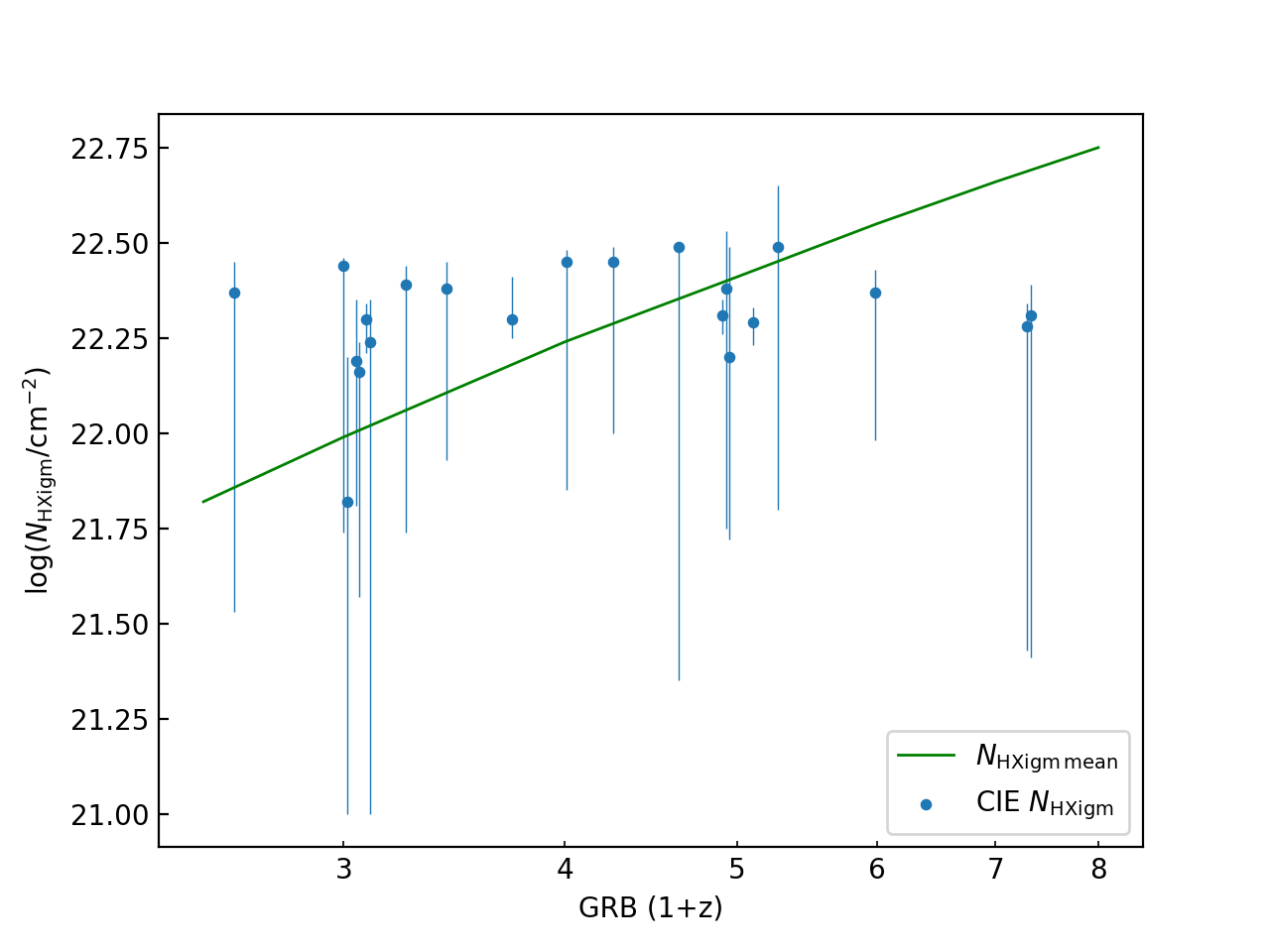} &
    \includegraphics[scale=0.3]{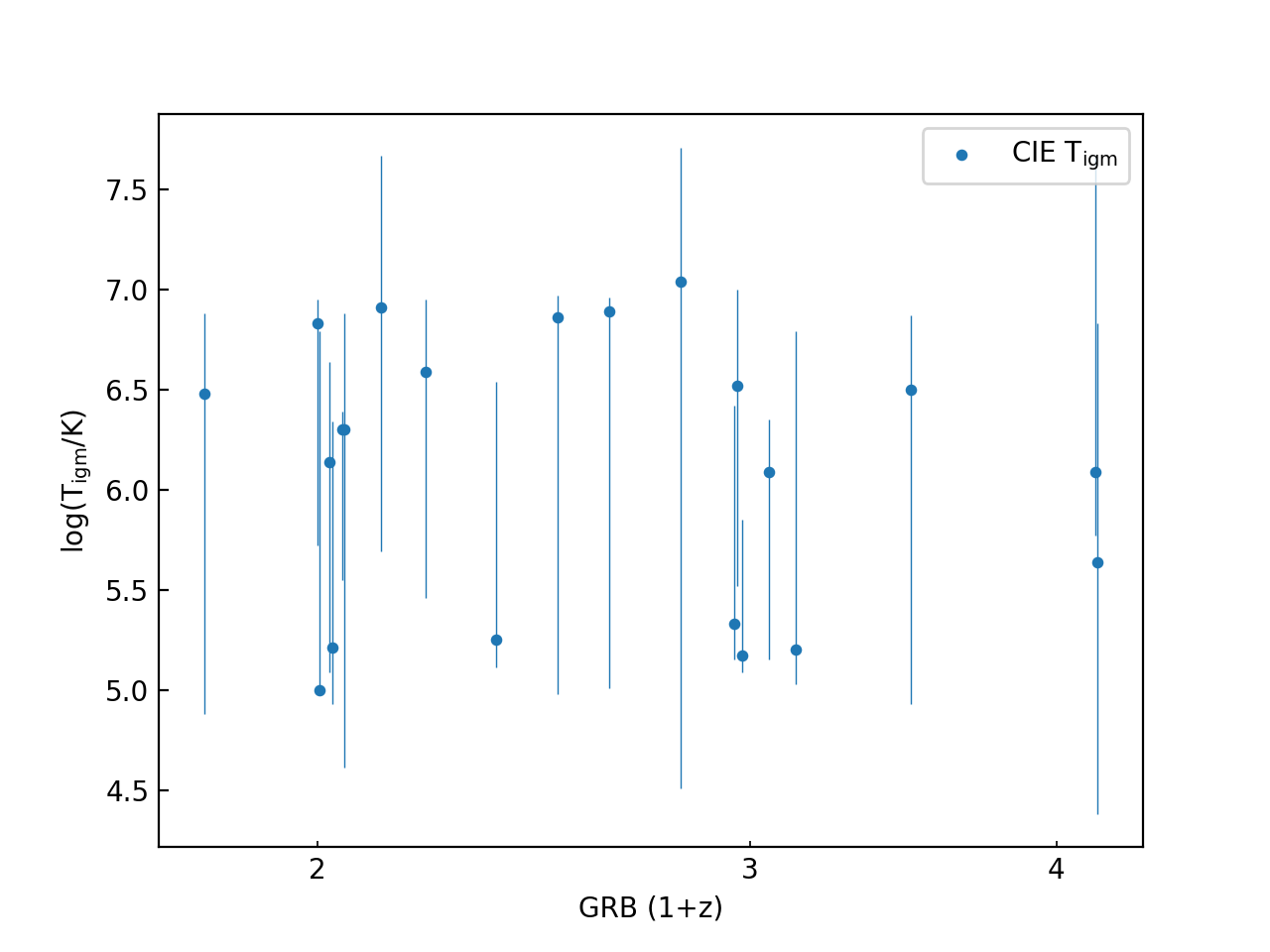} &
    \includegraphics[scale=0.16]{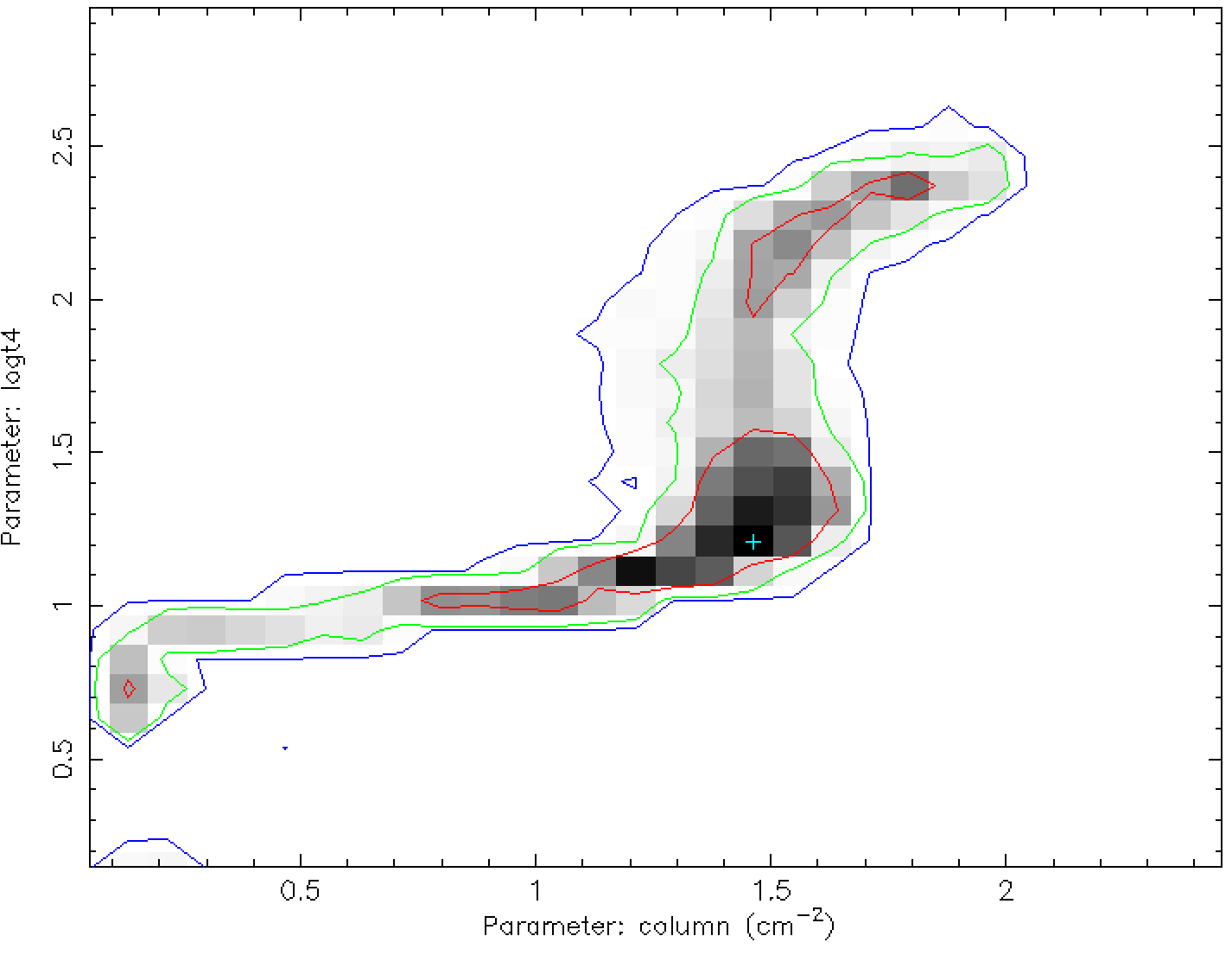}
     \\
    \end{tabular}
    \caption{Results for the IGM parameters using the CIE \textsc{hotabs} model with $Z = 0.01\sun$CIE Z=0.01. The error bars  are reported with a $90\%$ confidence interval. Left panel is $\textit{N}\textsc{hx}$ and redshift. The green line is the simple IGM model using the mean IGM density. Middle panel is temperature versus redshift. Right panel is an example of an integrated MCMC plot. The red, green and blue contours represent $68\%, 95\%$ and $99\%$ ranges for the two parameters respectively, with grey-scale showing increasing integrated probability from dark to light. On the y-axis in the bottom-left panel T4 means the log of the temperature is in units of 10$^4$ K.}
    \label{fig:CIEZ.01}
\end{figure*}
\graphicspath{ {./figurespaper2/}  }

\begin{figure*}
     \centering
     \begin{tabular}{c|c|c}
    \includegraphics[scale=0.3]{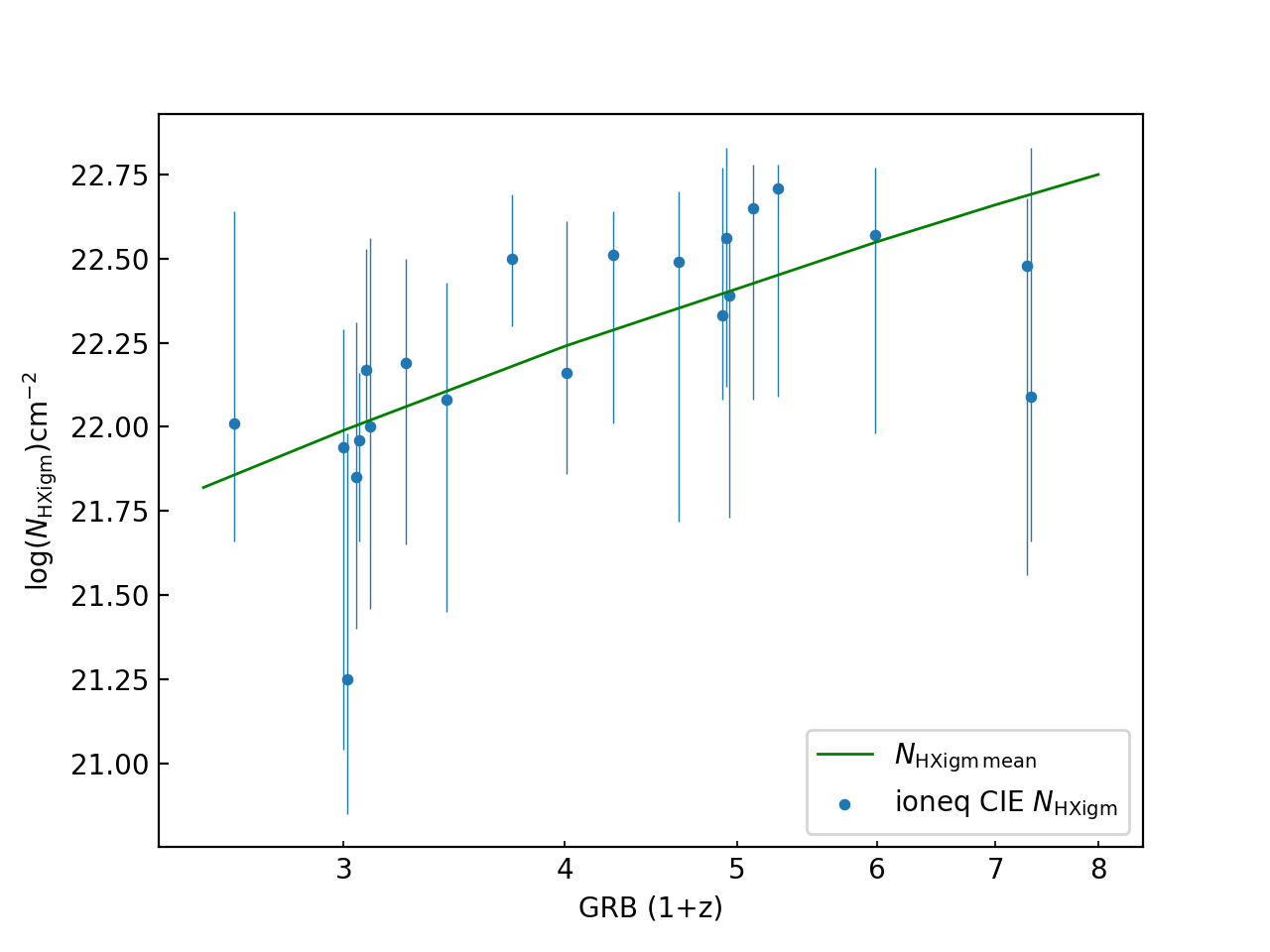} &
    \includegraphics[scale=0.3]{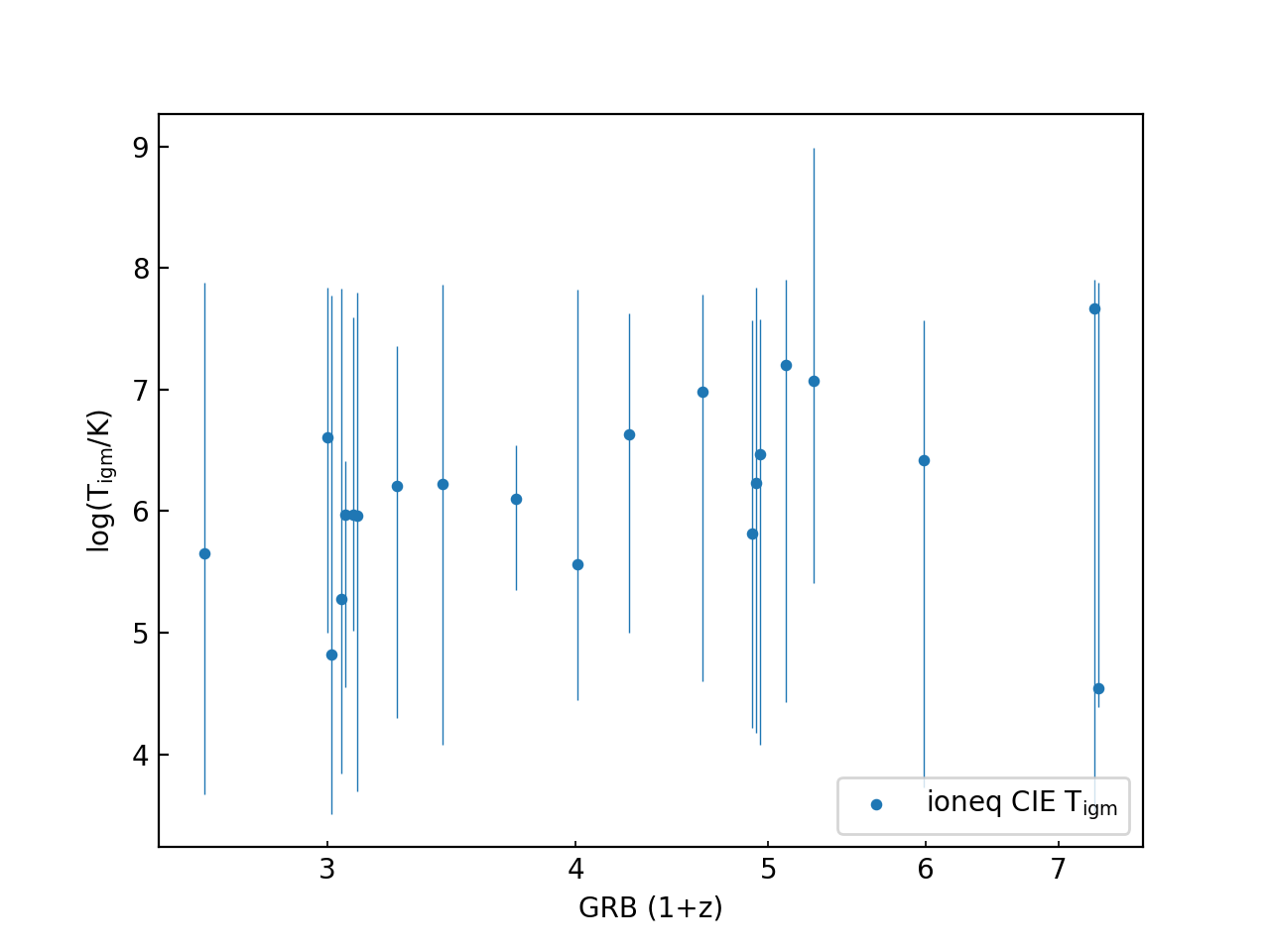} &
    \includegraphics[scale=0.16]{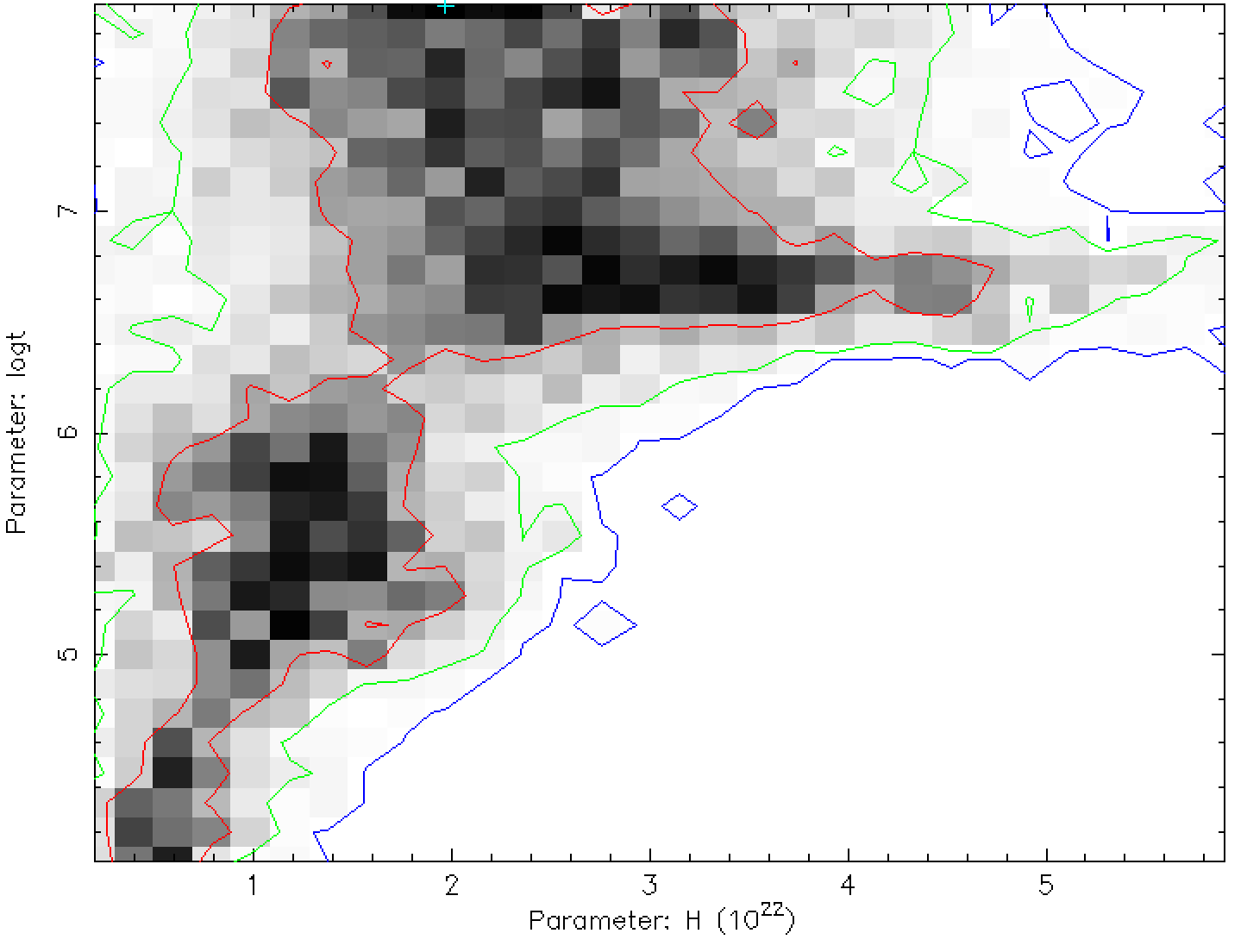}
     \\
    \end{tabular}
    \caption{Results for the IGM parameters using the \textsc{ioneq} CIE  model with IGM $Z = 0.01\sun$. The error bars  are reported with a $90\%$ confidence interval. Left panel is $\textit{N}\textsc{hx}$ and redshift. The green line is the simple IGM model using the mean IGM density. Middle panel is temperature versus redshift. Right panel is an example of an integrated MCMC plot. The red, green and blue contours represent $68\%, 95\%$ and $99\%$ ranges for the two parameters respectively, with grey-scale showing increasing integrated probability from dark to light.}
    \label{fig:ioneqCIEZ.01}
\end{figure*}

\graphicspath{ {./figurespaper2/}  }

\begin{figure*}
     \centering
     \begin{tabular}{c|c}
    \includegraphics[scale=0.33]{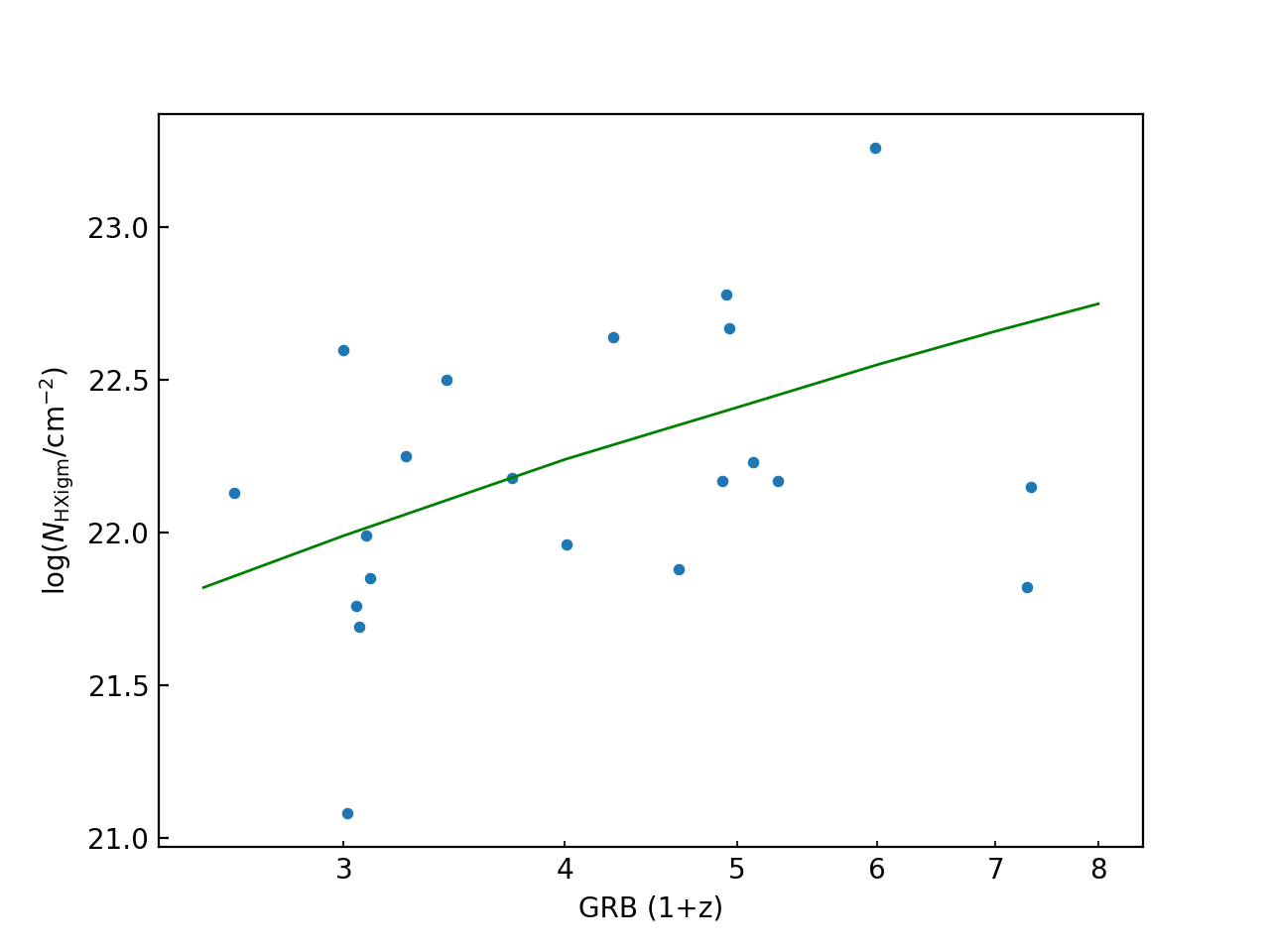} &
    \includegraphics[scale=0.33]{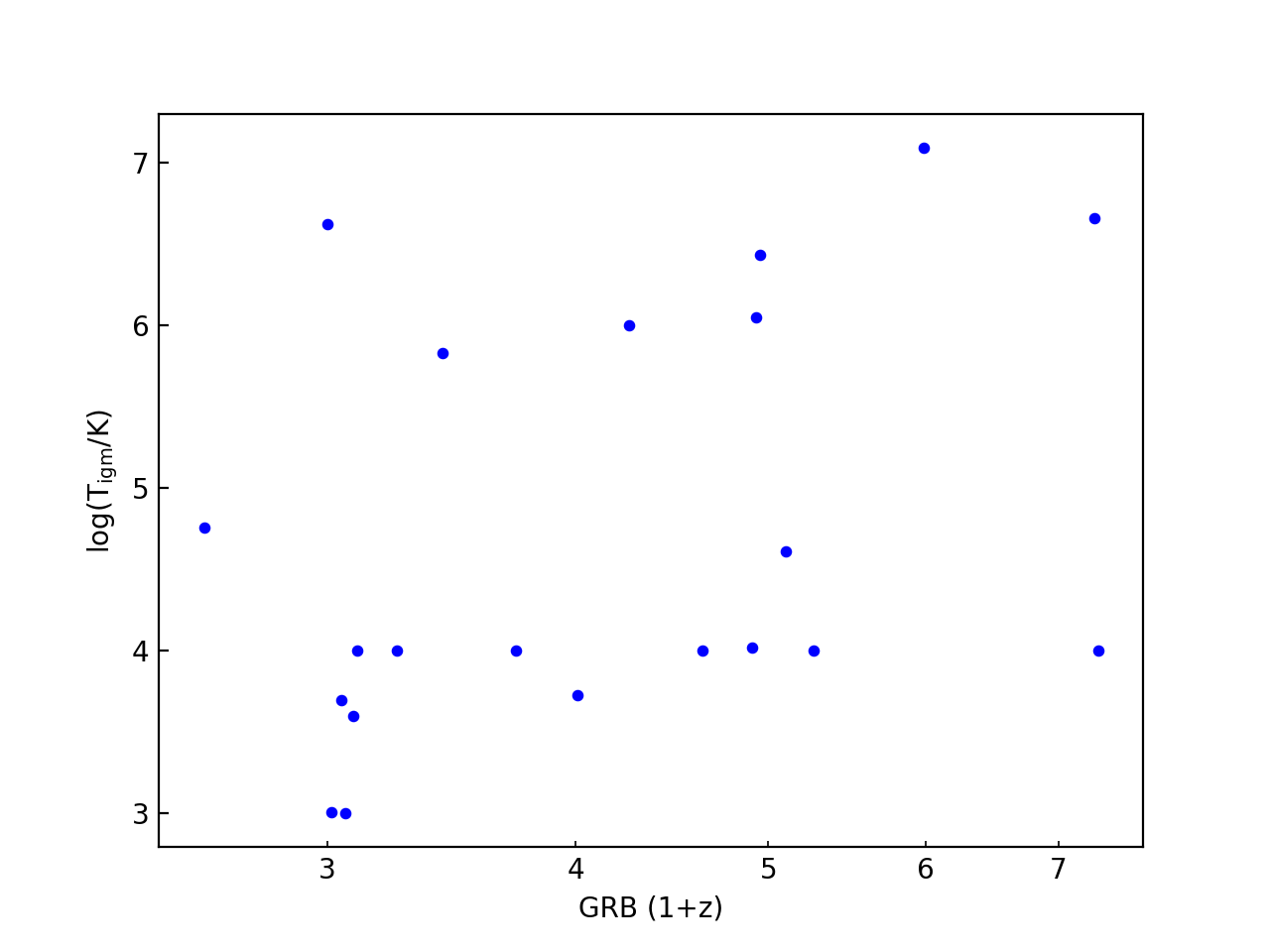}
     \\
     \end{tabular}
    \caption{Results for the IGM parameters using the \textsc{absori} CIE  model with the IGM $Z = 0.01\sun$. Left panel is $\textit{N}\textsc{hx}$ and redshift. The green line is the simple IGM model using the mean IGM density. Right panel is temperature and redshift. No error bars could be generated by $\textsc{xpec}$.}
    \label{fig:absoriCIEZ.01}
\end{figure*}

 To test the selected PIE and CIE models, we conducted trials freezing one key parameter at a time. We limited the sample data size to 15 covering the full redshift range $1.6 - 6.3$. For the first trial we froze the metallicity of the IGM to $Z = 0.01Z\sun$ as representative of the diffuse IGM. Most studies of the cooler PIE IGM in Ly$\alpha$ regions found virtually no evidence for metallicity evolution in the range z $\sim 2 - 4$ (e.g. S03, A08) so it is reasonable to test this scenario. The green line for all models is the mean density of the IGM based on the simple model from eq. \ref{eq:simpleIGM}.

\subsubsection{\textsc{warmabs} $(\mathrm{PIE})$}
In Fig. \ref{fig:PIEZ.01} left, the expected increase of $\mathit{N}\textsc{hxigm}$ with redshift does not arise with \textsc{warmabs} when the metallicity is frozen to $Z = 0.01Z\sun$. The low redshift $\mathit{N}\textsc{hxigm}$ is substantially higher than the expected mean IGM density, while at higher redshifts it is below the IGM mean density. This could indicate that a fixed metallicity assumption is unrealistic, or the PIE model is not appropriate for the LOS to the GRB. In Fig. \ref{fig:PIEZ.01} middle panel, we see a wide range of ionisation parameters with substantial error bars at lower redshift. There appears to be a negative trend with redshift, though this may be due to the metallicity being fixed.

Fig. \ref{fig:PIEZ.01} right panel shows an example of the a MCMC integrated probability plot for the \textsc{warmabs} PIE $\mathit{N}\textsc{hxigm}$ and $\xi$. Most MCMC integrated probability plots are reasonably consistent with $\textsc{steppar}$, indicating a good fit with low Cstat, but some are not. In this example, there are a few islands of high probability. As for several of the GRB, its shows that the best fit could have occurred at the low or high end of the confidence range.
In conclusion, primarily due to the result for column density, it is likely, that a fixed metallicity warmabs based PIE model for the IGM is not realistic.

\subsubsection{\textsc{hotabs} $(\mathrm{CIE})$}
Similar to the \textsc{warmabs} PIE model, the expected increase of $\mathit{N}\textsc{hxigm}$ with redshift in Fig. \ref{fig:CIEZ.01} left panel does not arise with \textsc{hotabs} when the metallicity is frozen to $Z = 0.01Z\sun$. This could indicate that a fixed metallicity assumption is unrealistic, or that the CIE model is not appropriate.  Again, at low redshift, the $\mathit{N}\textsc{hxigm}$ is much greater than the mean density model, while at high redshift it is much lower. The error bars are very large. In Fig. \ref{fig:CIEZ.01} middle panel, we see a wide range of temperatures with substantial error bars. The best fit data points appear to favour either the high or low end of the $90\%$ confidence range. Finally, Fig. \ref{fig:CIEZ.01} right panel shows an example of the MCMC integrated probability plot for the \textsc{hotabs} CIE $\mathit{N}\textsc{hxigm}$ and T. In this example, there is a characteristic S shape where, at high column density, a range of temperatures at a similar column density could fit, while at low temperature, there is a different range of column  densities that could fit. There is a single high maximum but there are a couple of islands of 1 sigma probability. 

In conclusion, it is likely, that a fixed metallicity \textsc{hotabs} based CIE model for the IGM is not realistic.

\subsubsection{\textsc{ioneq} $(\mathrm{CIE})$}
Modelling the IGM using \textsc{ioneq} with a fixed metallicity appears to present plausible results for $\mathit{N}\textsc{hxigm}$ in Fig. \ref{fig:ioneqCIEZ.01} left panel, showing a similar rise with redshift as the mean IGM density model, except at very high redshift. A power law fit to the $\mathit{N}\textsc{hxigm}$ versus redshift trend scales as $(1 + z)^{1.8\pm0.5}$ . We note that all metals included in the \textsc{ioneq} model, except O, Ne and Fe, are fixed to the solar abundance, an unrealistic value for the diffuse IGM. \textsc{ioneq} is currently being updated to allow all metals as free parameters but was not available for this paper (Gatuzz, E., private communication). As with \textsc{hotabs}, the error bars on temperature with redshift in Fig. \ref{fig:ioneqCIEZ.01} middle panel are substantial, but the best fits do not favour the high or low end of the confidence interval. 

Most MCMC integrated probability plots for \textsc{ioneq} fittings show large degeneracy as seen in the example in Fig. \ref{fig:ioneqCIEZ.01} right panel, with many local maxima. The Cstat fits to the GRB spectra are as good as for \textsc{warmabs} and \textsc{hotabs}. In conclusion, the plots suggest that a CIE IGM model with fixed metallicity of $Z = Z0.01\sun$ may be plausible. However, due to the MCMC showing substantial degeneracies, and the unrealistic solar metallicities, we have not used this model.

\subsubsection{\textsc{absori}}

Fig. \ref{fig:absoriCIEZ.01} left panel shows the results for $\mathit{N}\textsc{hxigm}$ using \textsc{absori} for the IGM absorption with metallicity fixed again at $Z = 0.01Z\sun$. In \textsc{absori}, only Fe is affected as the other 9 metals in the model are fixed to solar. The ionisation parameter was fixed at $\xi = 0$, so only temperature was allowed to vary as a CIE model. The fits were very poor, errors could not be generated in \textsc{xspec}, and the MCMC runs failed to generate plausible results. No apparent redshift correlation can be seen, similar to \textsc{warmabs} and \textsc{hotabs}. Due to the poor fits, it cannot be said whether this is due to the \textsc{absori} model being limited to 10 metals, having all metals, except Fe at solar, edge absorption only or the model not being self-consistent. Fig \ref{fig:absoriCIEZ.01} right panel shows the IGM temperatures from the fittings. As with all models, it shows a large scatter. In conclusion, \textsc{absori} is no longer an ideal model for IGM absorption.

In summary, \textsc{warmabs} and \textsc{hotabs} are the most sophisticated models, and the MCMC integrated probability plots were the most consistent with the $\textsc{steppar}$ results and have plausible integrated probability plots. Most show a single deep maximun, but there is degeneracy with several possible parameter fit solutions. Accordingly, we decided to proceed only with \textsc{warmabs} and \textsc{hotabs}, and not \textsc{ioneq} nor \textsc{absori} for the remaining tests. However, the fixing of metallicity for both PIE and CIE IGM models with redshift is not appropriate for any model.

\graphicspath{ {./figurespaper2/}  }

\begin{figure*}
     \centering
     \begin{tabular}{c|c}
    \includegraphics[scale=0.4]{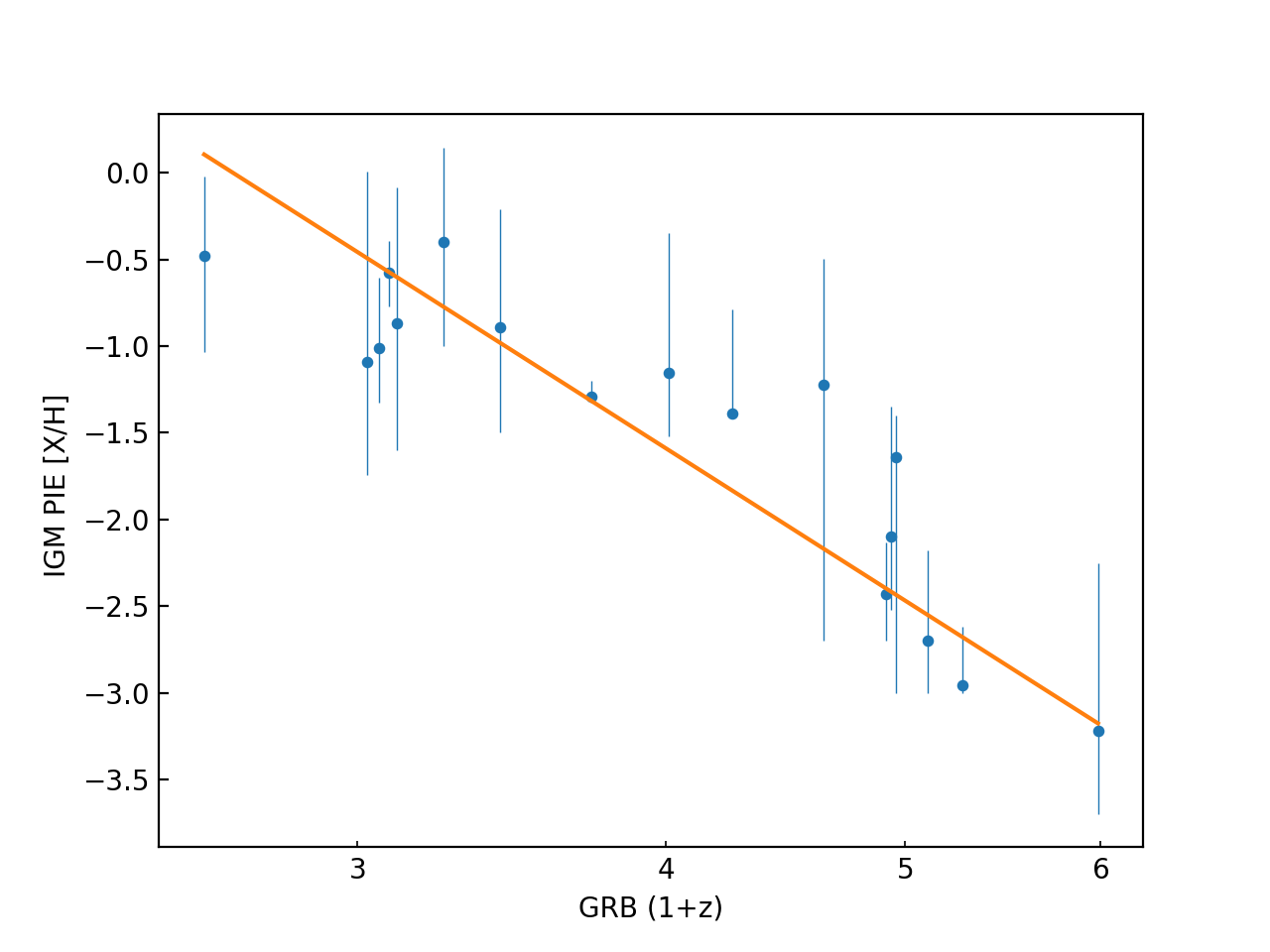} &
    \includegraphics[scale=0.4]{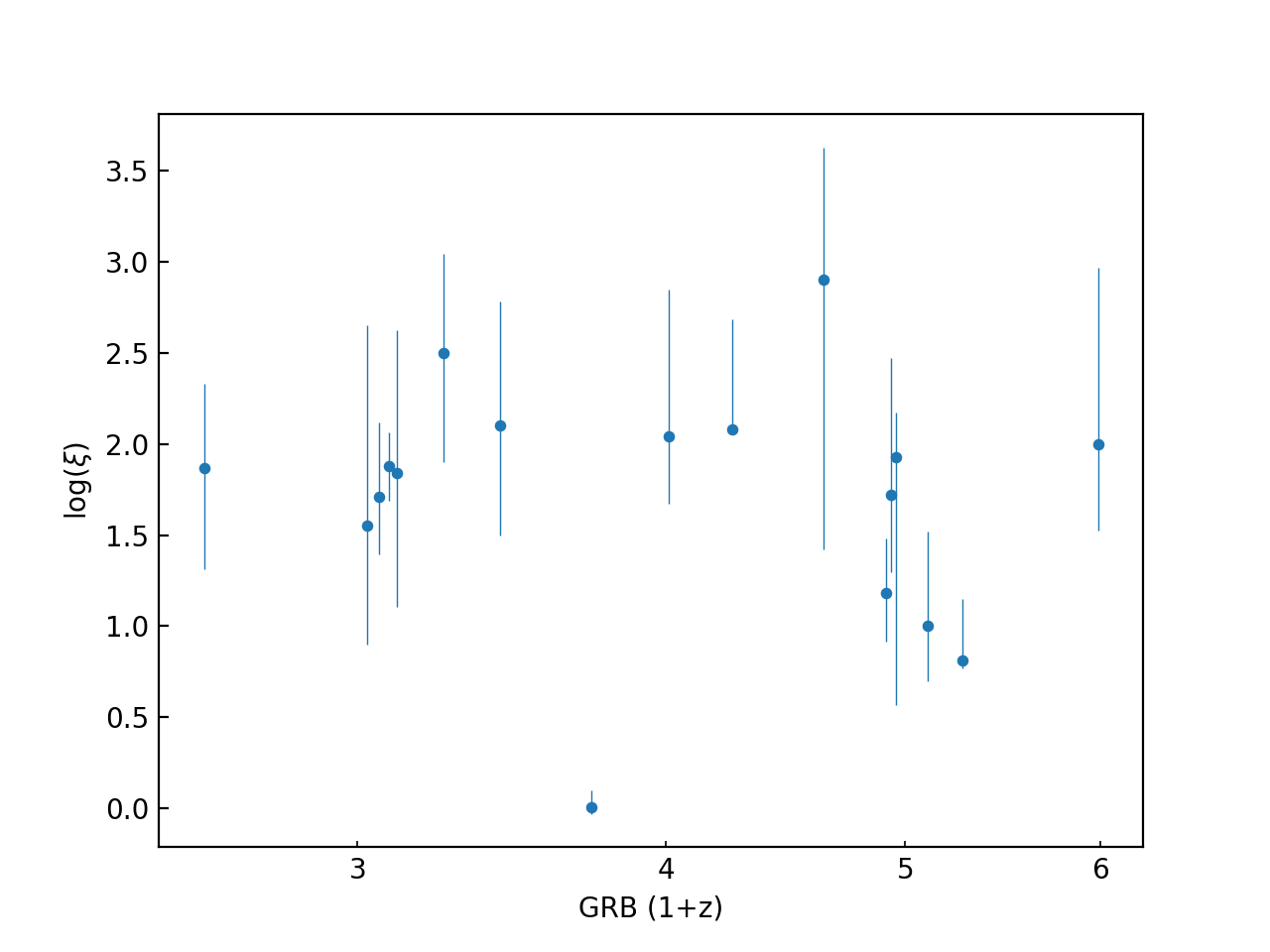} \\
    \includegraphics[scale=0.4]{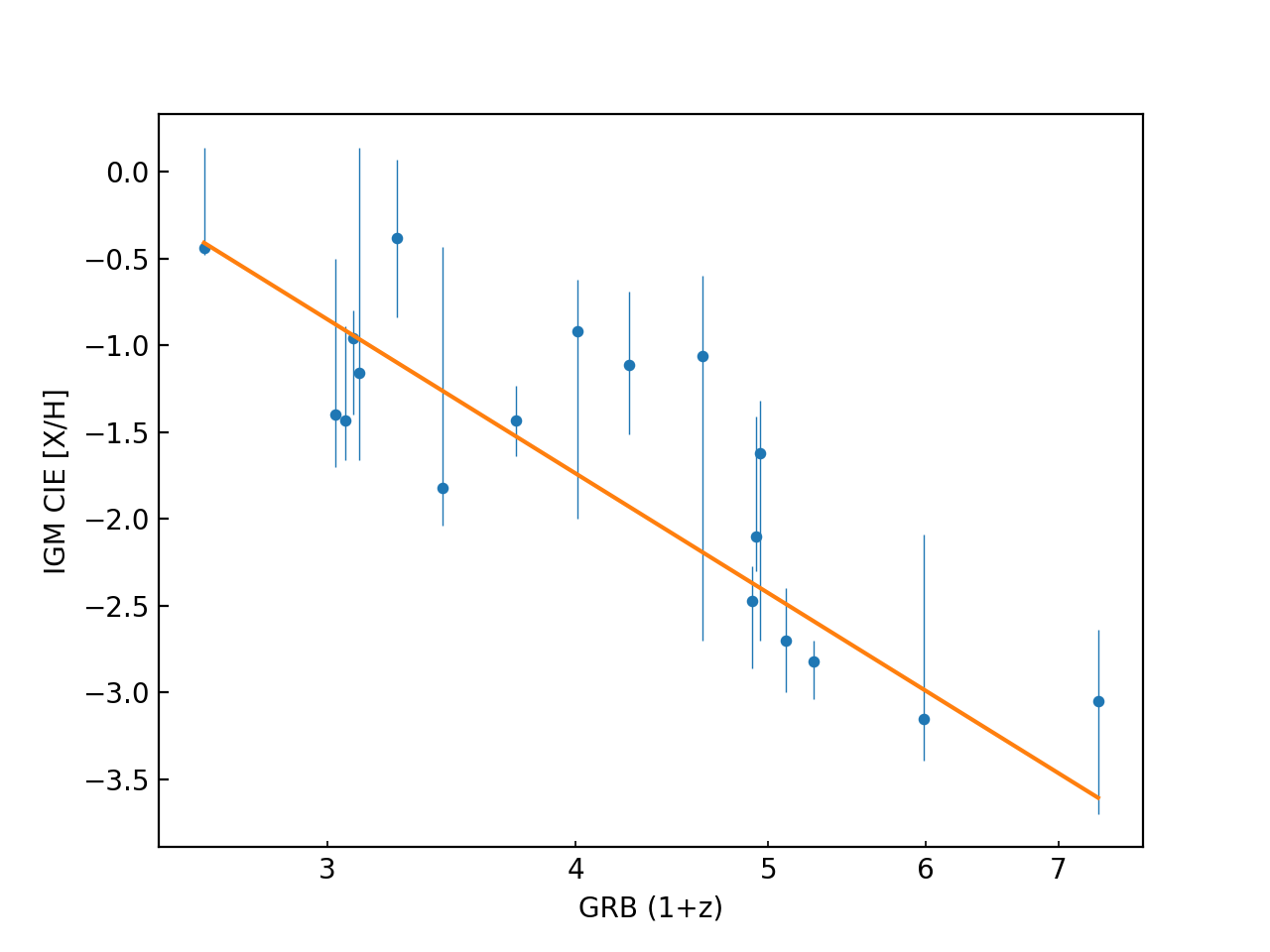} &
    \includegraphics[scale=0.4]{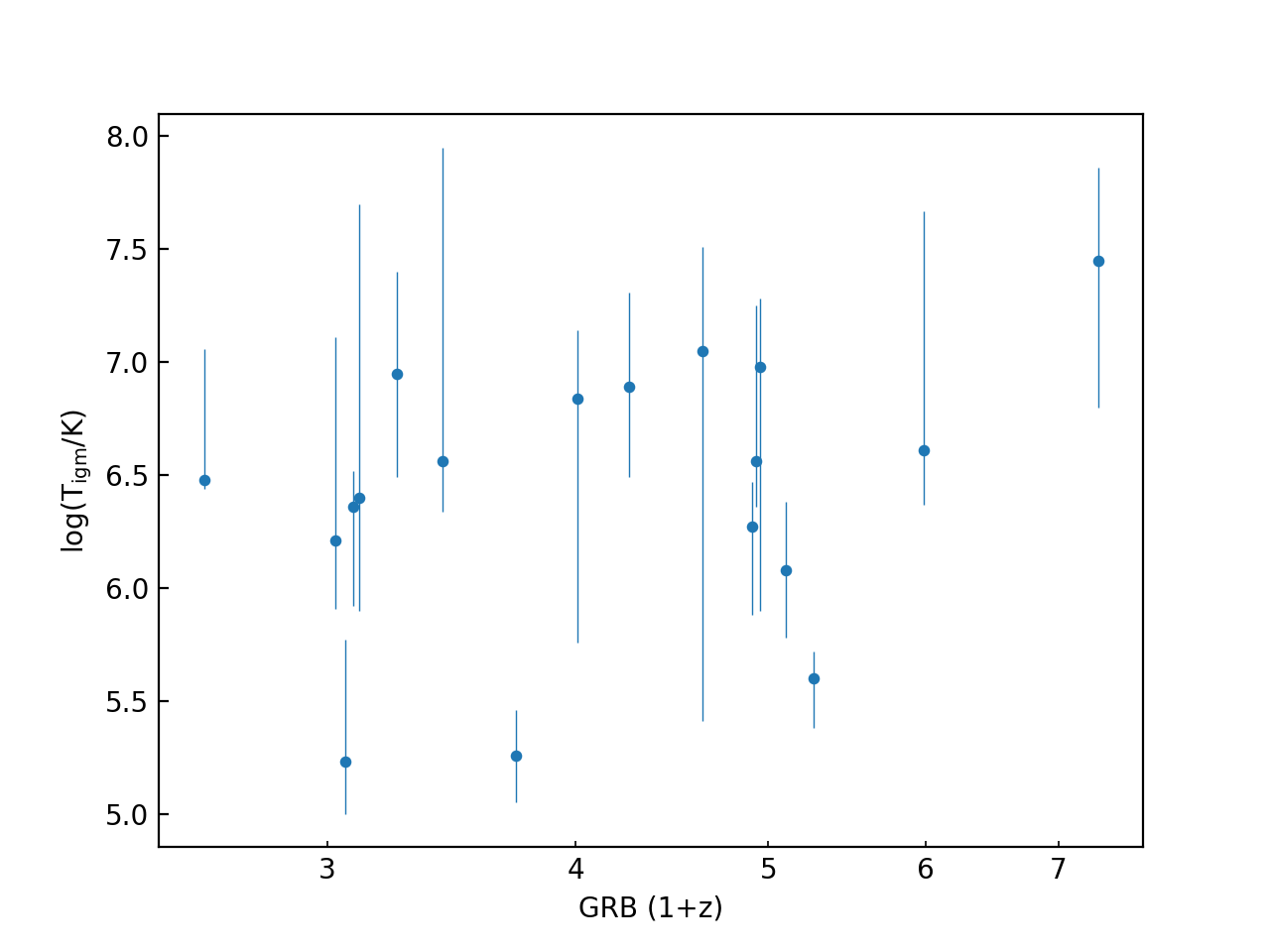} \\
    \end{tabular}
    \caption{Results for the IGM parameters  with $\mathit{N}\textsc{hxigm}$ fixed at the mean IGM density. The error bars are reported with a $90\%$ confidence interval. Top-left panel is $[X/H]$ versus redshift for the PIE model. Top-right panel is ionisation versus redshift for PIE. Bottom-left panel is the CIE model $[X/H]$ versus redshift. Bottom-right panel is log$(T/$K) versus redshift for CIE. The orange line is the $\chi^2$ fit. We do not include a $\chi^2$ curve in the temperature-redshift plot (bottom-right) as the fit had very large uncertainties.}
    \label{fig:nhxmean}
\end{figure*}

\graphicspath{ {./figurespaper2/}  }

\begin{figure*}
     \centering
        
    \includegraphics[width=0.6\columnwidth]{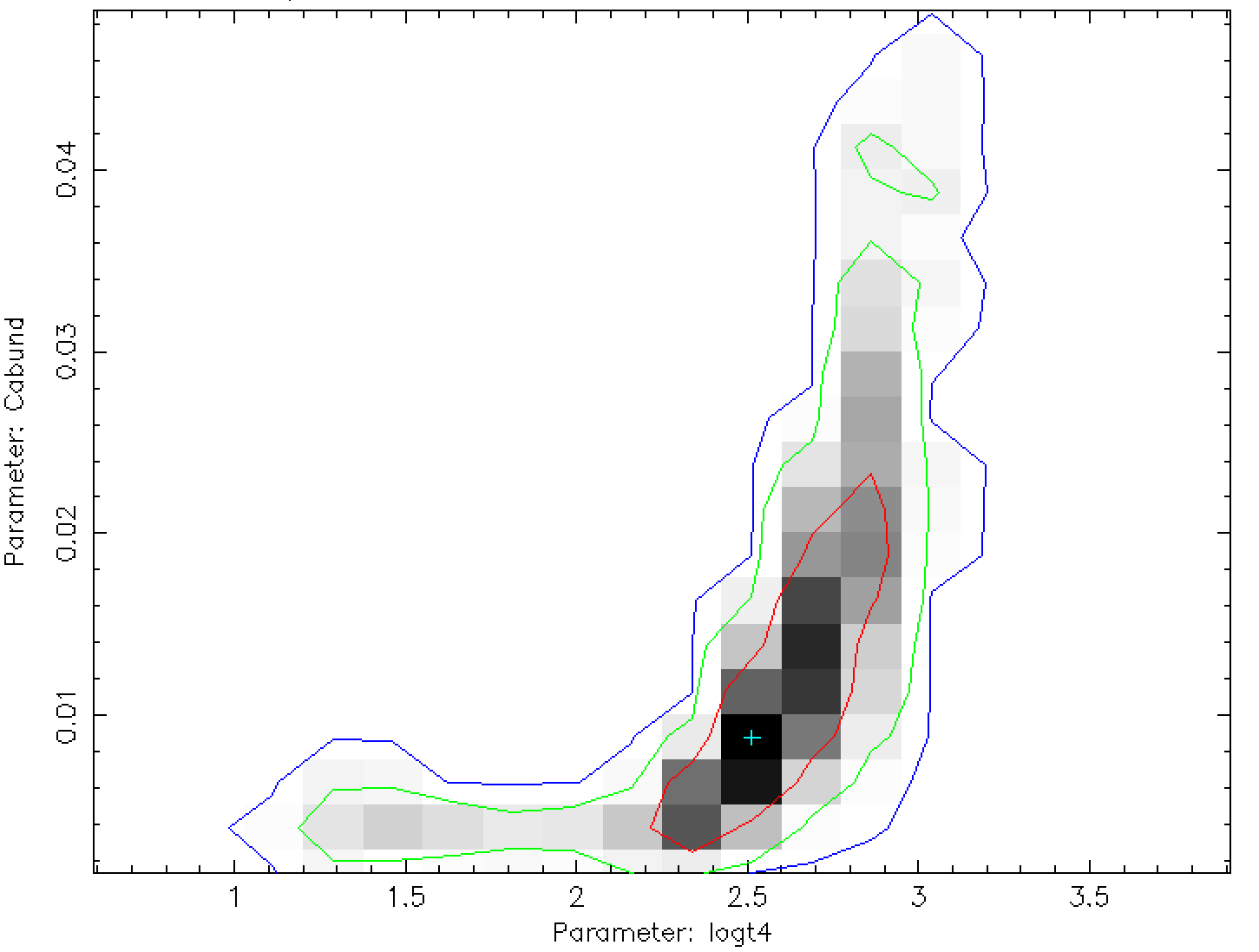}
    \includegraphics[width=0.6\columnwidth]{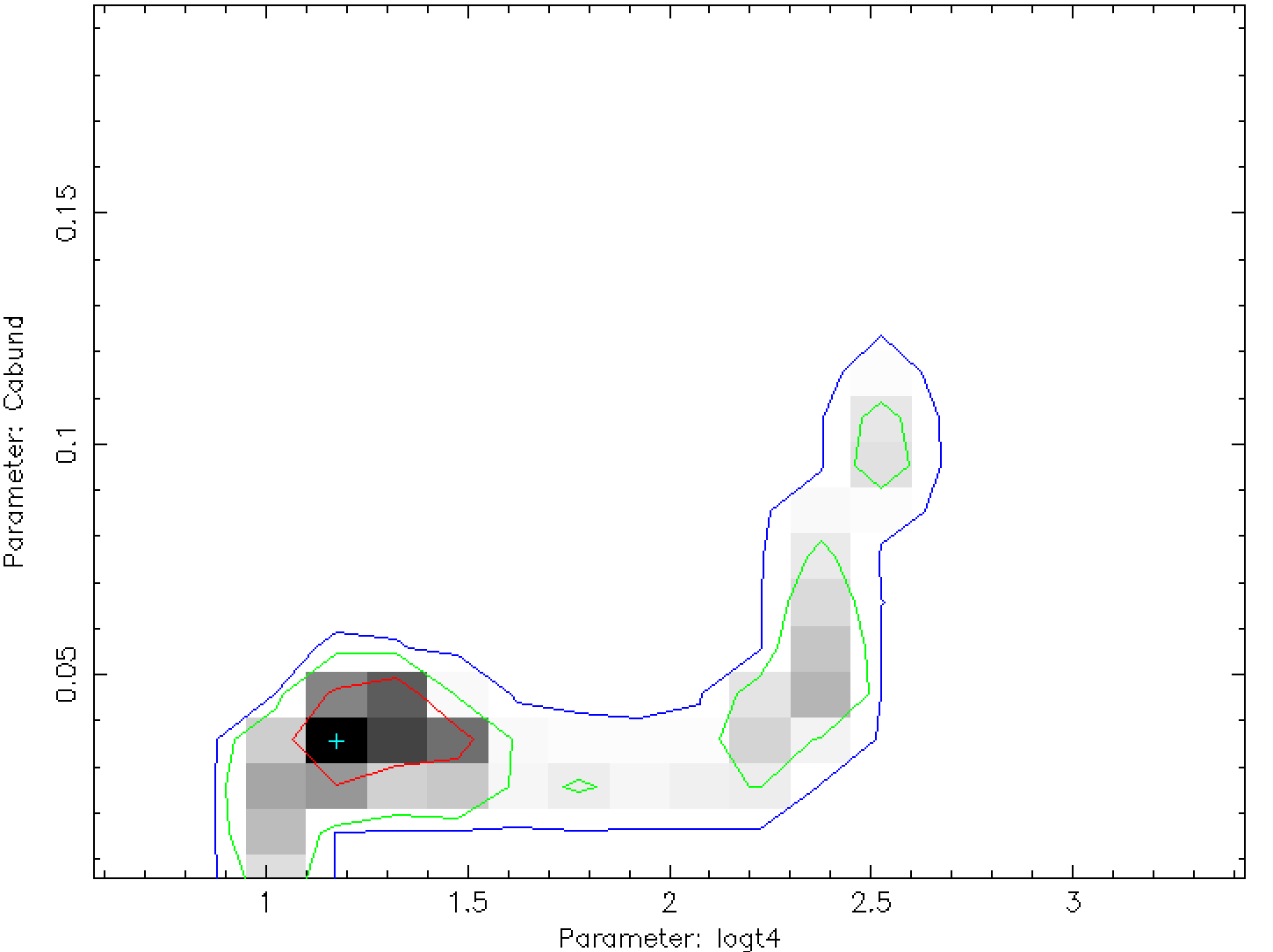}
    \caption{Sample MCMC integrated probability plots with hotabs CIE IGM for two GRB with $\mathit{N}_{\textsc{hxigm}}$ equal to the mean density. The red, green and blue contours represent $68\%, 95\%$ and $99\%$ ranges for the two parameters respectively, with grey-scale showing increasing integrated probability from dark to light. On the x-axis T4 means the log of the temperature is in units of 10$^4$ K. On the y-axis all metals are tied to the $Z/Z\sun$ for Carbon.}
    \label{fig:nhxmean_mcmc}
    
 \end{figure*}
 
 \newpage
 \subsection{Forcing \textit{N}\textsubscript{HXIGM} to equal the mean IGM density}
 
The next approach investigated was to freeze the $\mathit{N}\textsc{hxigm}$ parameter at the value for mean IGM density integrated to the GRB redshift using eq.\ref{eq:simpleIGM} . Metallicity and ionisation parameters (PIE) or temperature (CIE) were free.

For both PIE and CIE, nearly all fits were consistent with $\textsc{steppar}$ and showed good integrated probability plots. There is a requirement for strong metallicity evolution in both scenarios with power law fits to the $[X/$H$]$ versus redshift trend scaling as $(1 + z)^{-9.1\pm0.7}$ and $(1 + z)^{-7.1\pm0.8}$ for PIE (Fig. \ref{fig:nhxmean} top-left panel) and CIE (Fig. \ref{fig:nhxmean} bottom-left panel) respectively. The $90\%$ confidence range was much improved for the $[X/$H$]$ fits as compared with the fixed metallicity scenario fits for $\mathit{N}\textsc{hxigm}$. The ionisation parameter for the PIE fits varied widely between $0 <$ log$(\xi)\/\ < 3$ without any simple trend with redshift. The temperature parameter for the CIE fits also varied widely between $5 <$ log $(T/$K) $< 7.5$ without any simple trend with redshift.

Fig. \ref{fig:nhxmean_mcmc} shows two examples of MCMC integrated probability plots for the CIE scenario. Both show the patterns that most GRB showed in this scenario of $\mathit{N}\textsc{hxigm}$ fixed to the mean density where at high temperature, a range of metallicity could fit, while at low metallicity, there is a range of temperature that could fit.

In conclusion, if the scenario where the average density model of the IGM is valid for the GRB sight lines, it requires strong metallicity evolution for both CIE and PIE. It is not possible to determine which scenario (CIE versus PIE) is more plausible from the fits apart from the fact that the high redshift $z = 6.32$ GRB140505 was well fitted with CIE but not with PIE. The results support the Section 4 free parameter fit model scenarios for both PIE snd CIE IGM and could be interpreted as validity check.

 \newpage
 
 \subsection{Freezing temperature for CIE and ionisation parameter for PIE}
 
 \graphicspath{ {./figurespaper2/}  }

\begin{figure*}
     \centering
     \begin{tabular}{c|c}
    \includegraphics[scale=0.4]{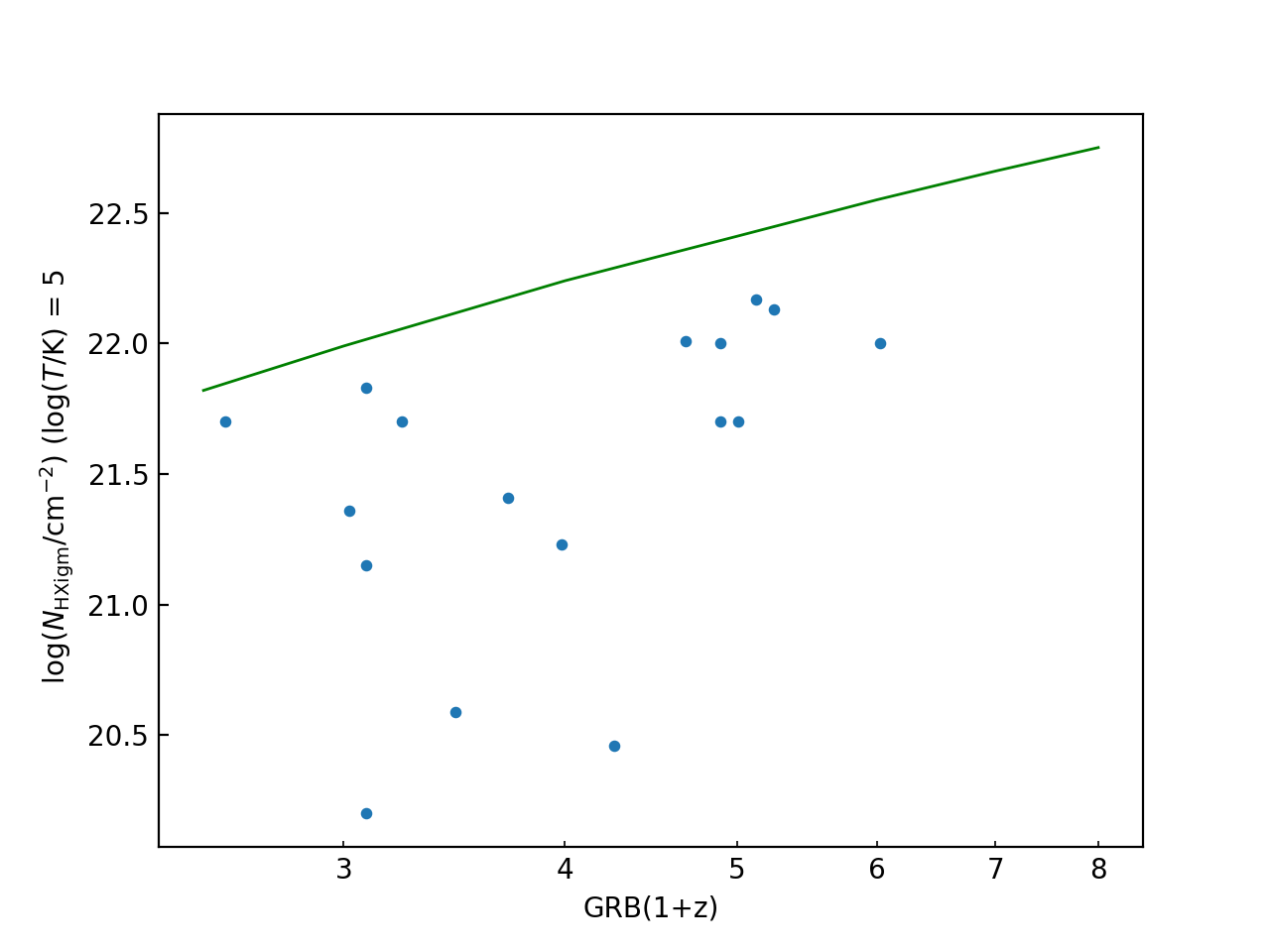} &
    \includegraphics[scale=0.4]{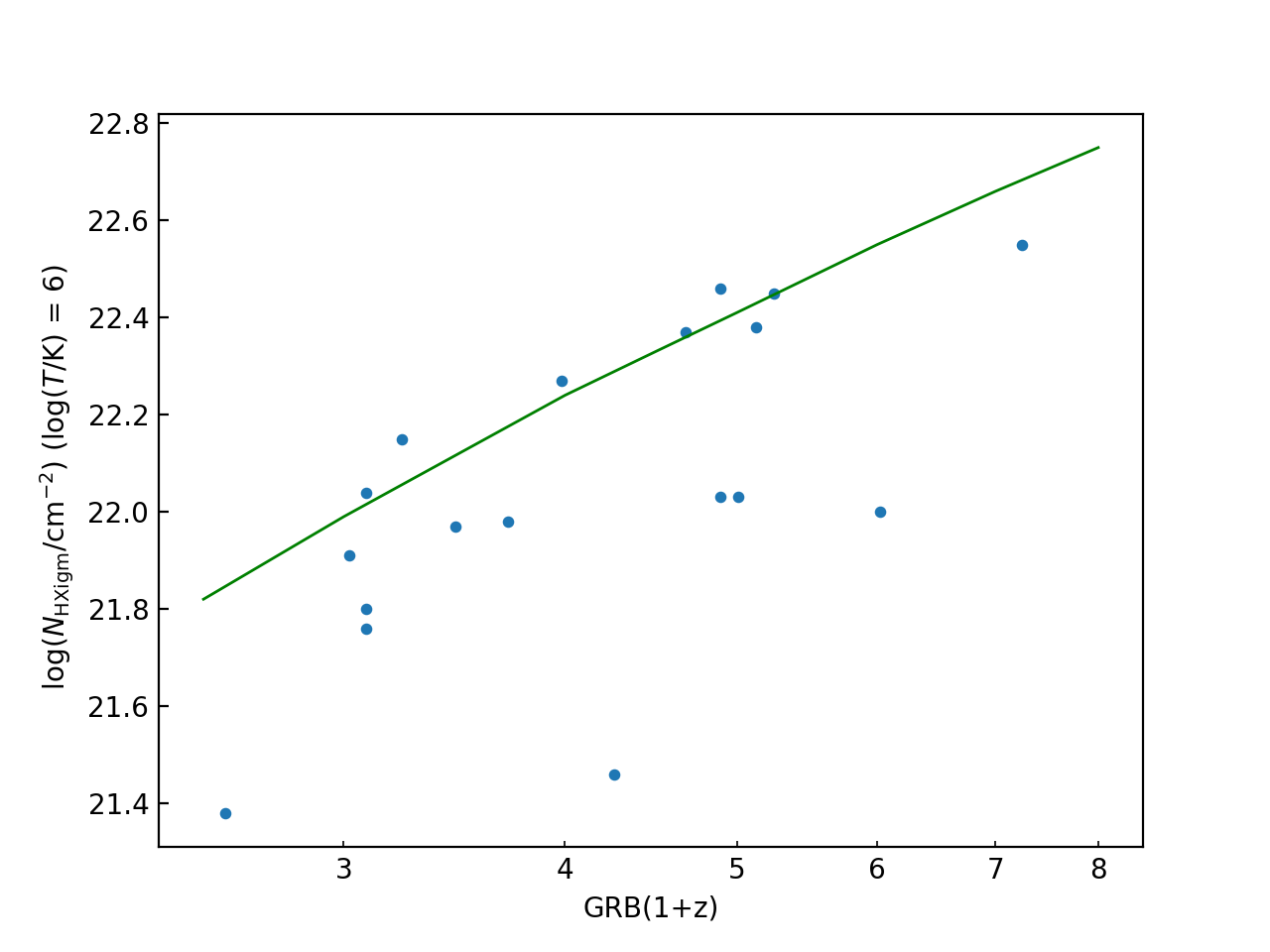}\\
    \includegraphics[scale=0.4]{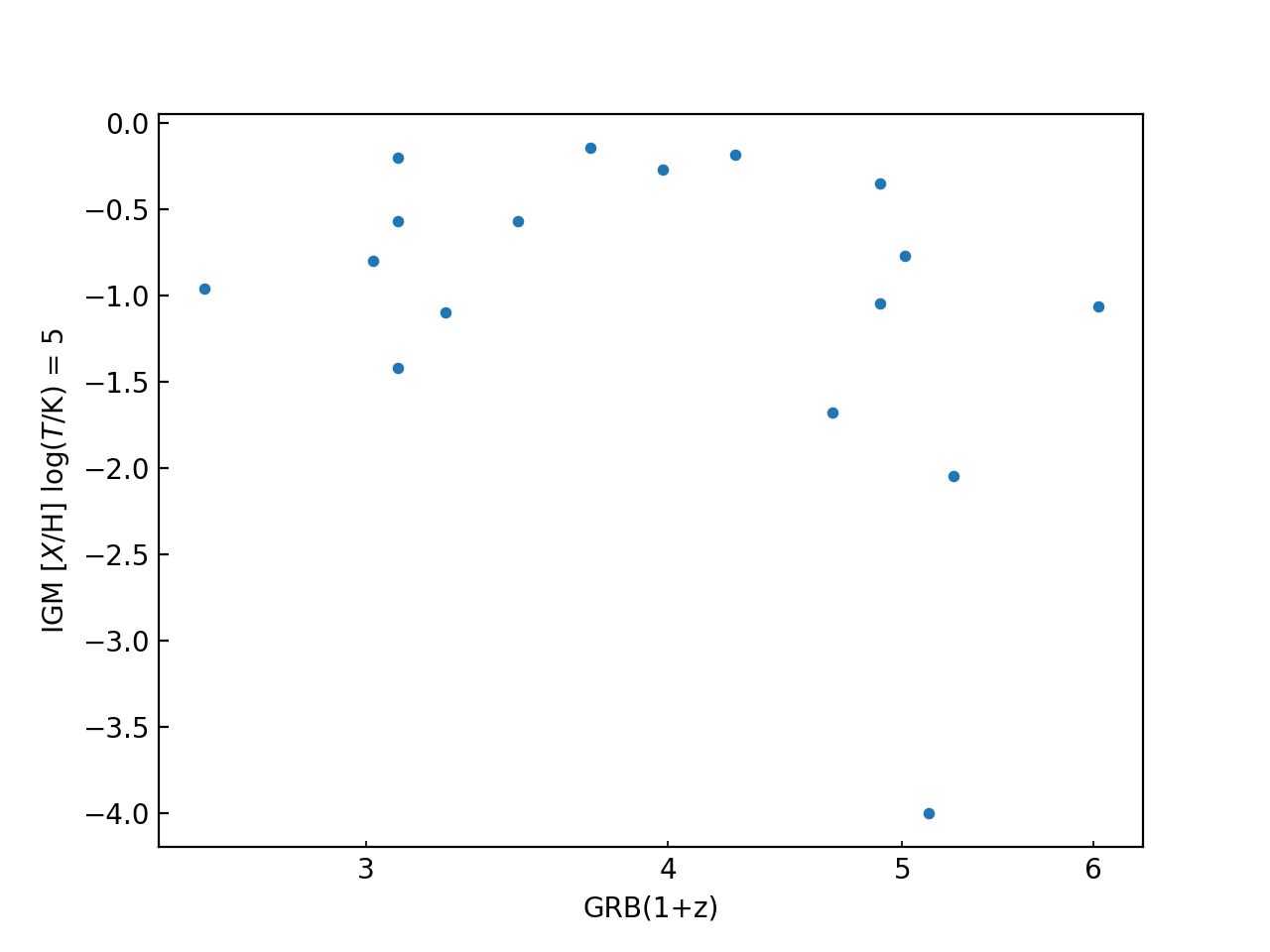} &
    \includegraphics[scale=0.4]{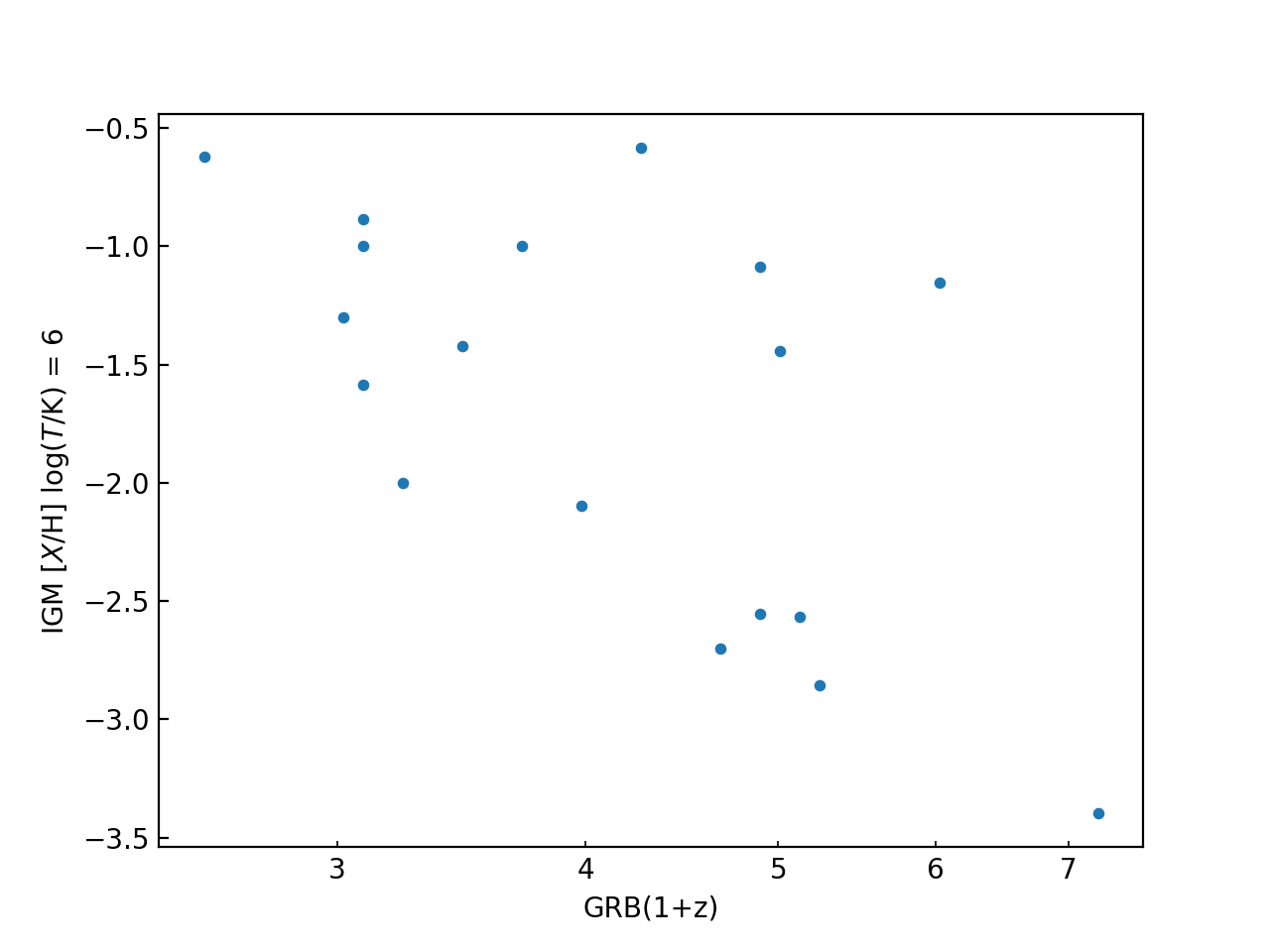}  \\
    \end{tabular}
    \caption{Results for the IGM parameters using the \textsc{hotabs} CIE  model with fixed temperatures. The green curve is the simple IGM model using the mean IGM density. Top-left panel is $\textit{N}\textsc{hxigm}$  versus redshift and bottom-left panel is $[X/$H$]$ versus redshift for the CIE model with log$(T/$K$) = 5$. Top-right panel is $\textit{N}\textsc{hxigm}$  versus redshift and bottom-right panel is $[X/H]$ versus redshift for the CIE model with log$(T/$K$) = 6$. We do not include error bars as the plots are not plausible models.}
    \label{fig:T fixed plots}
\end{figure*}

The next test was to freeze temperature for CIE and ionisation parameter for PIE and leave $\mathit{N}\textsc{hxigm}$ and metallicity free. For temperature in the CIE \textsc{hotabs} model, we froze temperature at log$(T$/K) = 5 and 6 as representative of the cooler and hotter CIE phases.

The fits for $\mathit{N}\textsc{hxgm}$ with temperature fixed at log$(T/$K$) = 5$, are much lower than the mean IGM model in Fig. \ref{fig:T fixed plots} top-left panel, with considerable scatter. In Fig. \ref{fig:T fixed plots} top-right panel with log$(T/$K$) = 6$, some fits are similar to the mean IGM model and show a suggestion of a rise with redshift with some outliers. However, several are well below the mean density. The metallicity plots for both fixed temperatures show no apparent relation with redshift. The higher temperature log$(T/$K$) = 6$ CIE model appears more realistic if the IGM mean density model is appropriate for the IGM. However, it is unlikely that a fixed average temperature approach is appropriate for our CIE IGM modelling.

\begin{figure*}
     \centering
     \begin{tabular}{c|c}
    \includegraphics[scale=0.4]{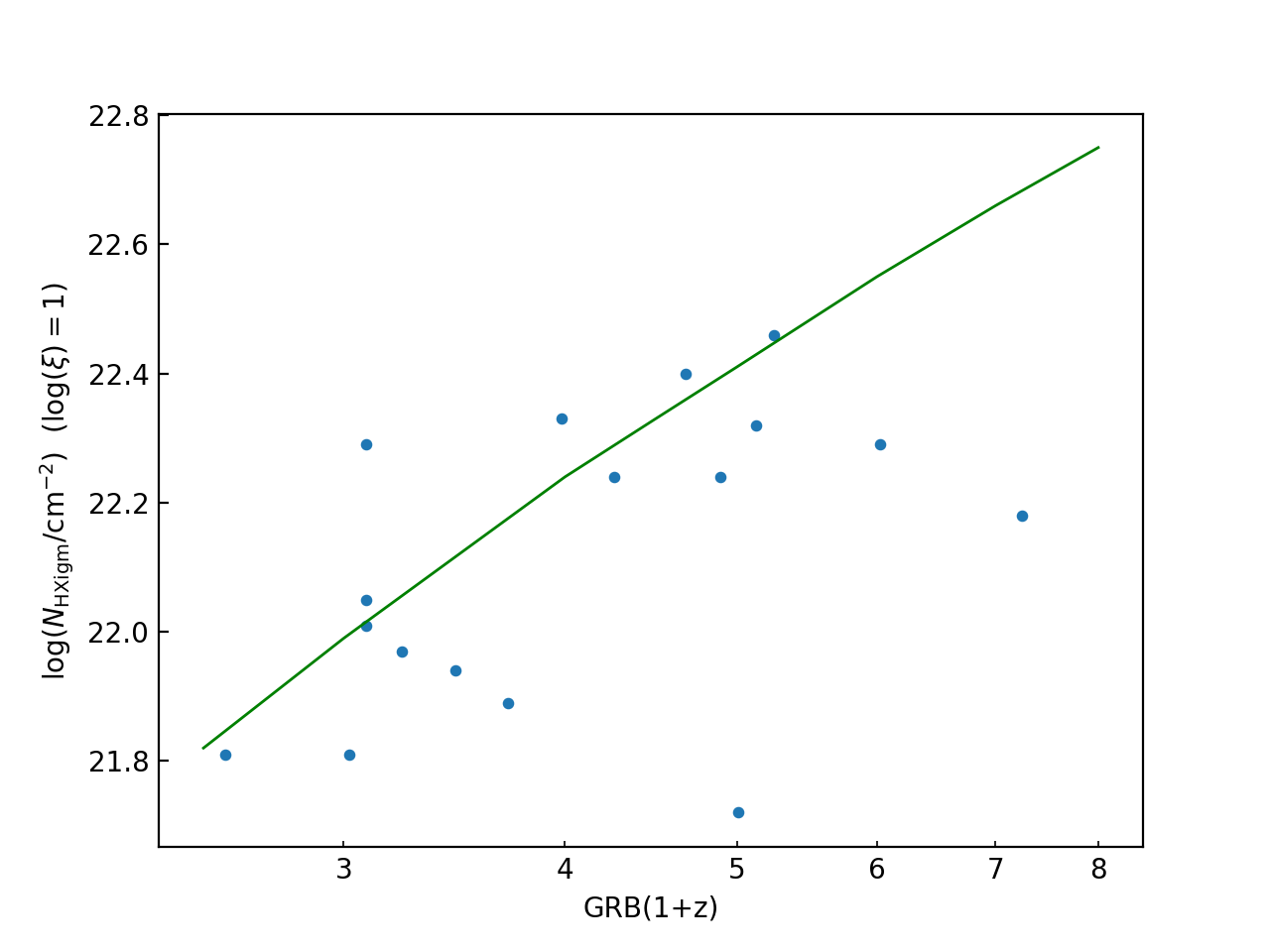} &
    \includegraphics[scale=0.4]{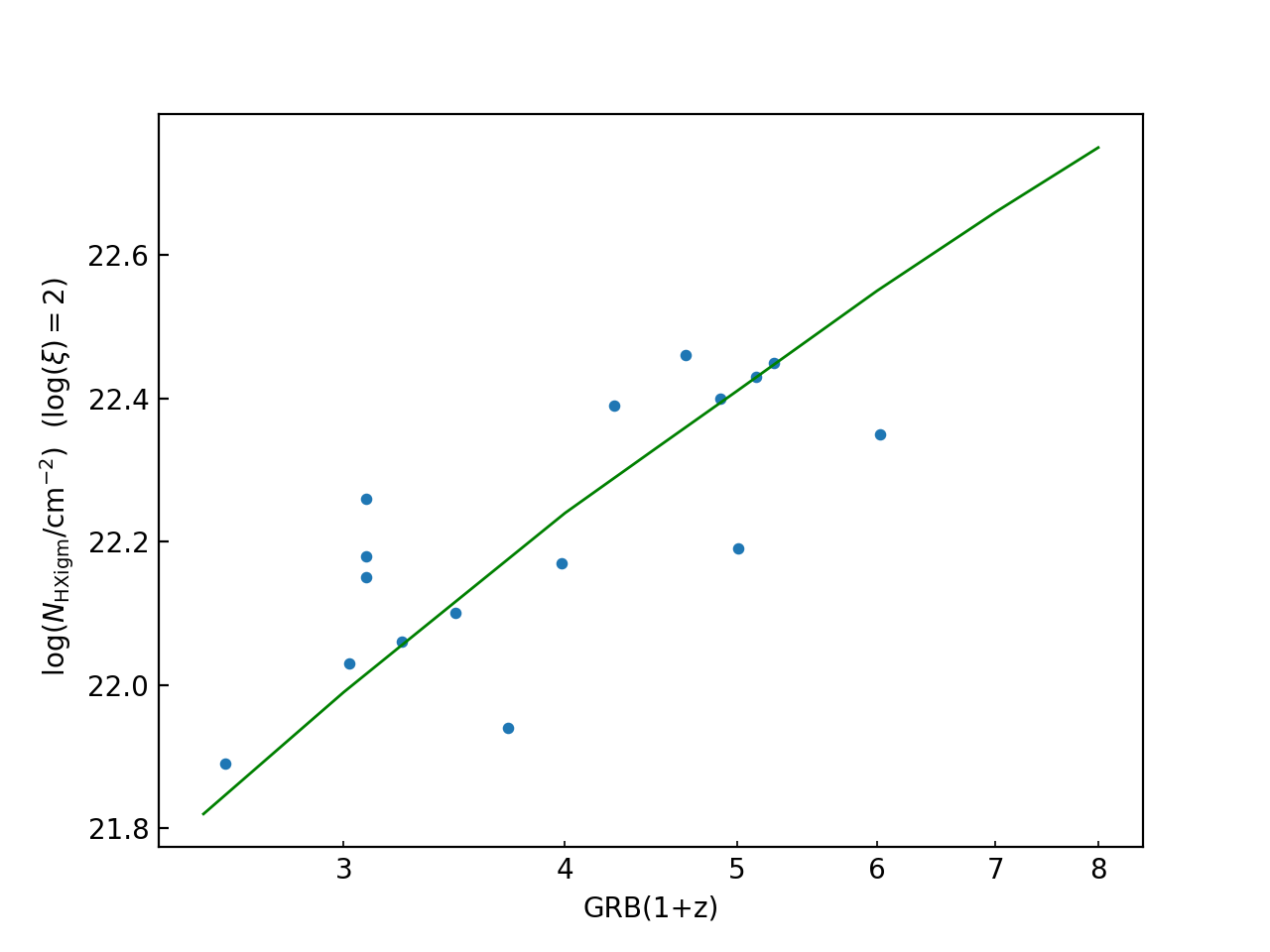}\\
    \includegraphics[scale=0.4]{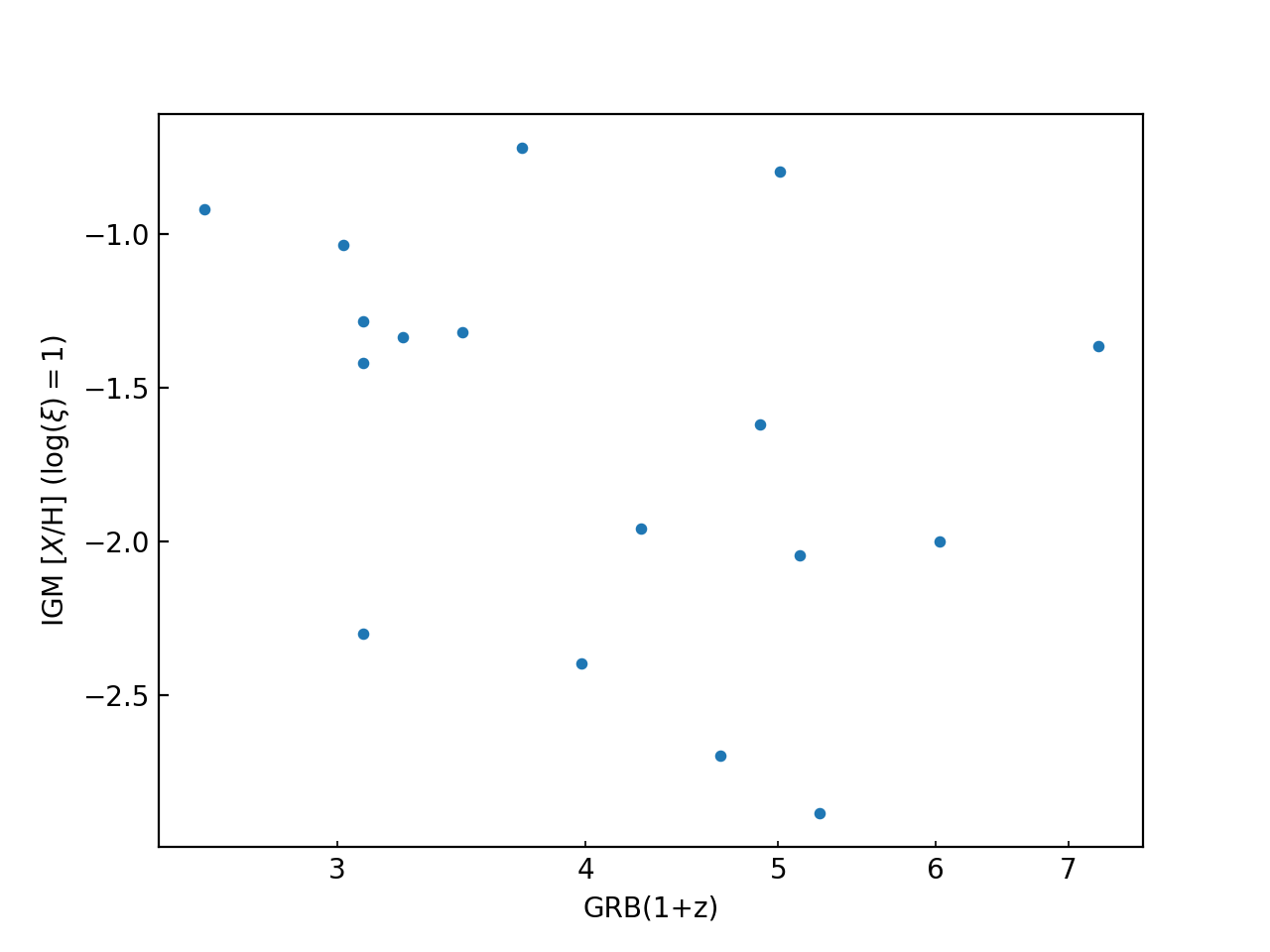} &
    \includegraphics[scale=0.4]{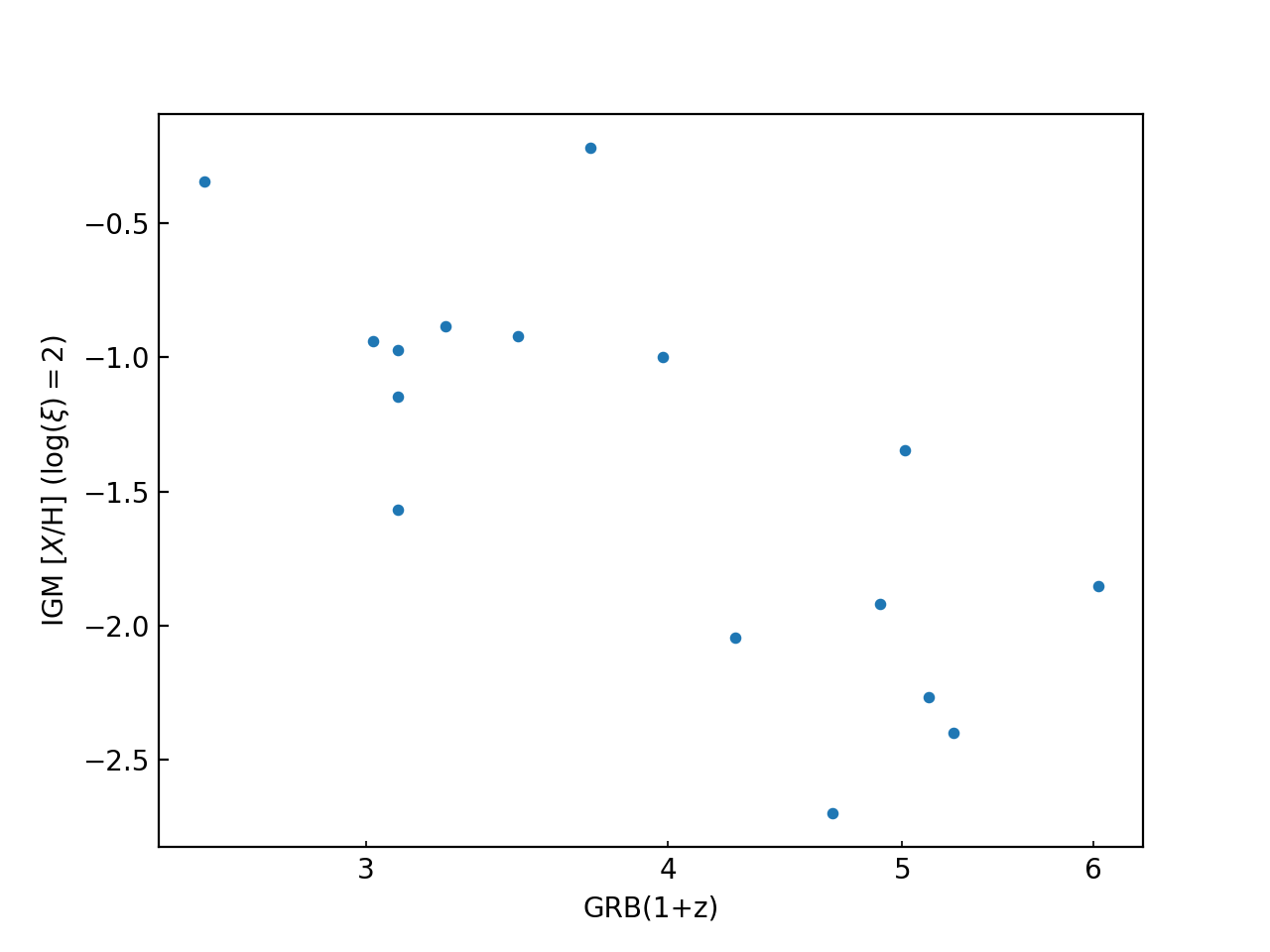}  \\
    \end{tabular}
    \caption{Results for the IGM parameters using the \textsc{warmabs} PIE  model with fixed ionisation parameters. The green curve is the simple IGM model using the mean IGM density. Top-left panel is $\textit{N}\textsc{hxigm}$  and redshift and bottom-left panel is $[X/$H$]$ and redshift for the PIE model with log$(\xi) = 1$. Top-right panel is $\textit{N}\textsc{hxigm}$  and redshift and bottom-right panel is $[X/H]$ and redshift for the PIE model with log$(\xi) = 2$. We do not include error bars as the plots are not meant to be representative of plausible models. }
    \label{fig:xi fixed plots}
\end{figure*}

For PIE, the ionisation parameter was frozen at log($\xi$) = 1 and 2. At both log($\xi$) = 1 and 2, there is a possible $\mathit{N}\textsc{hxigm}$ rise with redshift in Fig. \ref{fig:xi fixed plots} top-left and right panels. Further, the fits for both are similar to the mean density model, with log($\xi$) = 2 being closer. There is a suggestion of metallicity evolution at log($\xi$) = 2. It is not possible to say whether freezing ionisation parameter is a reasonable approach but the fits and overall results for log($\xi) = 2$ are better, with lower Cstat.

In summary, freezing the ionisation parameter gives somewhat more plausible results in the PIE scenarios than the CIE scenarios with fixed temperatures. However, overall, the scenarios with such fixed parameters are not preferred and therefore, we suggest that our free parameter IGM scenarios are more realistic in Section 4.

 The \textsc{warmabs} and \textsc{hotabs} models are more sophisticated than the current version of \textsc{ioneq} and \textsc{absori}, again supporting our model choices in Section 4.

 \graphicspath{ {./figurespaper2/}  }

\bsp	
\label{lastpage}
\end{document}